\documentclass[usegraphicx,usenatbib,usedcolumn,useAMS]{mn2e}

\usepackage{times}

\usepackage{aas_macros}

\newcommand{\sample}[1]{\emph{#1}}

\title[Galaxy Zoo: morphology and colour]%
{\vspace*{-25pt} Galaxy Zoo: the dependence of morphology and
  colour on environment%
  \thanks{This publication has been made possible by
    the participation of more than 100,000 volunteers in the Galaxy
    Zoo project.  Their contributions are acknowledged at
    http://www.galaxyzoo.org/Volunteers.aspx .}}
\author[S. P. Bamford et al.]{%
  Steven~P.~Bamford$^{1,2}$%
  \thanks{E-mail: steven.bamford@nottingham.ac.uk},
  Robert~C.~Nichol$^{1}$,
  Ivan~K.~Baldry$^{3}$,
  Kate~Land$^{4}$,\newauthor
  Chris~J.~Lintott$^{4}$,
  Kevin~Schawinski$^{4,5,6}$,
  An{\v z}e~Slosar$^{7}$,
  Alexander~S.~Szalay$^{8}$,\newauthor
  Daniel~Thomas$^{1}$,
  Mehri~Torki$^{1}$,
  Dan~Andreescu$^{9}$,
  Edward~M.~Edmondson$^{1}$,\newauthor
  Christopher~J.~Miller$^{10}$,
  Phil~Murray$^{11}$,
  M.~Jordan~Raddick$^{8}$,
  Jan~Vandenberg$^{8}$
\smallskip\\
  $^{1}$Institute of Cosmology and Gravitation, University of Portsmouth,
  Mercantile House, Hampshire Terrace, Portsmouth PO1 2EG\\
  $^{2}$Centre for Astronomy \& Particle Theory, University of Nottingham,
  University Park, Nottingham NG7 2RD\\  
  $^{3}$Astrophysics Research Institute, Liverpool John Moores University,
  Twelve Quays House, Egerton Wharf, Birkenhead CH41 1LD\\
  $^{4}$Astrophysics, University of Oxford, Denys Wilkinson Building,
  Keble Road, Oxford OX1 3RH\\
  $^{5}$Department of Physics, Yale University, New Haven, CT06511, USA\\
  $^{6}$Yale Center for Astronomy and Astrophysics, Yale University,
  P.O. Box 208121, New Haven, CT 06520, USA\\
  $^{7}$Berkeley Center for Cosmo. Physics,
  Lawrence Berkeley National Lab. \& Physics Dept.,
  Univ. of California, Berkeley CA 94720, USA \\
  $^{8}$Department of Physics and Astronomy, The Johns Hopkins University,
  Homewood Campus, Baltimore, MD 21218, USA \\
  $^{9}$LinkLab, 4506 Graystone Ave., Bronx, NY 10471, USA\\
  $^{10}$NOAO Cerro Tololo Inter-American Observatory,
  950 North Cherry Avenue, Tucson, AZ 85719, USA \\
  $^{11}$Fingerprint Digital Media, 9 Victoria Close, Newtownards, Co. Down,
  Northern Ireland, BT23 7GY
\vspace*{-15pt}
}

\begin{document}
  
\date{To appear in MNRAS}

\pagerange{\pageref{firstpage}--\pageref{lastpage}} \pubyear{2008}

\maketitle

\label{firstpage}

\begin{abstract}
  We analyse the relationships between galaxy morphology,
  colour, environment and stellar mass using data for over $10^5$ objects
  from Galaxy Zoo, the largest sample of visually classified
  morphologies yet compiled.  We conclusively show that colour and
  morphology fractions are very different functions of environment.
  Both colour and morphology are sensitive to stellar mass. However,
  at fixed stellar mass, while colour is also highly sensitive to
  environment, morphology displays much weaker environmental trends.
  Only a small part of both the morphology--density and colour--density
  relations can be attributed to the variation in the stellar mass
  function with environment.

  Galaxies with high stellar masses are mostly red, in all
  environments and irrespective of their morphology.  Low stellar-mass
  galaxies are mostly blue in low-density environments, but mostly red
  in high-density environments, again irrespective of their
  morphology.  While galaxies with early-type morphology do always
  have higher red fractions, this is sub-dominant compared to the
  dependence of red fraction on stellar mass and environment.  The
  colour--density relation is primarily driven by variations in colour
  fractions at fixed morphology, in particular the fraction of spiral
  galaxies that have red colours, and especially at low stellar
  masses.  We demonstrate that our red spirals primarily include
  galaxies with true spiral morphology, and that they constitute an
  additional population to the S0 galaxies considered by previous
  studies.  We clearly show there is an environmental dependence for
  colour beyond that for morphology.  The environmental transformation
  of galaxies from blue to red must occur on significantly shorter
  timescales than the transformation from spiral to early-type.

  We also present many of our results as functions of the distance to the
  nearest galaxy group.  This confirms that the environmental trends
  we present are not specific to the manner in which environment is
  quantified, but nevertheless provides plain evidence for an
  environmental process at work in groups.  However, the properties of
  group members show little dependence on the total mass of the group
  they inhabit, at least for group masses $\ga 10^{13}
  \mathcal{M}_{\odot}$.

  Before using the Galaxy Zoo morphologies to produce the above
  results, we first quantify a luminosity-, size- and
  redshift-dependent classification bias that affects this dataset,
  and probably most other studies of galaxy population morphology.  A
  correction for this bias is derived and applied to produce a sample
  of galaxies with reliable morphological type likelihoods, on which
  we base our analysis.

\end{abstract}

\begin{keywords}
galaxies: fundamental parameters --- galaxies: structure --- 
galaxies: statistics --- galaxies: evolution --- galaxies: clusters: general
\vspace*{-50pt}
\end{keywords}

\section{Introduction}
\label{sec:intro}
The discovery that galaxies may be naturally classified, simply by
their visual appearance, into two principal types: spiral and
elliptical, came even before these objects were firmly established as
external to our own Galaxy \citep{1922ApJ....56..162H}.  With further
study it became clear that the different galaxy types are very
differently distributed throughout space \citep{1931ApJ....74...43H}.
Over the intervening decades our understanding of galaxies has
improved dramatically.  Nevertheless, explaining the appearance of
galaxies, and how this varies with position, remains a central
concern.

The visual appearance, or morphology, of a galaxy is an indicator of
its current internal structure and kinematics, which in turn are a
result of the galaxy's developmental history.  Galaxies mainly
comprise two structures, a spheroid and a disc.  In the most basic
terms, elliptical galaxies are simply a spheroid, whereas other
morphologies generally comprise a central spheroid along with a larger
disc, which often contains spiral arms.  These disc and spheroid
components appear to develop in very different ways.  That the
distribution of morphologies changes as a function of position in the
universe suggests variability in these developmental processes.  This
may be the result of cosmological variations, or the interaction of
galaxies with their surroundings.  Galaxy morphologies therefore offer
valuable information with which to construct and constrain theories of
galaxy formation and evolution.

The relationships between a galaxy's local environment and its colour
and emission line strengths have been particularly well studied
recently.  Both of these observables are related to star-formation
history, and thus physical processes quite different from those that
determine a galaxy's morphology.  Comparing the behaviour of colour
and morphology versus environment may thus provide powerful clues to
the mechanisms through which galaxies have developed into the
population we see today.

One reason for the current popularity of classifying galaxies by
colour and, to a lesser extent, emission line strength, is the ease
and accuracy with which these quantities can be measured in modern
surveys, such as the Sloan Digital Sky Survey (SDSS). Traditional
morphological classification of galaxies, on the other hand, is
extremely time consuming - requiring a human visual inspection of each
object. Studies have thus far been limited to samples containing
several thousand galaxies
\citep{2004AJ....127.2511N,2007AJ....134..579F}, although the MOSES
project (Morphologically Selected Ellipticals in SDSS;
\citealt{2007MNRAS.382.1415S}) visually inspected nearly 50000
galaxies in order to identify a clean sample of ellipticals. 

Attempts at automatic morphological classifications have been made,
with varying success.  The most common, and arguably most useful, of
these are simple measurements which quantify the radial light profile
of a galaxy, such as concentration and Sersic index
\citep[e.g.][]{2003MNRAS.341...33K,2003ApJ...594..186B}. These
quantities measure the dominance of a spheroid over any disc component
present. However, a spiral galaxy with a bright bulge or nucleus is
still a spiral galaxy. Concentration is more strongly related to
luminosity than morphology, with more luminous galaxies having more
concentrated profiles \citep{2000A&A...361..863G}. These quantities
are therefore not true measures of morphology in the traditional and
most discriminating sense (as discussed in the next paragraph).
Indeed, \citet{2008arXiv0801.1995V} finds very different mass and
environment dependences for galaxy type-fractions based on
concentration versus those derived from an indicator more closely
related to visual morphology.

There are several avenues to more sophisticated automatic
morphologies, some which use physical insight or statistical methods
to naturally classify galaxies, while others aim to directly reproduce
classifications by professional astronomers
\citep[e.g.][]{1995Sci...267..859L,2002ApJS..142....1S,2003ApJS..147....1C,2003MNRAS.346..601G,2004MNRAS.348.1038B,2006MNRAS.371....2A,2006ApJ...644...30B,2006MNRAS.373.1389C,2008MNRAS.383..907B,2007ApJ...658..898P}.
While a number of these approaches are very promising, they do not yet
provide a direct equivalent to traditional visual morphology,
generally relying instead on the correlations between true morphology
and other parameters, such as luminosity, colour and concentration.
The human eye has consistently proved better than computational
techniques at identifying faint spiral structure in noisy images, the
appearance of which is a primary indicator of morphology, along with
the relative luminosity of bulge and disk. The presence and form of
galaxy spiral arms has important physical implications. These patterns
are related to density waves propagating around the disk, and indicate
the dynamical state of the galaxy. They are also an important, though
not exclusive, mechanism for inducing star formation, and thus provide
information on the process by which a galaxy is currently forming it's
stars. In addition, as the visibility of spiral structures may persist
for some time after star formation ceases, their appearance gives an
indication of the timescale of any decline in star formation. Finally,
to have confidence in any automated technique we must compare with a
large number of visually classified objects that cover the full range
of galaxy appearance.

The Galaxy Zoo project was born out of the need for reliable,
visual morphologies for a large sample of SDSS galaxies.  For the
reasons mentioned above, the presently available automated methods were
deemed insufficient for the task.  Our approach was to enlist the
public's help to visually morphologically classify all the galaxies in
SDSS Data Release 6 \citep[DR6,][]{2007arXiv0707.3413A} which were
targeted for spectroscopy; nearly one million objects.  Further
details of the Galaxy Zoo project, including its motivation, design
and the initial stages of the data reduction, are given in
\citet{Chris_GZ}.


A wide range of science is possible with the Galaxy Zoo dataset.  One
initial aim of the project, the investigation of a population of rare,
blue, star-forming, early-type galaxies, is considered in
\citet{Kevin_GZ}.  Another early aim was to measure the statistical
properties of spiral galaxy `spin' orientations, investigated in
detail in \citet{Kate_GZ}.  This present paper makes a start in
exploiting the Galaxy Zoo data to study the dependence of morphology
on a host of galaxy properties.  Here we concentrate on the dependence
of morphology on local galaxy environment, how this compares with
the dependence of colour on environment, and the role of stellar mass.

\subsection{Structure of this paper}

Before describing our own study, we set the scene with a brief
discussion of previous, related work in Sec.~\ref{sec:previous_work}.
Following this, in Sec.~\ref{sec:data}, we provide details of the data
products and sample definitions (\ref{sec:basicdata}), stellar
masses (\ref{sec:stellarmasses}), visual morphologies
(\ref{sec:morphs}) and environmental measures (\ref{sec:data_env})
that we employ in this paper.  Though not essential for understanding
our main results, an important component of this work is a
quantification of the biases present in the morphological
classifications.  These are discussed whilst describing our
morphologies in Sec.~\ref{sec:morphs}, but the full details are
deferred to Appendix~\ref{sec:bias}.

The main results of this paper are presented in
Sec.~\ref{sec:morph-env} \& Sec.~\ref{sec:morph-colour-env}.  Firstly,
in Sec.~\ref{sec:morph-env}, we consider in detail the local
relationships between galaxy visual morphology, environment and
stellar mass.  Then, in Sec.~\ref{sec:morph-colour-env}, we compare
these morphology relations with those derived using colour to divide
the galaxy population, and explore the origins of the differences we
find.  Our conclusions are summarised in Sec.~\ref{sec:conclusions}.

\subsection{Previous studies of morphology and colour versus
  environment}
\label{sec:previous_work}

The classic papers studying the relationships between visual
morphology and environment in the local universe date from the 1980's
and early 1990's
\citep{1980ApJ...236..351D,1984ApJ...281...95P,1991ApJ...367...64W,1993ApJ...407..489W}.
The study of traditional, visual morphology for large samples of local
galaxies has since stood still for the past decade, due to the great
effort required to perform such measurements.  In contrast, as
discussed above, it is now straightforward to measure colours and
simple structural parameters for large galaxy samples, and these
quantities have therefore taken precedence.  It is crucial that we
link recent results on large surveys, such as SDSS, back to the wealth
of earlier, traditional morphology studies on which much of our
understanding is founded.

In order to efficiently sample a wide range of galaxy environments,
many studies have concentrated on the regions in and around rich
galaxy clusters.  In particular, \citet{1980ApJS...42..565D} measured
the morphology of approximately 6000 galaxies in 55 nearby, rich
clusters.  Studying this dataset, \citet{1980ApJ...236..351D} found a
strong relation between local galaxy surface density and morphological
type fractions.  As local density increases, the spiral fraction was
found to fall steadily, the S0 fraction rises in a corresponding
manner, and the elliptical fraction increases sharply at the highest
densities.

The same dataset was revisited by \citet{1984ApJ...281...95P}, who
found that the morphology--density relation extended smoothly to
galaxy group environments identified in the CfA redshift survey
\citep{1983ApJS...52...89H} and by \citet{1982ApJ...257..423H} (see
Fig.~\ref{fig:pg84}).  Both \citeauthor{1980ApJ...236..351D} and
\citeauthor{1984ApJ...281...95P} identified local density, rather than
distance to the centre of the nearest group, as being more closely
related to morphology.  \citet{1991ApJ...367...64W} and
\citet{1993ApJ...407..489W} again reanalysed the
\citet{1980ApJS...42..565D} data, but contrastingly claimed that
groupocentric distance was a better indicator of morphology than local
galaxy density.  These opposing results have yet to be resolved.  We
aim to address this with the Galaxy Zoo dataset in a forthcoming
paper.

Studying the dependence of an automated measure of morphology on both
local density and groupocentric distance in the SDSS Early Data
Release, \citet{2003MNRAS.346..601G} found both relations to be
strong.  Their most striking result is evidence that the fraction of
galaxies with intermediate morphological types (early-type spirals,
Sa-b) increases with galaxy density before falling at the highest
densities.  Late-type spirals (Sc-d), on the other hand, steadily
decrease in fraction with local density, whereas the elliptical
fraction increases sharply at the highest densities.  These findings
were interpreted as a suggestion that multiple mechanisms were at work
in shaping the morphology--density relation.

Many mechanisms have been proposed for the transformation of spiral
galaxies to earlier morphological types in dense environments;
see \citet{2006PASP..118..517B} for a thorough recent review of the
proposed mechanisms and the evidence for them.  The mechanisms may be
crudely divided into those which simply stop star formation, which
indirectly affects morphology by removing the appearance of spiral
arms and reducing the prominence of the disk, and those which affect
the stellar kinematics of galaxies as well as leading to a cessation
of star formation.  While both categories of mechanism can form S0s
from spirals, only the latter can create elliptical galaxies.

With the launch of the \emph{Hubble Space Telescope}, studies of
morphology versus environment have focused on measuring the evolution
with redshift
\citep{1997ApJS..110..213S,1997ApJ...490..577D,1999ApJ...518..576P,2003ApJ...591...53T,2005ApJ...620...78S,2005ApJ...623..721P}.
Trends in morphological fractions versus environment are found at $z
\sim 0.5$--$1$ that are similar to those measured locally.  The
principal difference is a reduced fraction of galaxies with S0
morphology in distant intermediate- and high-density cluster
environments, balanced by a higher spiral fraction.  Neither
elliptical galaxies nor the field population show significant
evolution.  The static nature of the massive elliptical populations
has been confirmed by dedicated studies of luminous, red galaxies
(LRGs), which find very little evolution in their number density or
properties since $z \sim 1$ \citep{2006MNRAS.372..537W}.

As mentioned above, environmental trends in galaxy colours and
star-formation rates have received much attention recently
\citep{2002MNRAS.331..333P,2002MNRAS.334..673L,2003ApJ...584..210G,2004MNRAS.348.1355B,2004AIPC..743..106B,2006MNRAS.373..469B,2004MNRAS.353..713K,2005ApJ...629..143B,2006MNRAS.366....2W,2007ApJ...664..791B}.
The general result is that the fraction of blue (star-forming)
galaxies decreases with local density, in favour of red (passive)
objects.  Interestingly, the properties of the individual
galaxy sub-populations do not appear to change substantially, while
their fractions vary greatly.  Studies at intermediate redshift find
that the fraction of blue, star-forming objects in clusters has
decreased substantially since $z \sim 0.5$--$1$
\citep{1984ApJ...285..426B,1992ApJS...78....1D,1999ApJ...518..576P}.
This is often associated with the decline of the cluster spiral
population \citep{1998ApJ...497..188C}.

The advent of the halo model \citep{2002PhR...372....1C} has led to a
focus on disentangling the properties of galaxies at the centre of
their dark matter halo and satellites orbiting within a larger halo
\citep[e.g.][]{2008arXiv0805.0002V}.  The models frequently assume
that all satellite galaxies have ceased forming stars.  However, a
recent halo-model analysis of the colour-marked correlation function
has found that a non-negligible fraction of blue satellites, which
depends on satellite mass, is required to reproduce SDSS observations
\citep{2008arXiv0805.1233S}.  Another interesting finding of such
studies is a conformity between the morphologies of satellites and
central galaxies \citep{2006MNRAS.366....2W,2008arXiv0805.0637A}.  In
this paper we do not attempt to separate central and satellite
galaxies, but consider the population as a whole.

Together, all these results suggest that more than one physical
process is responsible for the observed dependence of morphology and
colour on environment: one which only acts at early times and is
responsible for the excess of ellipticals in dense regions, and
ongoing processes that prevent further star formation in ellipticals
and which transform disc galaxies to early-type morphologies.
However, many of the details remain uncertain and speculative.  Even
so, our current models of galaxy formation are unable to
simultaneously reproduce many of the general features of the
observational picture.

\section{Data}
\label{sec:data}

\subsection{Basic galaxy properties}
\label{sec:basicdata}
The basis of the data used in this paper is SDSS DR6
\citep{2007arXiv0707.3413A}. For our main analysis we consider only
those galaxies in the Main Galaxy Sample, extended objects with
$r_{\rmn{Petro}} < 17.77$ \citep{2002AJ....124.1810S}. In addition we
only use galaxies with measured spectroscopic redshifts (with $z >
0.01$). Our sample is therefore incomplete in high density
environments due to fibre collisions \citep{2007MNRAS.379..867V}; the
spectroscopic completeness in rich clusters is estimated to be $\sim
65$\% \citep{2007MNRAS.379..867V,2008ApJS..176..414Y}. Two SDSS
spectrograph fibres cannot be placed closer than $55$~arcsec. However,
the fibre assignments were based solely on target positions, with no
consideration of other galaxy properties, and in cases where multiple
targets could only have a single fibre assigned, the target selected
to be observed was chosen randomly. Incompleteness due to fibre
collisions is therefore independent of galaxy properties. As our
analysis considers only trends in the fractions of objects versus
environment, rather than absolute numbers, fibre collisions have no
significant effect on our analysis. We refer to this sample as our
\sample{full sample}; it contains $565798$ objects and is used in
demonstrating the classification bias in Sec.~\ref{sec:morphs} and
deriving corrections for it in Appendix~\ref{sec:bias}.

We further restrict our main analysis to a redshift range $0.03 < z <
0.085$. We refer to this sample as our \sample{magnitude-limited
  sample}, containing $192960$ objects. The lower redshift limit
ensures that the morphological classification bias correction is
stable, as explained in Sec.~\ref{sec:morphs} and
Appendix~\ref{sec:bias}. The upper redshift limit is a compromise
between the number of galaxies in the sample, the luminosity range
over which we are volume-limited ($M_r < -20.17$~mag), and the
reliability of the local density estimates.

In order to combine or compare galaxies across a range of redshifts we
must account for the redshift dependent selection biases. To remove
selection bias from the analyses in this paper we restrict the
galaxies considered to those that would meet our apparent magnitude,
size and surface brightness criteria if they were located at the upper
limit of the redshift range considered. As measures of galaxy size
and surface brightness we use the radius containing 50 per cent of the
Petrosian flux, $R_{50}$, and the average surface brightness within
this radius, $\mu_{50}$, all from the $r$-band imaging.  Given the
upper redshift-limit we adopt, $z < 0.085$, the Main Galaxy Sample
limits ($r < 17.77$~mag, $R_{50} \ga 1$~arcsec and $\mu_{50} \la
23.0$~mag~arcsec$^{-2}$), and our assumed cosmology (see below), we
thus limit to the subsample of \sample{magnitude-limited sample}
galaxies with $M_r < -20.17$~mag, $R_{50} > 1.6$~kpc and absolute
surface brightness $\mu_{50} < -13.93$~mag~kpc$^{-2}$. We refer to
this sample, which contains $125923$ objects, as our
\sample{luminosity-limited sample}.  Figure~\ref{fig:z_dist} (in
Appendix~\ref{sec:bias}) shows the redshift distribution for our
\sample{luminosity-limited sample}, compared with that of
spectroscopically observed objects in the SDSS Main Galaxy
Sample. This sample is used for much of the analysis in this paper,
with the exception of when we wish to explore down to low stellar
masses. In this case we use the \sample{magnitude-limited sample} but
either apply a bin-dependent upper redshift-limit to ensure we are
complete in each mass-bin, or use a $1/V_{\rmn{max}}$ weighting. These
mass-limited samples are fully defined in the following section.

Our photometry is from the SDSS DR6 Ubercal
\citep{2007astro.ph..3454P}. Where subscripts are omitted, we use
model magnitudes for colours and Petrosian magnitudes
otherwise. All magnitudes are on the AB zero-point system and
corrected for Galactic extinction. Absolute magnitudes are determined
using \textsc{kcorrect v4\_1\_4} \citep{2007AJ....133..734B}.

Throughout this paper we assume a
Friedmann-Lema{\^\i}tre-Robertson-Walker cosmology with $\Omega_m =
0.3$, $\Omega_{\Lambda} = 0.7$, $H_0 = 70$ km~s$^{-1}$~Mpc$^{-1}$.

\subsection{Stellar masses}
\label{sec:stellarmasses}

We determine stellar masses from $r_{\rmn{Petro}}$ and
$(u-r)_{\rmn{model}}$ using the relation given in figure~5 of
\citet{2006MNRAS.373..469B}, which is calibrated on the
spectrally-determined stellar masses of \citet{2003MNRAS.341...33K}
and \citet{2004Natur.430..181G}.  Being based on a single colour,
these stellar masses may be expected to be of limited accuracy.  They
display a 0.13 dex scatter around the spectrally-derived masses on
which they are calibrated.  However, this is comparable to the quoted
uncertainties on the spectral measurements \citep{2003MNRAS.341...33K}
and the scatter and systematic differences between different spectral
mass estimates \citep{2005MNRAS.362...41G}.  On average, our stellar
masses therefore have a comparable accuracy to spectrally-derived
estimates.  In any case, our colour-based estimates provide a reliable
way to rank galaxies, and thus examine trends, with respect to stellar
mass.

With the adopted limits of our \sample{luminosity-limited sample}, $z <
0.085$ and $M_r < -20.17$~mag, we become incomplete for red galaxies
below a stellar mass of $\sim 10^{10.3} \mathcal{M}_{\odot}$. In order
to explore a wider range when binning by stellar mass we must ensure a
high level of completeness in each bin. We achieve this, only when
considering bins of stellar mass, by lifting the sample absolute
magnitude limit, i.e. using the \sample{magnitude-limited sample}, but
further limiting the redshift range of objects which contribute to
each bin, such that $\la 0.1$ per cent of galaxies in each stellar
mass range are omitted due to their being fainter than the survey
apparent magnitude limit. For example, in the bin with
$\log(\mathcal{M}_{\ast}/\mathcal{M}_{\odot}) = 9.9$--$10.1$, 99.9 per
cent of galaxies are bluer than $(u-r) = 2.96$, and hence have
$\log(\mathcal{M}_{\ast}/L_{r}) < 0.37$ in solar units (from the
calibration of \citealt{2006MNRAS.373..469B}). For
$\log{\mathcal{M}_{\ast}/\mathcal{M}_{\odot}} = 9.9$ this corresponds
to an absolute magnitude limit of $M_r < -19.18$. The survey apparent
magnitude limit of $r < 17.77$ and assumed cosmology thus require that
we restrict objects in this stellar mass bin to $z < 0.055$. We also
ensure completeness in terms of size and surface-brightness in the
same manner as for the \sample{luminosity-limited sample}, by limiting
these quantities to the range of values observable at the highest
redshift considered, $z < 0.085$. Given the low redshift limit of our
\sample{magnitude-limited sample}, $z \ge 0.03$, we are able to study
galaxies with stellar masses down to
$\log{\mathcal{M}_{\ast}/\mathcal{M}_{\odot}} = 9.5$ in statistically
useful numbers. We refer to this as our \sample{binned mass-limited
  sample}.

Occasionally we will wish to produce a relation for a sample which is
complete down to a low stellar mass limit, but for which limiting the
redshift range as described above would leave us with too few
high-mass galaxies. In this case we use the \sample{magnitude-limited
  sample}, but weight each object by $1/V_{\rmn{max}}$, where
$V_{\rmn{max}}$ is the volume in which it would have been possible to
observe that object given its absolute magnitude, size, and surface
brightness, and given the survey limits on the corresponding apparent
quantities. With this weighting we can limit to
$\log{\mathcal{M}_{\ast}/\mathcal{M}_{\odot}} > 9.8$ without excessive
uncertainty, and refer to this as our \sample{$V_{\rmn{max}}$-weighted
  mass-limited sample}.

\subsection{Galaxy Zoo morphologies}
\label{sec:morphs}

The morphologies utilised in this paper are derived from
classifications by over 80,000 members of the international public as
part of the Galaxy Zoo project.  This project is described in detail
by \citet{Chris_GZ}.  Briefly, each galaxy received several,
independent morphological classifications, each by a different user.
The four possible classifications were labelled as: `elliptical',
`spiral', `don't know', and `merger'.  The `elliptical' class, in
addition to containing galaxies with elliptical morphology, also
contains the majority of S0 galaxies, as will be shown later in this
section.  We therefore refer to it henceforth as the early-type class.
The merger class is mainly comprised of interacting pairs, generally
with tidal features, which may or may not be expected to eventually
merge.  It also suffers from some degree of contamination from galaxy
pairs that are overlapping in projection, but not physically
related.  This is of no consequence to the present paper, however, as
only a small fraction ($< 1$ per cent) of objects are identified as a
merger by a majority of classifiers, and these classifications are not
specifically considered further herein.  The spiral classification was
subdivided into `clockwise', `anti-clockwise', and `edge-on/unsure',
referring to the direction and orientation of the spiral arms.  This
was primarily for use in studies of galaxy spins
\citep{Kate_GZ,Anze_GZ}.  However, it also provides us with an
indication of whether a galaxy was classified as spiral due to either
noticeable spiral structure or a disky, edge-on appearance.

The median number of classifications per object is 34, with 98 per
cent of our \sample{full sample} having at least 20 classifications
each. These classifications were processed into raw `likelihoods' for
a galaxy being of a given morphological type, directly from the
proportion of classifications for each type. We denote these as
$p_{\rmn{el}}$, $p_{\rmn{sp}}$, $p_{\rmn{dk}}$ and $p_{\rmn{mg}}$ for
early-types, spirals, `don't know' and mergers, respectively.

A choice that must be made is how to go from the measured
morphological-type `likelihoods' to individual morphologies and type
fractions.  This choice depends upon the intended usage of the data.
If one requires reliable morphologies for individual galaxies, or
samples containing just one type, then assigning classifications to
galaxies with type likelihoods above a certain threshold is
appropriate.  The downside of this is that it results in a significant
fraction of the sample being unclassified, these being objects with a
spread of likelihood between two or more types.  The choice of
threshold is somewhat arbitrary, depending on the `cleanness' and size
of the required sample.  The characteristics of samples defined using
Galaxy Zoo thresholded classifications are considered in more detail
by \citet{Chris_GZ}.

\begin{figure*}
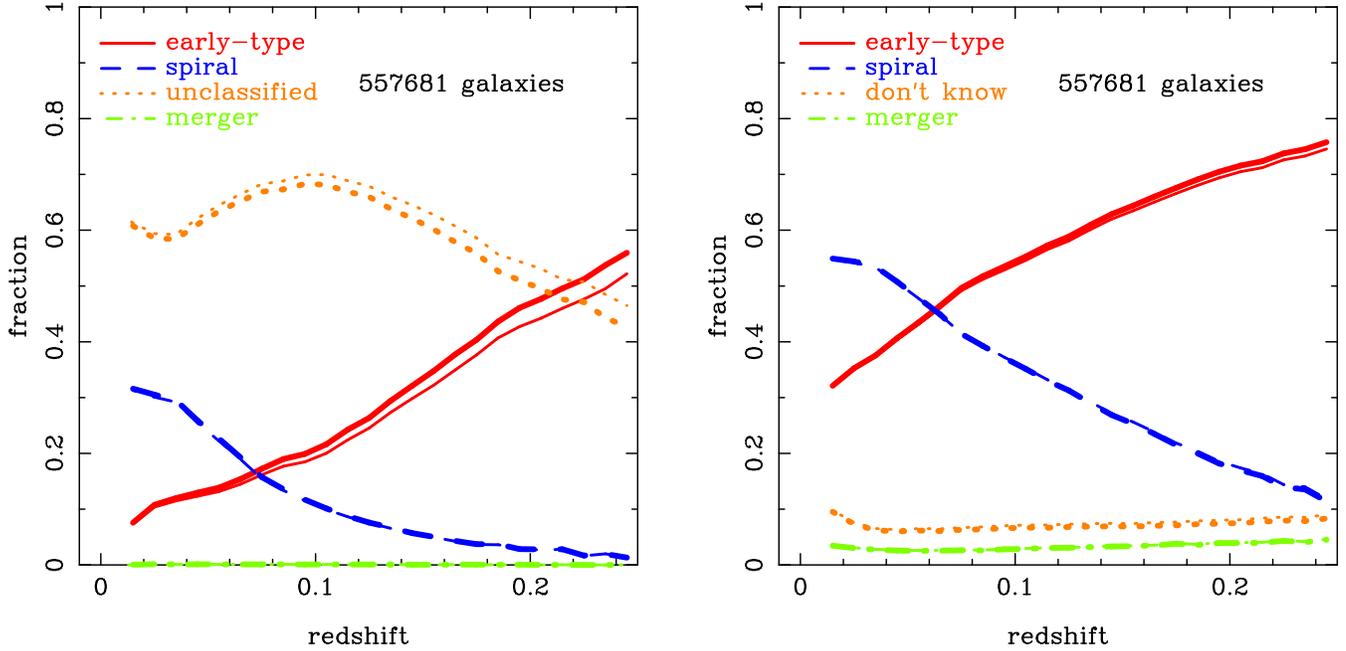

\centering
\includegraphics[height=0.475\textwidth,angle=270]{fig1_left.ps}
\hfill
\includegraphics[height=0.475\textwidth,angle=270]{fig1_right.ps}
\caption{\label{fig:zbin_counts_weights} Morphological type fractions
  from our \sample{full sample} of raw Galaxy Zoo likelihoods, plotted as a function of
  redshift. Classifications are based (left) on a likelihood threshold
  of $p > 0.8$, and (right) directly on the likelihoods themselves.  The
  thick and thin lines corresponds to the weighted and unweighted
  likelihoods, respectively.  "Unclassified" refers to all objects
  which do not meet the threshold of $p > 0.8$ for any class,
  whereas "don't know" refers to the likelihood of an object
  belonging to the "don't know" class, i.e. the fraction of times the
  classifiers clicked the "don't know" button.
  \vspace*{3ex}}
\end{figure*}

Alternatively, if one is interested in type fractions, then the
likelihoods themselves can be used in the statistics. This retains
more information than thresholding, and all galaxies can be included,
but it cannot provide classifications for individual objects.
Figure~\ref{fig:zbin_counts_weights} shows the Galaxy Zoo type
fractions as a function of redshift, determined by both thresholding
($p > 0.8$) and directly using the raw measured likelihoods. Assuming
that there is negligible evolution in the galaxy population over the
redshift interval considered, and that the survey contains a similar
distribution of environments at each redshift, then the true type
fractions should be constant with redshift. As these two assumptions
are expected to be reliable, any trends in the observed GZ type
fractions with redshift may be attributed to a combination of two
biases. The first is selection bias, due to variation in the size and
luminosity distribution of galaxies in our \sample{magnitude-limited
  sample}, resulting from the apparent selection limits. The second is
classification bias, the tendency for otherwise identical galaxies
viewed at different distances to receive different morphological type
classifications, due to signal-to-noise and resolution effects.

Figure~\ref{fig:zbin_counts_weights} also illustrates the difference
in type fractions when weighted or unweighted type likelihoods
are used \citep[see][]{Chris_GZ}.  The effect of the weighting is to
further increase the apparent fraction of early-type galaxies with
increasing redshift.  However, the effect is small and only
significant when thresholded classifications are used.  Throughout the
rest of this paper we use the unweighted type likelihoods, without
thresholding.

\begin{figure*}
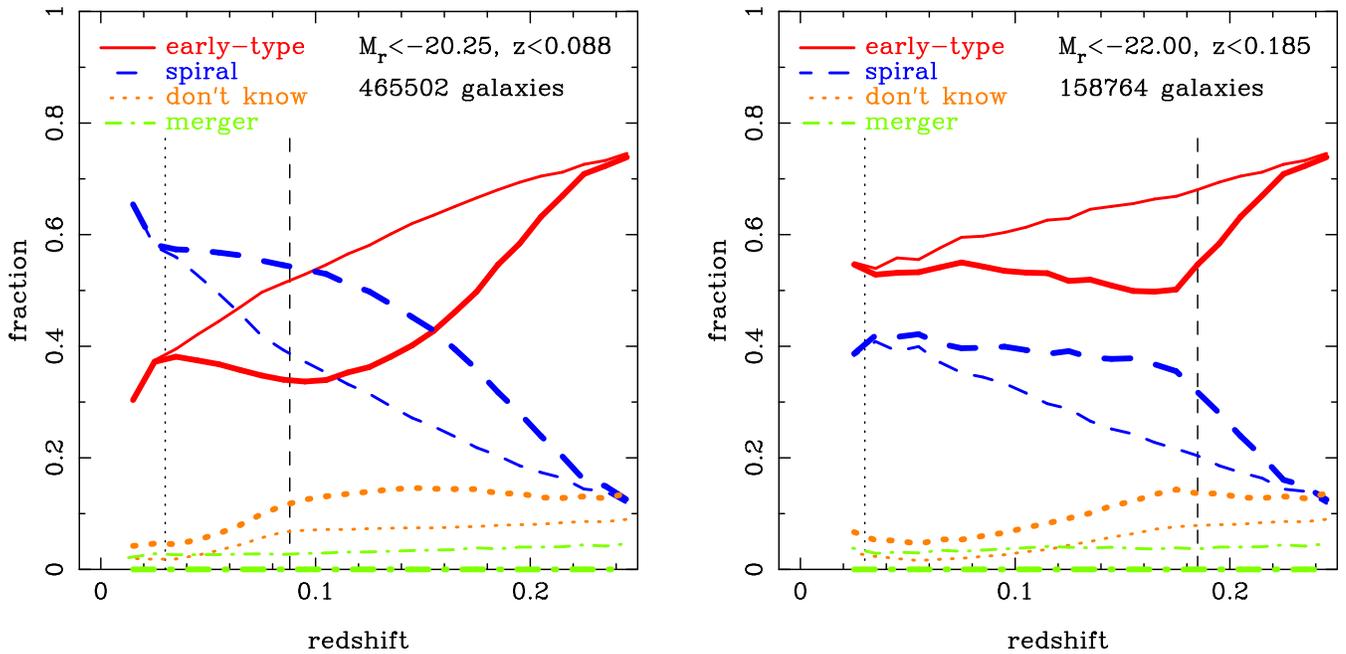

\centering
\includegraphics[height=0.475\textwidth,angle=270]{fig2_left.ps}
\hfill
\includegraphics[height=0.475\textwidth,angle=270]{fig2_right.ps}
\caption{\label{fig:zbin_weights_corrected} Morphological type
  fractions, from our \sample{full sample} of unweighted Galaxy Zoo
  likelihoods, as a function of redshift. The thick and thin lines
  correspond to the de-biased and raw likelihoods, respectively.
  Selection effects have been removed below (left) $z=0.088$ and
  (right) $z=0.185$ by imposing the magnitude limits given in each
  panel. These redshifts are indicated by the vertical dashed lines. The vertical
  dotted line in each panel indicates the low redshift limit ($z \ge
  0.03$) applied to all our main analysis samples. Note that the ratio of
  de-biased early-type and spiral fractions is roughly constant over
  the redshift range for which selection effects are not present. }
\end{figure*}

As trends are clearly present in Fig.~\ref{fig:zbin_counts_weights},
the raw type likelihoods are, in general, biased.  In
Fig.~\ref{fig:zbin_weights_corrected} we attempt to remove selection
effects below given redshifts by limiting in absolute magnitude.  The
thin lines in this figure correspond to the raw type likelihoods.
They still show strong trends with redshift, indicating that
classification bias is present in the raw Galaxy Zoo data.

For otherwise morphologically-identical galaxies, apparently fainter
and smaller objects are more likely to be classified as early-type due
to the diminished spatial resolution and signal-to-noise.  However,
this classification bias can be statistically corrected for, by
applying an adjustment to the raw type likelihoods.  These adjusted
likelihoods, $p_{\rmn{el,adj}}$ and $p_{\rmn{sp,adj}}$, may then be
used to generate type fractions that are unbiased with respect to
galaxy luminosity, size and redshift.  The thresholding approach may
also be based on these de-biased likelihoods, or limited to
considering galaxies in regions of parameter space where the biases
are shown to be small.  The classification bias and our procedure for
removing it are explained in detail in Appendix~\ref{sec:bias}.  For
the remainder of this paper the type fractions we consider are
estimated by averaging the de-biased type likelihoods over bins of
galaxy parameter space.

The performance of our de-biasing procedure may be judged from the
thick lines in Fig.~\ref{fig:zbin_weights_corrected}. These indicate
the type fractions after application of the de-biasing procedure. In
the redshift range that is free from selection effects the de-biased
type fractions are approximately flat and, in particular, the ratio of
early-types to spirals is constant. This indicates that the de-biasing
procedure is working as anticipated. The downturn in early-type
fraction at low redshift for the sample with the fainter limit is
likely due to two effects. The small volume at low redshifts leads to
a reduced occurrence of bright, and hence preferentially early-type,
objects. In addition, the de-biasing procedure relies on the lowest
redshift bins to establish a baseline luminosity--size distribution
against which higher redshifts are compared. Therefore, the lowest
redshift bins cannot be entirely corrected by the procedure we adopt.
To avoid this issue with our bias correction we limit our main
analysis samples to $z \ge 0.03$. This conservative cut reduces the
number of objects in our \sample{luminosity-limited sample} by only 3 per cent.

\begin{figure}
\centering
\includegraphics[width=0.45\textwidth]{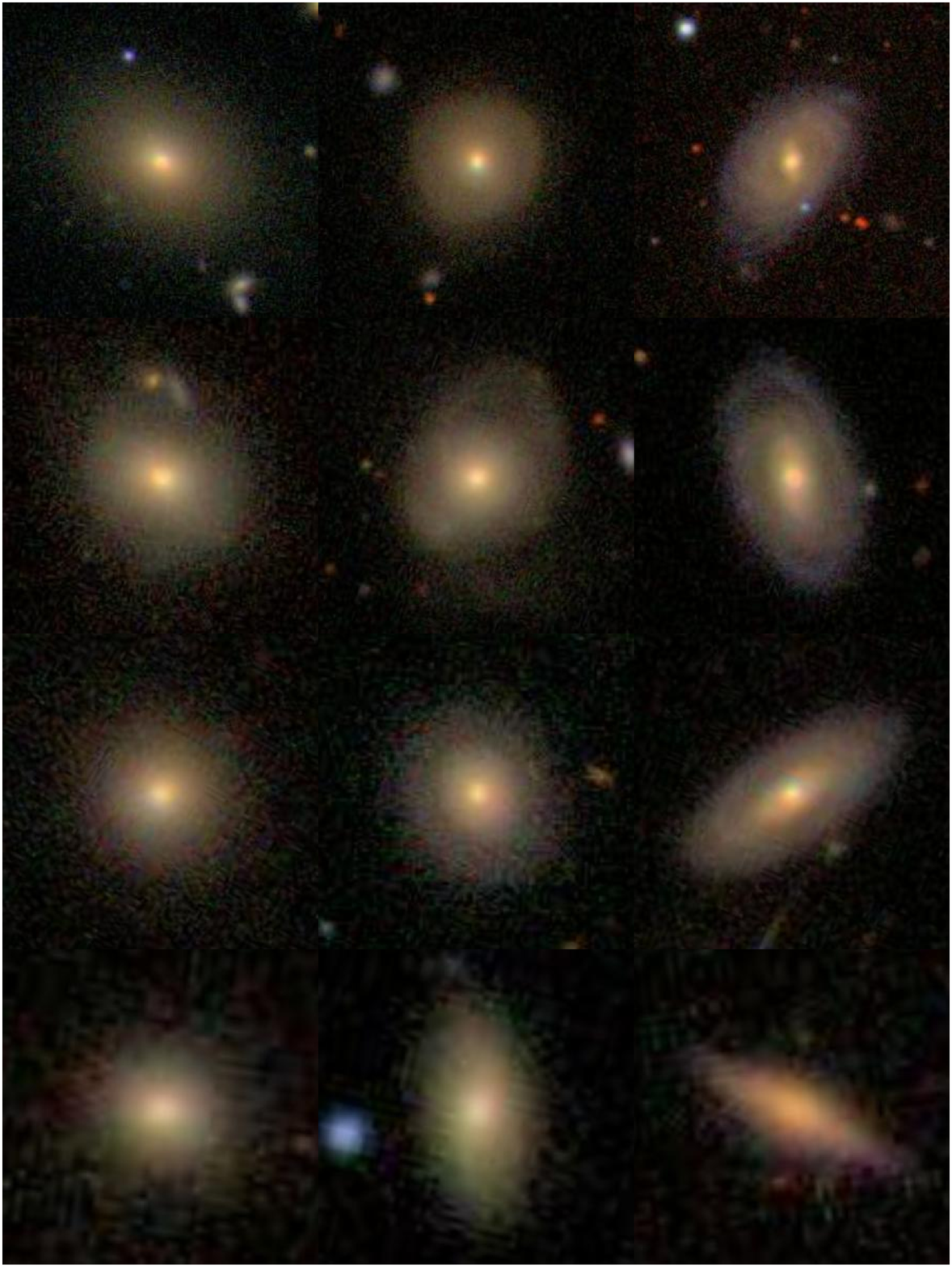}
\caption{\label{fig:examples} Example images illustrating the
  appearance of galaxies with different morphological-type
  likelihoods. Objects in the left, centre and right columns have
  $p_{\rmn{sp}} \sim 0.1$, $0.5$ and $0.9$, respectively.  All objects
  are at a redshift $z \sim 0.05$, but each row corresponds to a
  different luminosity--size bin, from top to bottom: $(M_r, R_{50}) =
  (-22.625, 7.75), (-22.375, 5.75), (-21.875, 4.25), (-20.375, 2.25)$,
  corresponding to the white dots in
  Fig.~\ref{fig:ratio_baseline_fit}.  These bins lie in the
  transition region between early-types and spirals in
  luminosity--size space, and are therefore not representative of our
  general sample, but provide a convenient comparison of otherwise
  similar objects with different morphological-type likelihoods.  Each
  image is scaled to $10 R_{50}$ on a side.}
\end{figure}

To give an impression of the how the galaxy images translate into
morphological-type `likelihoods', Fig.~\ref{fig:examples} shows
examples for $p_{\rmn{sp}} \sim (0.1, 0.5, 0.9)$ for four different
luminosity--size bins at fixed redshift. One can see that the
appearance of spiral or disk features become clearer with increasing
$p_{\rmn{sp}}$.

Finally, we highlight that S0 galaxies are not considered as a
separate class in this work, due to the difficulty in discriminating
them from the other classes and our initially conservative
expectations of the abilities of our classifiers.  Face-on S0s are
difficult to distinguish from elliptical galaxies, whilst edge-on S0s
may appear the same as edge-on spirals.  Even those who are highly
experienced in morphological classification struggle with these
distinctions.

\begin{figure*}
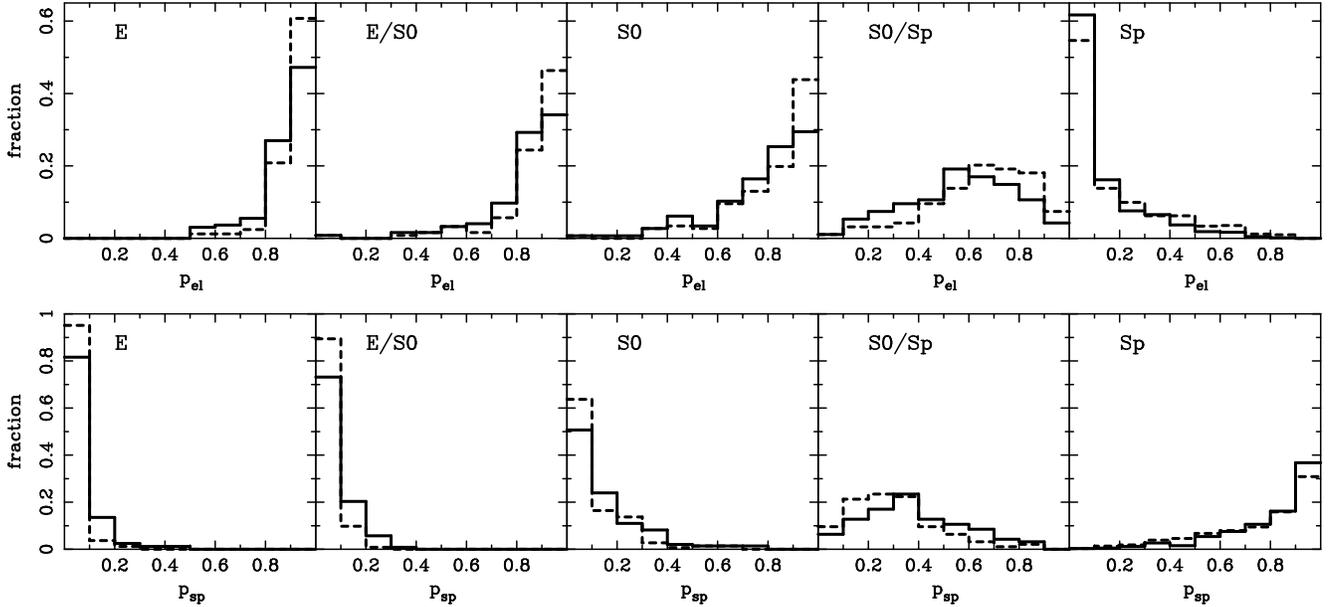

\centering
\includegraphics[height=0.99\textwidth,angle=270]{fig4_top.ps}\\[5pt]
\includegraphics[height=0.99\textwidth,angle=270]{fig4_bottom.ps}
\caption{\label{fig:p_distribs} Distribution of (top) early-type and
  (bottom) spiral morphological type likelihoods for galaxies in our
  \sample{luminosity-limited sample} with classifications from
  \citet[][F07]{2007AJ....134..579F}.  Each panel contains objects
  classified by F07 as having the indicated morphological type. The
  dashed and solid lines correspond to the raw and de-biased type
  likelihoods, respectively.}
\end{figure*}

The Galaxy Zoo classification scheme specifies that objects with
visible spiral arms or an edge-on appearance should be classified as
spirals, the remainder should generally be assigned to the early-type
classification.  The `don't know' option is available for those who
decide the image quality is insufficient for them to make the
distinction, or for objects that do not appear to be galaxies.
However, this resort was rarely chosen, partly because the Galaxy Zoo
interface does not explicitly indicate the image resolution, and
under-resolved galaxies of any type often appear to have elliptical
morphology. Face-on or moderately inclined S0s will thus be found in
the Galaxy Zoo early-type class.  We can test this by examining the
Galaxy Zoo type-likelihood distributions for objects classified by
experts.  For this we use the largest sample of
morphologically-classified SDSS galaxies that attempts to distinguish
S0s from the other types, provided by \citet[][hereafter
F07]{2007AJ....134..579F}.  The early-type and spiral type-likelihood
distributions are shown in Fig.~\ref{fig:p_distribs} (this reproduces
Fig.~8
from \citet{Chris_GZ} with the addition of histograms showing
the de-biased type likelihoods used in this paper).  We show
distributions for the F07 ellipticals, spirals, and S0s, as well as
ambiguous objects that are classified as E/S0 or S0/Sa.

The majority of objects that F07 classify as E/S0 or S0 have high
$p_{\rmn{el}}$ and low $p_{\rmn{sp}}$ (both raw and de-biased).  The
S0/Sa galaxies span a range of intermediate $p_{\rmn{el}}$ and
$p_{\rmn{sp}}$.  We might expect edge-on S0s to be assigned to the
spiral class in Galaxy Zoo.  However, Fig.~\ref{fig:p_distribs} shows
that extremely few of the F07 E/S0, S0 or S0/Sa galaxies receive high
$p_{\rmn{sp}}$.  For the F07 sample, despite there being one S0 for
every four spirals, S0s contribute just 3 percent to the Galaxy Zoo
spiral fraction.  Either edge-on S0s identified by F07 appear more
like ellipticals than spirals to the Galaxy Zoo classifiers, or F07
and Galaxy Zoo both classify edge-on S0s as spirals, or perhaps a
combination of the two.  \citet{Chris_GZ} demonstrate that this is
also true when thresholded classifications based on the raw type
likelihoods are used.  In any case, the uncertain distribution of true
S0s between the classes henceforth referred to `early-type' and
`spiral' should be borne in mind when considering the results of this
paper.

\subsection{Measuring environment}
\label{sec:data_env}
\subsubsection{Local galaxy density}
\label{sec:data_density}

We estimate local galaxy density by following exactly the detailed
prescription in \citet{2006MNRAS.373..469B}, but extended from DR4 to
DR6. The local density for a galaxy is given by $\Sigma_N = N/(\pi
d_N^2)$ where $d_N$ is the projected distance to the $N$th nearest
neighbour that is more luminous than $M_r=-20$ (with a small evolution
correction, see below). In our analysis we use an estimate of local
galaxy density, $\Sigma$, determined by averaging $\log\Sigma_N$ for
$N = 4$ and $5$. In addition, the area $\pi d_N^2$ is modified close
to a photometric edge; and $\Sigma$ is determined by averaging the
density determined using spectroscopic neighbours ($|c\, \Delta z| <
1000$~km~s$^{-1}$) with that determined using both spectroscopic and
photometric neighbours ($|\Delta z| < z_{\rmn{err,95}}$, the 95 per
cent confidence interval, for galaxies with only photometric redshifts
determined by \citealt{2006MNRAS.373..469B}). The highly continuous
imaging coverage in DR6 thus increases the number of galaxies with
accurate local density estimates significantly beyond that expected
simply from the increase in area since DR4.

The \sample{density defining population} is constructed independently of
the other samples used in this paper. It comprises all galaxies with
$r<18$ in DR6 and with $M_r < M_{r,\rmn{limit}} - Q(z - z_0)$, where
$M_{r,\rmn{limit}} = -20$ and $Q = 1.6, z_0=0.05$ account for
evolution as determined by \citet{2003ApJ...592..819B}. With these
parameters, local density measurements are reliable for $z < 0.085$,
the redshift to which the density defining sample is luminosity-limited.
For density defining galaxies without spectroscopic redshifts, $M_r$
is determined using the redshift of the galaxy whose environment is
being measured. We determine the potential range in $\log\Sigma$ that
may result from both edge effects and the use of galaxies with only
photometric redshifts. Galaxies for which the full range of potential
$\log\Sigma$ is larger than 0.75 dex are rejected from the analysis
sample. This amounts to $\sim 10$ per cent of objects, reducing our
\sample{luminosity-limited sample} to 113579 objects when we are considering
local density. This sample has a $\log\Sigma$ uncertainty range of less
than 0.4 and 0.15 for 75 and 50 per cent of galaxies, respectively.
For 25 per cent of our \sample{luminosity-limited sample} the density is based on only
spectroscopic galaxies far from survey edges, and thus the uncertainty
range attributable to these issues is zero.

As noted by \citet{2007ApJ...670..206V} it may be more meaningful to
limit the density-defining population in terms of stellar mass, rather
than luminosity.  However, as \citeauthor{2007ApJ...670..206V} show,
while this may change the absolute values and distribution of local
density, it does not significantly affect the ranking of galaxy
environments, and thus would have no effect on our conclusions.

\subsubsection{Group properties}
In the standard $\Lambda$CDM cosmology, the properties of galaxies are
thought to have a strong dependence on the characteristics of the dark
matter halo in which they reside \citep{2002PhR...372....1C}.  Even
without assuming this background model, the empirical clustering of
galaxies into groups suggests a natural scale on which to measure
environment.  The local density around a galaxy, as measured by the
distance to its fifth-nearest neighbour, may fluctuate significantly
and rapidly compared with the timescales for galaxy evolution.
Therefore, recent environmental influences on the properties of a
galaxy may not be reflected by the local density measured for that
galaxy now.  On the scale of galaxy groups, changes in environment
will generally be slower and smoother.  It is unclear, however, which
environmental scales play a significant role.  Determining this is
complicated by strong correlations between different environmental
measures.

To probe the influence of galaxy groups we utilise the C4 catalogue of
galaxy groups and clusters \citep{2005AJ....130..968M}. The C4
catalogue was constructed by identifying galaxy over-densities
simultaneously in both three-dimensional position and four-dimensional
colour-space.  This method greatly reduces the twin problems of
projection effects and redshift-space distortions in identifying
physical galaxy groups.  Tests indicate that the C4 catalogue is $\sim
90$ per cent complete and $\sim 95$ per cent pure above
$\mathcal{M}_{\rmn{C4}} = 10^{14} h^{-1} \mathcal{M}_{\odot}$ and
within the redshift range considered here \citep{2005AJ....130..968M}.
The method for estimating the virial radius, and hence virial mass,
for the C4 groups has been refined beyond those given in the original
catalogue \citep{Torki_C4}.

At the time of analysis the C4 catalogue was only available for
DR5. Whenever considering group quantities we thus limit our sample to
objects with spectroscopy in DR5, though we continue to use DR6
measurements. We will employ two estimates of environment based on the
C4 groups: the distance from a galaxy to its nearest C4 group, and the
mass of the nearest C4 group.

In order to determine the distance from a galaxy to its nearest group
we require both the line-of-sight and projected distances between
them. A first approach is to convert redshifts, via an assumed
cosmology, directly to line-of-sight distances between the galaxies
and their nearest C4 group, $d_{\rmn{C4}}$.  However, this neglects
the peculiar motions of galaxies. These motions are small for isolated
galaxies, but increase in denser regions. They comprise two
effects. First, there is a coherent infall in the vicinity of groups
\citep{1987MNRAS.227....1K}, but the velocities associated with this
are generally low and produce a fairly small effect on the inferred
galaxy distances. Secondly, there are the random motions of virialised
group members. These virial motions are significant compared with
cosmological velocities, particularly at low redshift. The effect is
such that galaxies which are truly group members are inferred to be at
very different line-of-sight distances: the `Fingers of God'
effect. To correct for this one must identify galaxies that are likely
to be group members, and put them at the same line-of-sight distance
as their parent group. Our procedure for doing this is detailed in
Appendix~\ref{sec:FoG}, resulting in corrected distances,
$D_{\rmn{C4}}$, which are additionally normalised by the virial radius
of the nearest C4 group.

When using groupocentric distances, we limit the galaxies considered
to those which are closer to a C4 group than they are to an edge of
the DR5 spectroscopic survey region.  We also permit a galaxy's
nearest C4 group to lie beyond the redshift limits of our
\sample{magnitude-limited sample}.  We thus avoid the issue of clusters possibly
located just beyond the survey and sample boundaries. Recall that our
redshift limits are $0.03 < z < 0.085$, and so our
\sample{magnitude-limited sample} is
restricted to redshifts where the C4 group catalogue is reasonably
complete ($\sim 90$ per cent: \citealt{2005AJ....130..968M}).

\begin{figure*}
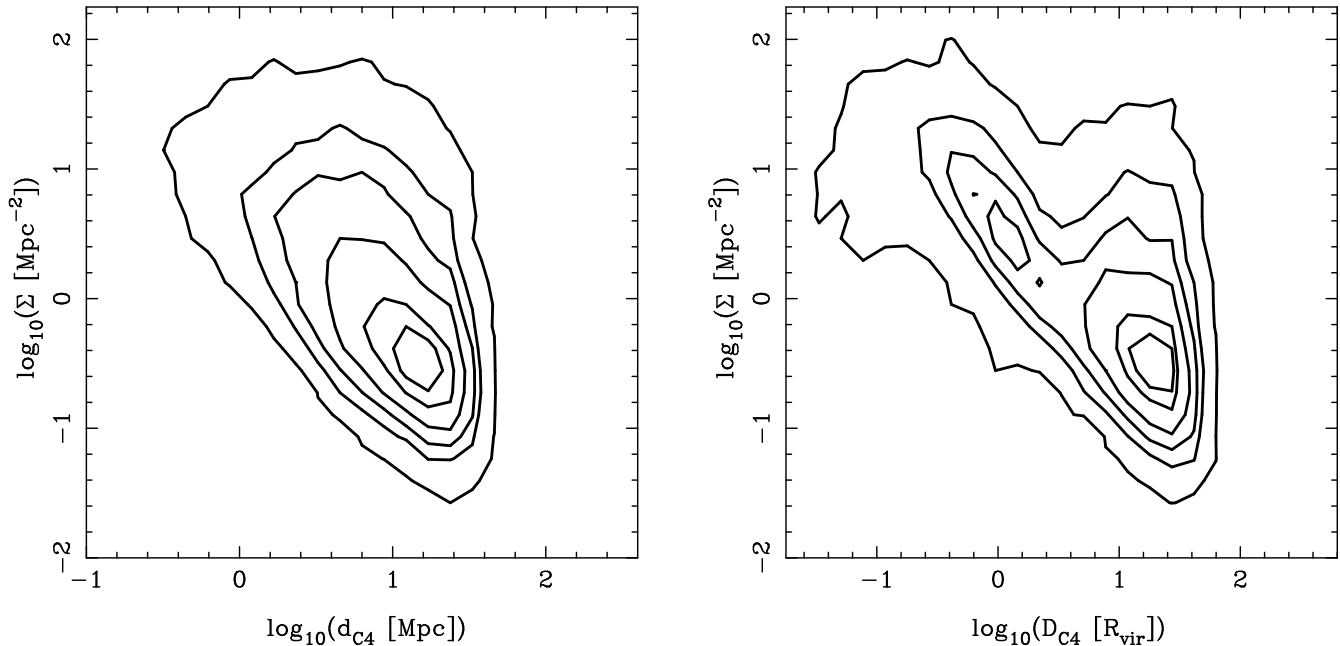

\centering
\includegraphics[height=0.475\textwidth,angle=270]{fig5_left.ps}
\hfill
\includegraphics[height=0.475\textwidth,angle=270]{fig5_right.ps}
\caption{\label{fig:sigma_c4dist} Local galaxy density
  versus (left) raw and (right) corrected distance to the nearest C4
  group.  Contours enclose 10, 25, 50, 75, 90 and 99 per cent of
  galaxies in the \sample{luminosity-limited sample}.}
\end{figure*}

Although they measure different quantities, the environmental measures
based on groups correlate strongly with local
density. Figure~\ref{fig:sigma_c4dist} shows the relationship between
local galaxy density and our two measures of distance from a C4 group.
At large distances from a C4 group ($\ga 10$ Mpc or $R_{\rmn{vir}}$)
galaxies are found in the full range of local densities, although
predominantly these densities are low. However, approaching a C4 group
the minimum local density steadily increases, such that all galaxies
in the vicinity of C4 groups are also in regions of high local
density.

Correcting for the `Fingers of God' effect clearly improves the
correlation between distance to a C4 group and local density. There
remain a large number of objects in regions of fairly high local
density which are distant from a C4 group. There are several potential
reasons for this: (i) even though density and group distance may be
strongly related, the C4 catalogue forms an incomplete census of
galaxy groups; (ii) the `Fingers of God' suppression is not fully
effective, it has a reduced efficiency for objects at intermediate
densities -- for which we cannot distinguish between true group
members and nearby, but gravitationally unbound, objects -- and is
designed to be conservative and under-correct in such circumstances,
preferring incompleteness over contamination of the group members
sample; (iii) the redshift interval over which local density is
measured is large enough to allow nearby groups to influence the local
density estimate for some isolated objects, but note that we do limit
the local density uncertainty, see Sec.~\ref{sec:data_density}; (iv)
high local densities may occur in structures which do not meet the
requirements to be selected as C4 groups, e.g. filaments, too few
members, inhomogeneous member colours, etc.  All of these issues are
likely together responsible for the spread in local density at large
distances from C4 groups and the bimodality visible along the line of
(anti-)correlation between density and group distance.

Close to C4 groups, within a few Mpc or virial radii, we expect a
spread in local density due to substructure -- azimuthal variations in
local density.  Even so, local density is confined to higher values
than seen in the more typical environments.  For substructures
containing more than five density-defining galaxies the local density
estimate will be well localised on the sky.  However, low density
measurements around groups may be precluded by the method's limited
line-of-sight resolution.  Nevertheless, the overall correlation
between local density and C4 group distance must be a real effect.

Local galaxy density and distance to a C4 group are thus useful
alternative estimates of environment, with different caveats for their
interpretation.  The agreement of results when using either of these
methods will therefore be a good sign that the results are robust with
respect to the general concept of `environment'.  Any contrasts in the
results from these two methods may point to different effects at work in
classical groups and in more general dense regions.

\begin{figure*}
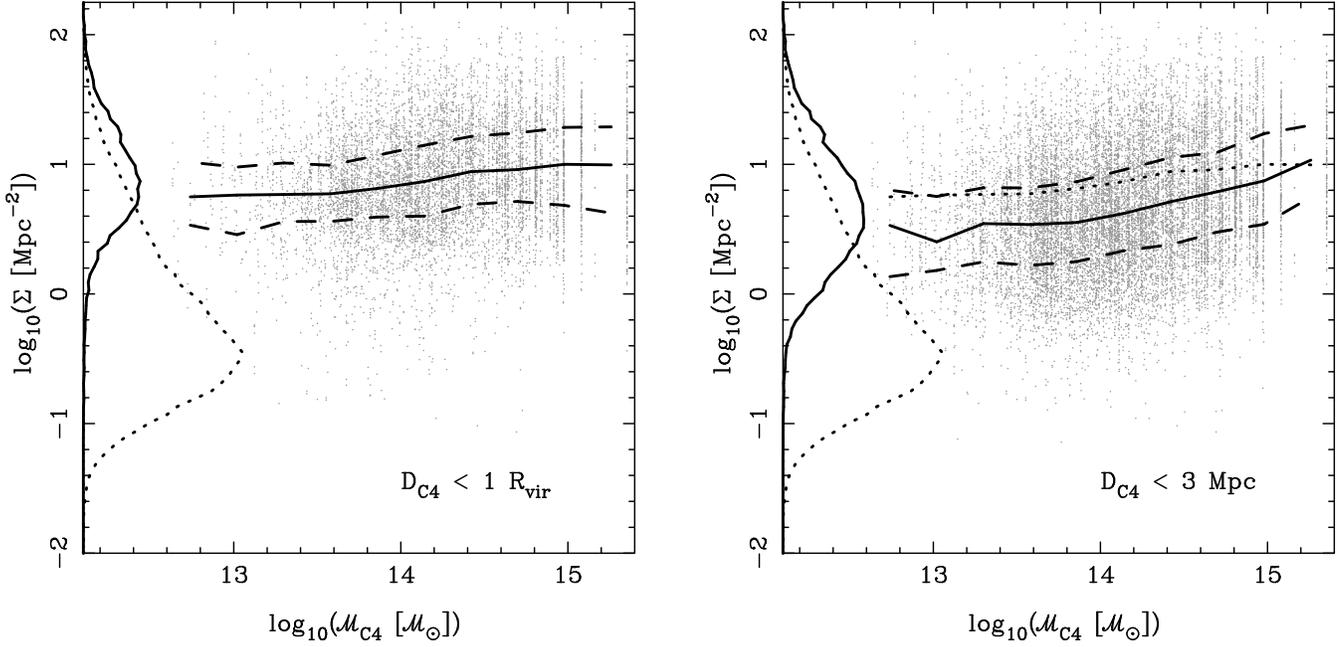

\centering
\includegraphics[height=0.475\textwidth,angle=270]{fig6_left.ps}
\hfill
\includegraphics[height=0.475\textwidth,angle=270]{fig6_right.ps}
\caption{(Left) \label{fig:sigma_c4mass} points indicate local galaxy
  density versus C4 group mass for members of C4 groups with
  $D_{\rmn{C4}} < 1 R_{\rmn{vir}}$.  The solid and dashed lines
  through the points indicate the median and quartiles of
  $\log{\Sigma}$ in bins of $\mathcal{M}_{\rmn{C4}}$.  Along the
  left-hand side of the plot appear histograms comparing the
  $\log{\Sigma}$ distribution for C4 group members (solid) with that
  for all galaxies (dashed).  The frequencies for the histogram of C4
  group members have been multiplied by four to improve visibility.
  (Right) as the left panel, but now for
  objects within a constant 3 Mpc radius of each C4 group
  ($D_{\rmn{C4}} < 3$~Mpc).  The additional dotted line indicates the
  median relation from the left panel.}
\end{figure*}

For galaxies within the influence of a C4 group, the mass of that
group may well be expected to determine the magnitude of any
environmental effect.  The local density for members of C4 groups is
shown in the left panel of Fig.~\ref{fig:sigma_c4mass} as a function
of C4 group mass.  As Fig.~\ref{fig:sigma_c4dist} suggests, galaxies
within the virial radius of C4 groups are confined to high local
galaxy densities.  This is clearly seen by comparing the histograms
for cluster members and all galaxies along the side of the left panel
in Fig.~\ref{fig:sigma_c4mass}.  However, there is almost no trend
in the distribution of local galaxy density versus the mass of the group.

The right panel of Fig.~\ref{fig:sigma_c4mass} shows the same plot
as the left, but now containing all galaxies within a fixed physical
radius, 3 Mpc, of a C4 group.  A trend with group mass is now much
more apparent.  Comparing the left and right panels of
Fig.~\ref{fig:sigma_c4mass} illustrates that the primary role of group
mass is in determining the scale of the region, through the virial
radius, over which local galaxy density is elevated, rather than the
determining the absolute degree of density enhancement.

\section{Morphology versus environment}
\label{sec:morph-env}

\subsection{Local galaxy density}

\begin{figure*}
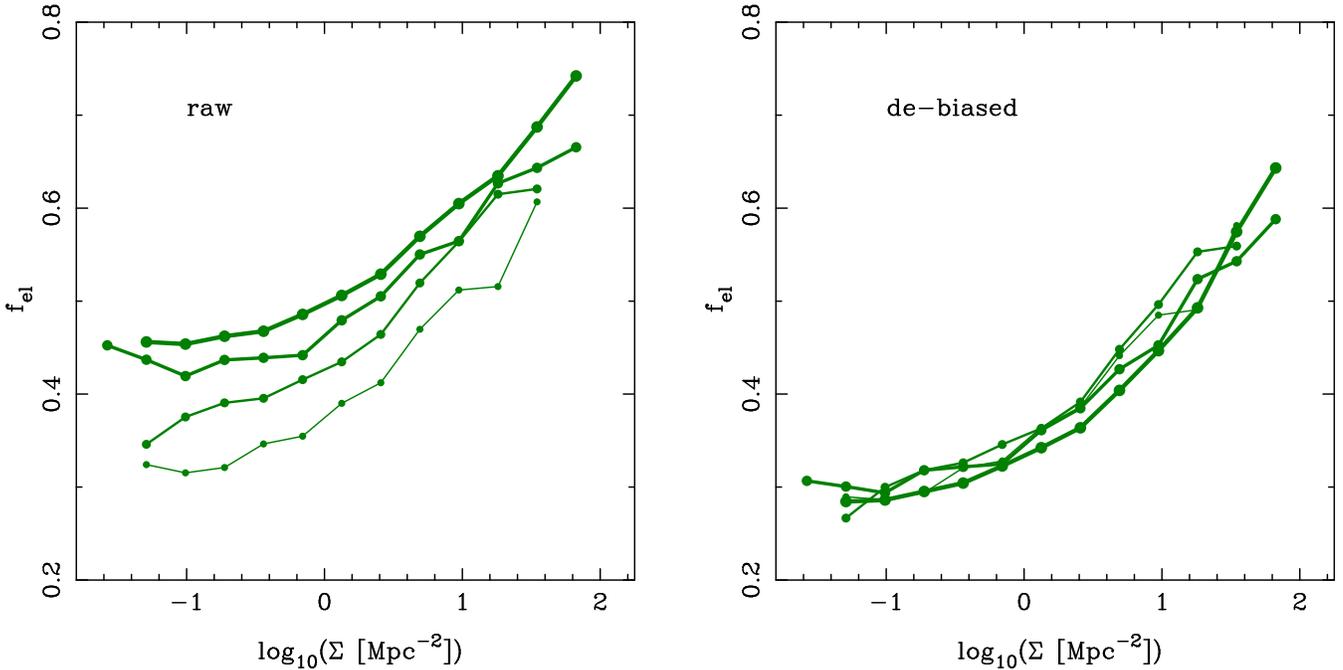

\centering
\includegraphics[height=0.475\textwidth,angle=270]{fig7_left.ps}
\hfill
\includegraphics[height=0.475\textwidth,angle=270]{fig7_right.ps}
\caption{\label{fig:morph-density_zbins} Early-type fraction versus
  local galaxy density for four redshift bins ($z \sim 0.037, 0.051,
  0.064, 0.078$). The left and right panels show the fractions
  determined from the raw and de-biased morphological type
  likelihoods, respectively.}
\end{figure*}

As a first look at the Galaxy Zoo morphology--density relation, and a
final demonstration of the removal of classification bias, in the left
panel of Fig.~\ref{fig:morph-density_zbins} we plot the early-type
fraction based directly on the raw type likelihoods, as a function of
local galaxy density.  This is shown for objects in four redshift
slices.  Note the offsets between the redshift slices: the average raw
early-type fraction increases with increasing redshift, due to
classification bias. However, in each redshift bin a clear relation
can be seen, with early-type galaxies favouring higher density
environments.

\begin{figure*}
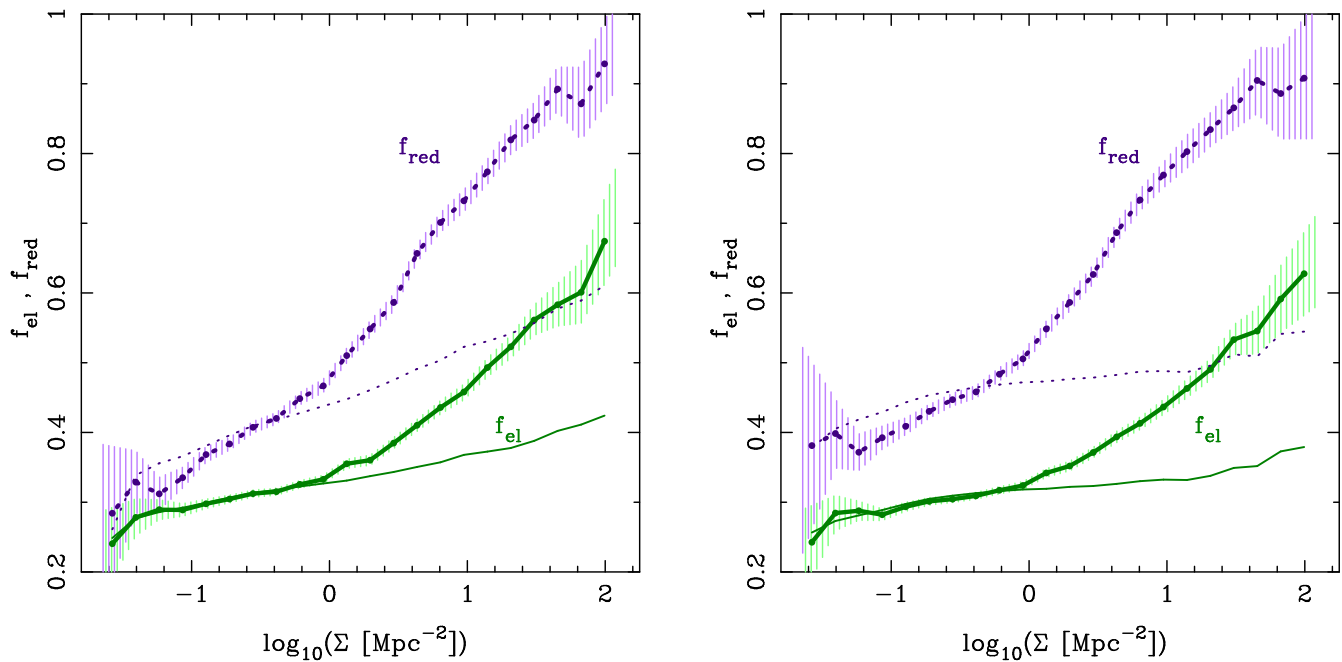

\centering
\includegraphics[height=0.475\textwidth,angle=270]{fig8_left.ps}
\hfill
\includegraphics[height=0.475\textwidth,angle=270]{fig8_right.ps}
\caption{\label{fig:morph-colour-density} Early-type (green, thick,
  solid line) and red (purple, thick, dotted line) fractions versus
  local galaxy density for galaxies in (left) our
  \sample{luminosity-limited sample} and (right) our
  \sample{$V_{\rmn{max}}$-weighted mass-limited sample}, complete for
  stellar masses $\log(\mathcal{M}_{\ast}/\mathcal{M}_{\odot}) > 9.8$,
  and constructed by applying $1/V_{\rmn{max}}$ weightings to our
  \sample{magnitude-limited sample} (see Sec.~\ref{sec:stellarmasses}).
  The early-type fraction is determined from the de-biased type
  likelihoods, as described in Sec.~\ref{sec:morphs}. The shaded
  regions indicate the 2-sigma statistical uncertainties on each
  equally-spaced $\log{\Sigma}$ bin. Thin lines show the contribution
  to each relation attributable to variation in the stellar mass
  function with environment (see Sec.~\ref{sec:sigma-massdep}). The
  early-type fraction is discussed alone in Sec.~\ref{sec:morph-env},
  and compared with the red fraction in
  Sec.~\ref{sec:morph-colour-env}.}
\end{figure*}

To produce a single, reliable, local morphology--density relation we
use the de-biased type likelihoods described in Sec.~\ref{sec:morphs}
and derived in Appendix~\ref{sec:bias}. The resulting
morphology--density relations are shown in the right panel of
Fig.~\ref{fig:morph-density_zbins} for the same four redshift slices
as the left panel. Clearly the de-biasing procedure works well, as the
relations from independent redshift ranges now match. We can thus
combine galaxies over the full redshift range we are considering,
$0.03 < z < 0.085$. The morphology--density relation for our full
\sample{luminosity-limited sample}, containing information on 113579
galaxies, is shown in Fig.~\ref{fig:morph-colour-density}. We
determine the uncertainties on the mean type fraction in each bin from
the standard deviation of the type likelihoods divided by the
square-root of the number of objects in each bin. The reliability of
these uncertainties has been confirmed by bootstrap resampling. In all
plots we show $2\sigma$ uncertainty ranges. (The colour--density
relation is also shown in Fig.~\ref{fig:morph-colour-density} and
subsequent figures, for which red galaxies are defined as those on the
red side of the $u-r$ versus stellar mass bimodality using the divider
determined by \citet{2006MNRAS.373..469B}. However, we defer
discussion of this relation until Sec.~\ref{sec:morph-colour-env}.)

Our morphology--density relation is extremely well defined. It is
consistent with a monotonic function, with a significant gradient over
almost the whole range of galaxy densities that occur, $0.1 \la \Sigma
\la 25$ galaxies Mpc$^{-2}$. Even outside this range, there is no
evidence for the relation flattening off completely. While more
complicated representations could be acceptable, the data suggests
that the relation is most appropriately described by a single, smooth
function.

\begin{figure}
\centering
\includegraphics[height=0.475\textwidth,angle=270]{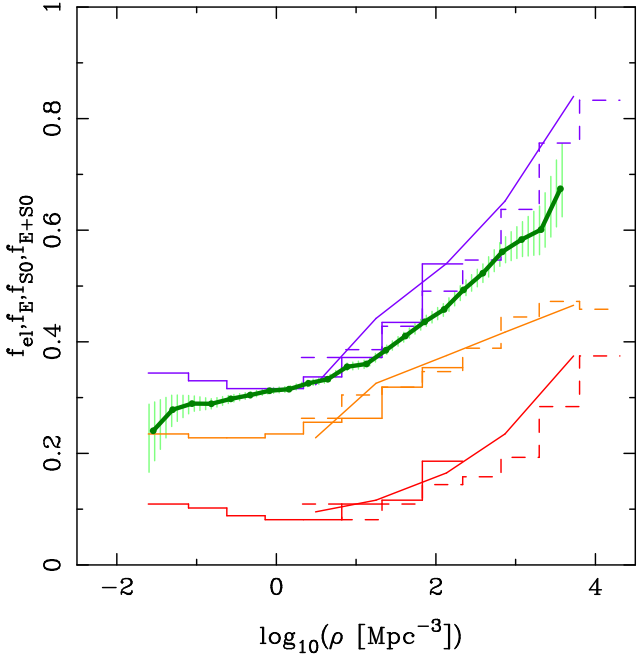}
\caption{\label{fig:pg84} The thin lines reproduce the morphological
  type fractions versus local galaxy volume density from figure 1 of
  \citet{1984ApJ...281...95P}.  From low to high, the groups of lines
  correspond to ellipticals (red), S0 galaxies (orange), and
  ellipticals and S0 galaxies combined (purple).  In each group, the
  solid and dashed histograms represent data from CfA groups and
  clusters, respectively \citep{1983ApJS...52...89H}, and the line
  indicates the converted cluster relation of
  \citet{1980ApJS...42..565D}.  For comparison, the points joined by a
  thick green line show the Galaxy Zoo early-type fraction versus local
  surface density relation from Fig.~\ref{fig:morph-colour-density}
  converted to the volume density scale of
  \citet{1984ApJ...281...95P}, as described in the text.  }
\end{figure}

To compare our morphology--density relation to previous studies, in
Fig.~\ref{fig:pg84} we reproduce the relations from \citet[][hereafter
PG84]{1984ApJ...281...95P}, which contains a combination of data for
galaxies in clusters, from \citet{1980ApJS...42..565D}, and groups,
from the CfA redshift survey \citep{1983ApJS...52...89H} and
\citet{1982ApJ...257..423H}.  In PG84 the local density is measured in
a rather different manner to that employed in this paper.  However, to
qualitatively compare our relation to these classic results we attempt
to convert our measured local surface densities ($\Sigma$) to the
local volume density scale ($\rho$) of PG84.  We correct for the
different depths of the density defining populations ($M_B < -17.5$
versus $M_r \la -20.0$) by empirically determining the number density
ratio given these selection limits, as a function of local surface
density.  We convert from surface to volume densities assuming the
overdensity to be spherical.  Finally we adjust to $H_0 = 100$
km~s$^{-1}$~Mpc$^{-1}$.  The resulting relation is shown by the thick
green line in Fig.~\ref{fig:pg84}.

As discussed in Sec.~\ref{sec:morphs}, S0 galaxies are not explicitly
separated from the other types in the Galaxy Zoo data.  However, the
trend in the combined PG84 elliptical + S0 fraction with local density
is very similar to our relation.  A comparison of the type fractions
suggests that the Galaxy Zoo early-type class contains the majority
of the S0 population, as also indicated by our earlier comparison with
F07 in Fig.~\ref{fig:p_distribs}.  We defer the separation of the
elliptical and S0 populations to a future Galaxy Zoo project.

The remaining differences between our early-type fraction versus local
density relation and that of PG84 may be attributed to a number of
issues, but the most important is the differing selection functions of
the Galaxy Zoo and PG84 samples. As shall shortly be demonstrated, the
morphology--density relation is strongly dependent upon the stellar
mass (or luminosity) of the galaxies considered, and thus different
sample selections will naturally produce different relations.

\subsubsection{Stellar mass dependence}
\label{sec:sigma-massdep}

A significant advantage of the Galaxy Zoo sample, in comparison with
previous visual morphology catalogues, is its size.  It is possible to
subdivide our sample in terms of various properties, and still retain
sufficient galaxies in each subsample to make reliable inferences.  In
particular, we can consider morphology trends in narrow bins of
stellar mass.

\begin{figure*}
\centering
\includegraphics[height=0.475\textwidth,angle=270]{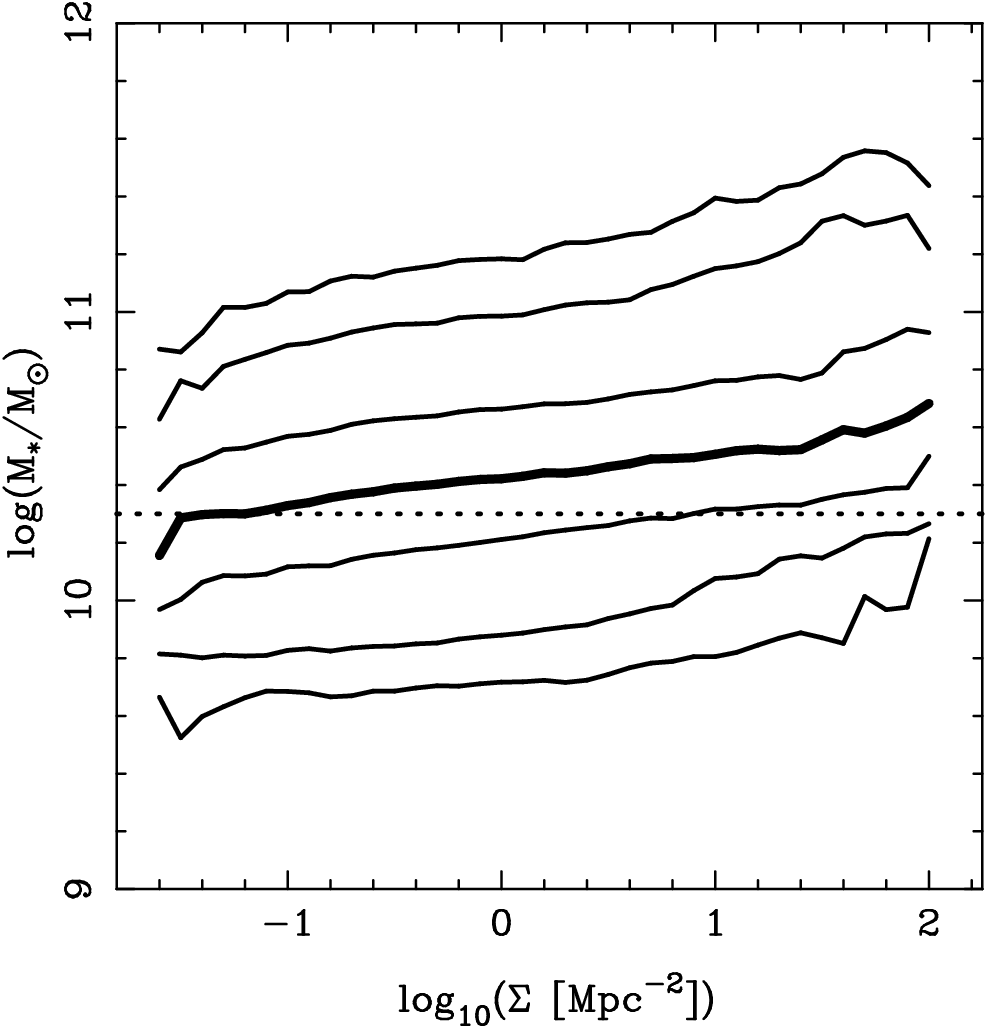}
\hfill
\includegraphics[height=0.475\textwidth,angle=270]{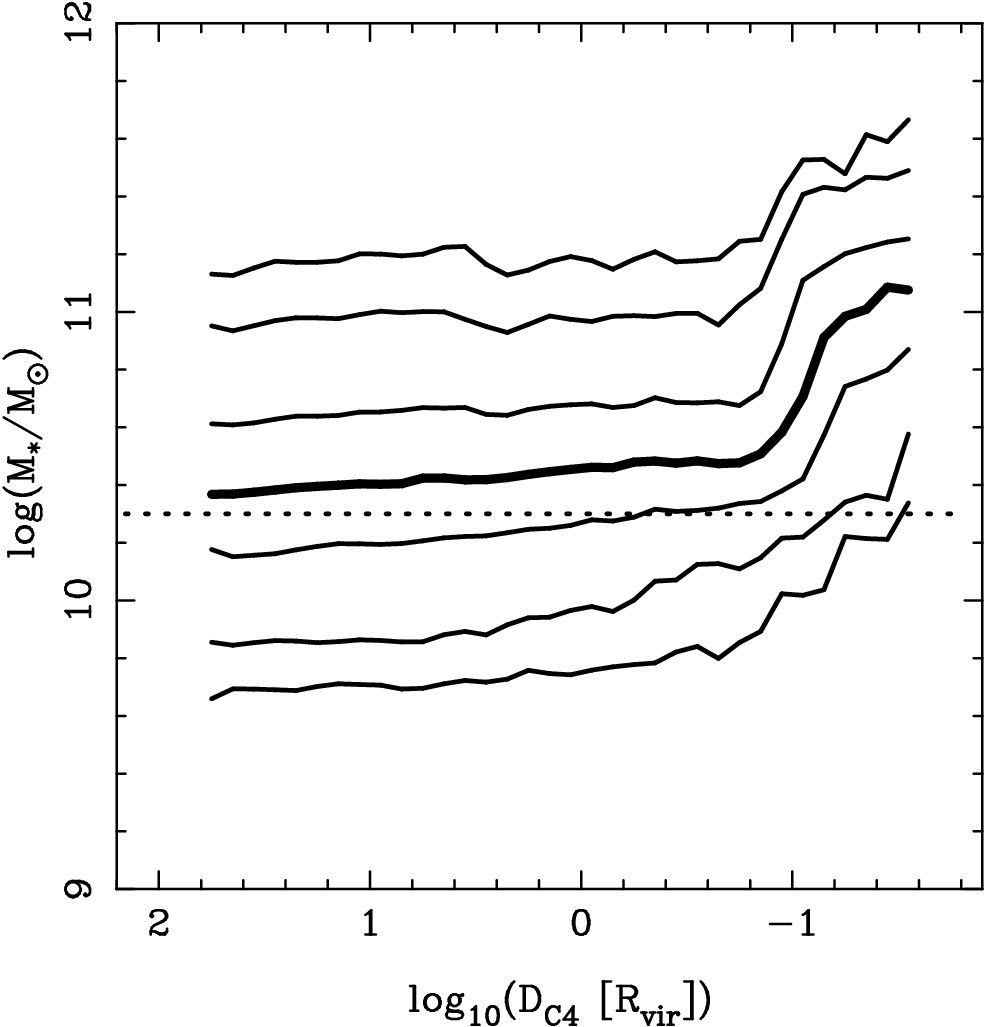}
\caption{\label{fig:mstar-density-c4normdist} The distribution of
  stellar mass in our \sample{luminosity-limited sample} as a function of (left)
  local galaxy density and (right) groupocentric distance.  The lines
  trace the 1, 5, 25, 50 (thick), 75, 95 and 99th-percentiles of the
  stellar mass distribution in bins of environment.  The stellar mass
  distribution shifts steadily versus local density, while versus
  groupocentric distance most of the change in the stellar mass
  distribution occurs within $\sim 0.2 R_{\rmn{vir}}$.  The dotted
  horizontal lines indicate a stellar mass of $10^{10.3}
  \mathcal{M}_{\odot}$, below which our \sample{luminosity-limited
  sample} becomes incomplete for red galaxies.}
\end{figure*}

It is now well known that the stellar mass function of galaxies varies
with environment \citep{2001ApJ...557..117B,2002ApJ...571..172Z}.
This can be, to some extent, theoretically understood by the biasing
of the dark matter halo mass function varying with the large-scale
density field
\citep{1999MNRAS.308..119S,2002MNRAS.336..112M,2004MNRAS.349..205M,2006MNRAS.369...68S},
although there are still many unsolved issues in connecting haloes
with galaxies. In the left panel of
Fig.~\ref{fig:mstar-density-c4normdist} we show how the distribution
of stellar mass depends on local galaxy density in our
\sample{luminosity-limited sample}. The distribution of stellar masses
steadily shifts to higher masses with increasing local density.

The mass function of galaxies is also expected to depend on
morphological type \citep{2008arXiv0801.4568M}, as it varies with
other galaxy properties \citep{2005MNRAS.356.1155C,
  2006MNRAS.373..469B}.  We shall see this clearly later in this
section.  In dense large-scale environments we expect a greater
fraction of high-mass haloes, and such haloes are preferentially
inhabited by galaxies with high stellar mass and early-type
morphology.  We must therefore consider whether the observed
morphology--density relation may simply be a consequence of these
effects.

One way we can test this is by determining the morphology--density
relation that would be present in our sample on the assumption that
the early-type fraction is only a function of stellar mass. To do so,
we directly measure the dependence of the early-type fraction on
stellar mass for galaxies in our \sample{luminosity-limited sample} that
reside in low-density environments ($\log\Sigma < 0$). Then, for each
bin of environment, we determine the morphological fraction predicted
by applying this relation to the distribution of stellar masses in
that particular environmental bin. The result is a relation between
early-type fraction and local density that is due only to the
empirical dependence of stellar-mass distribution on environment and
morphology on stellar mass, without any direct dependence of
morphology on environment. This is shown by the thin, green line in
the left panel of Fig.~\ref{fig:morph-colour-density}.

\begin{figure}
\centering
\includegraphics[height=0.475\textwidth,angle=270]{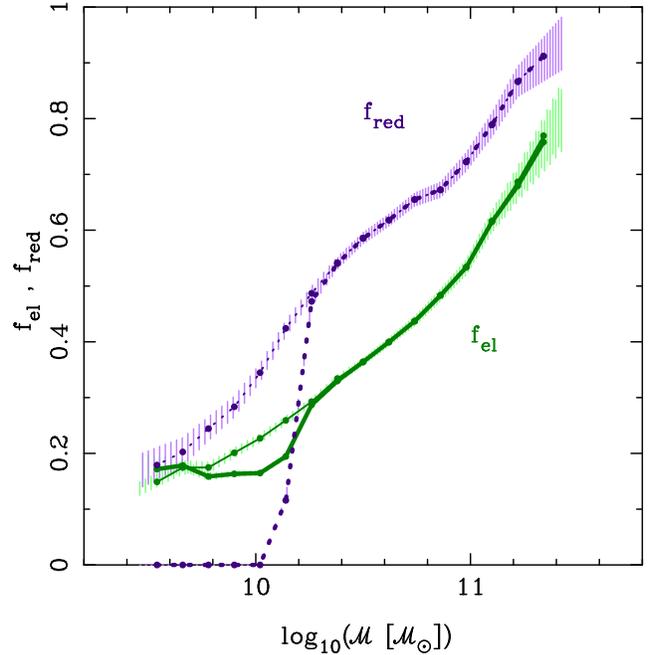}
\caption{\label{fig:morph-colour-Mstar} Early-type (green, solid
  lines) and red (purple, dotted lines) fractions versus stellar mass
  for galaxies in low-density environments ($\log\Sigma < 0$).  In
  order to fairly sample to low stellar masses we limit the redshift
  range contributing to each point to ensure completeness.  The thick
  lines show the relation for our \sample{luminosity-limited sample}, while the
  thin lines give the relation for our \sample{$V_{\rmn{max}}$-weighted
  mass-limited sample}.  Note that below
  $\log(\mathcal{M}_{\ast}/\mathcal{M}_{\odot}) = 10.3$ the absolute
  magnitude limit of our \sample{luminosity-limited sample} results in an absence
  of red galaxies, and hence a deficit of early-types. The shaded
  regions indicate the 2-sigma statistical uncertainties on each
  equally-spaced $\log{\mathcal{M}_{\ast}}$ bin.}
\end{figure}

The variation of the stellar mass function with environment therefore
accounts for $\sim 40$ per cent of the morphology--density relation in
our \sample{luminosity-limited sample}. However, this effect has been
enhanced by the fact that this sample is limited in terms of absolute-magnitude. At
low stellar masses all galaxies in the sample are blue, and
preferentially spiral. As we shall see later in this paper, such
galaxies exist primarily at low densities. If, rather than an
absolute-magnitude limited sample, we use a stellar-mass limited
sample, then the contribution to the morphology--density relation due
to the varying stellar mass function is reduced. This is demonstrated using our
\sample{$V_{\rmn{max}}$-weighted
  mass-limited sample} (limited to
$\log(\mathcal{M}_{\ast}/\mathcal{M}_{\odot}) > 9.8$ and with
completeness ensured by a $1/V_{\rmn{max}}$ correction) in the right
panel of \ref{fig:morph-colour-density}. Figure
\ref{fig:morph-colour-Mstar} shows the relationship of early-type
fraction versus stellar mass in low-density environments for both our
\sample{luminosity-limited sample} and \sample{$V_{\rmn{max}}$-weighted
  mass-limited sample}. It is these relations, along with the dependence of
stellar mass on local density as illustrated in the left panel of
Fig.~\ref{fig:mstar-density-c4normdist}, that are used above in
determining the contribution of the varying stellar-mass function to
the morphology-density relation.

\begin{figure*}
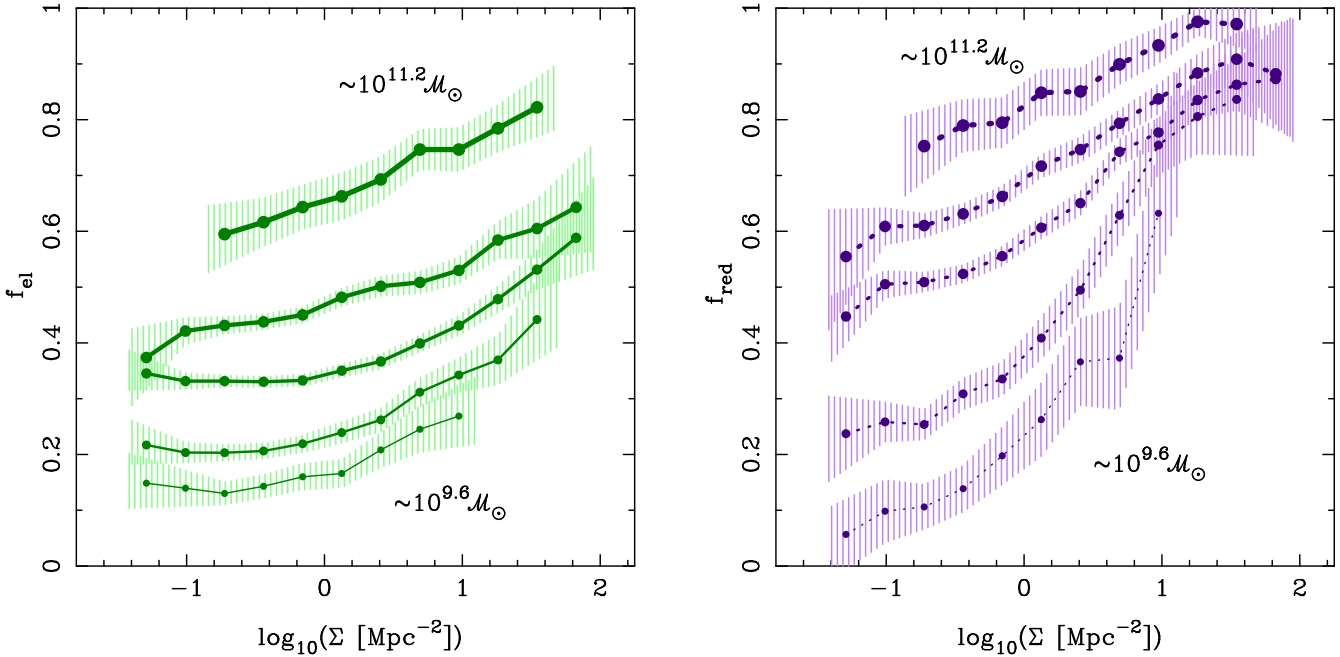

\centering
\includegraphics[height=0.475\textwidth,angle=270]{fig12_left.ps}
\hfill
\includegraphics[height=0.475\textwidth,angle=270]{fig12_right.ps}
\caption{\label{fig:morph-colour-density_mstarbins} (Left) early-type
  and (right) red fraction versus local galaxy density for galaxies in
  selected narrow bins of stellar mass:
  $\log(\mathcal{M}_{\ast}/\mathcal{M}_{\odot}) =$ 9.5--9.7,
  9.9--10.1, 10.3--10.5, 10.7--10.9, 11.1--11.3, in order from the
  lowest to highest average $f_\rmn{el}$, $f_\rmn{red}$.  The shaded
  regions indicate the $2\sigma$ statistical error on each
  $\log{\Sigma}$ bin.}
\end{figure*}

For the \sample{$V_{\rmn{max}}$-weighted
  mass-limited sample}, only approximately 32 per cent of the
morphology-density results from the variation of the stellar mass
function with environment. The remainder must be due to changes in the
early-type fraction with environment at fixed stellar mass. For both
samples the morphology-density relation at low densities ($\Sigma \la
1$ Mpc$^{-2}$) is entirely explainable by the environmental variation
in the stellar mass function. At greater densities it appears that
some process must increase the early-type fraction at fixed stellar
mass.

To investigate this in more detail, we utilise our \sample{binned
  mass-limited sample} to divide our sample into several complete bins
of stellar mass (see Sec.~\ref{sec:stellarmasses}). We make these bins
as narrow as reasonable given the uncertainties on the stellar masses.
They thus have width 0.2 dex in stellar mass.  We only plot alternate
bins for clarity. The left panel of
Fig.~\ref{fig:morph-colour-density_mstarbins} plots the
morphology--density relation for several stellar-mass bins. We see
that the relations are offset from one another, with the overall
fraction of early-types increasing with stellar mass as expected from
Fig~\ref{fig:morph-colour-Mstar}. As we have seen, some of the
morphology--density relation is due to this effect, in combination
with the varying stellar-mass function.

However, a significant trend of morphological fraction with local
density is present in each stellar mass bin.  The morphology--density
relation is not simply a product of a morphology--mass relationship
and varying stellar-mass function; it exists even at fixed stellar
mass.  This confirms the recent result of \citet{2008arXiv0801.1995V},
which is based on an automatic measure of morphology
\citep{2006ApJ...644...30B}, and highlights the difference between
morphology and concentration; concentration shows little
environmental dependence at fixed stellar mass.

Surprisingly, the change in early-type fraction with environment is
very similar at each stellar mass, being approximately an increase of
$0.2$ in $f_{\rmn{el}}$ over the local density range probed
($\log\Sigma = -1$--$1.5$).  For low stellar masses this corresponds
to a doubling of the early-type fraction, while for high masses it is a
fractional increase of around a third.

There is a hint that for lower masses the morphology--density relation
is flat in low-density environments.  For the whole
range of masses studied here, $\mathcal{M}_{\ast} >
10^{9.5}\mathcal{M}_{\odot}$, a correlation between early-type
fraction and local density is clearly present for $\Sigma \ga 1$
Mpc$^{-2}$, but these is little evidence for a correlation below
this density.

\subsection{Group properties}

In the previous subsection we considered the behaviour of
morphological type fraction versus the density of neighbouring
galaxies.  However, the relative velocities of galaxies in groups and
clusters implies that local density may change rapidly, on timescales
potentially shorter than those for morphological transformation.  The
average local density a galaxy experiences is related to the mass of
the group halo it resides within.  In addition, some proposed
mechanisms for environmental galaxy evolution, such as gas starvation
and tidal effects, relate directly to global halo properties rather
than the local galaxy density.  Several studies have thus considered
whether galaxy morphology is more fundamentally related to global
environmental properties, such as distance to the centre of the
nearest galaxy group \citep[groupocentric
distance,][]{1991ApJ...367...64W,1993ApJ...407..489W,2003ApJ...584..210G,2007ApJ...664..791B}.
It would also improve our confidence in our results if the trends can
be shown to be independent of the detailed way in which environment is
characterised.

To examine the influence of galaxy groups on morphological type
fraction we use the C4 group catalogue.  The range of local galaxy
densities, $\Sigma$, covered in groups of varying mass,
$\mathcal{M}_{\rmn{C4}}$, is illustrated by the left panel of
Fig.~\ref{fig:sigma_c4mass}.  Clearly members of C4 groups reside in
regions of high local galaxy density, but the distribution of $\Sigma$
varies very little across the range of C4 group masses.  In
Fig.~\ref{fig:sigma_c4dist} we show the variation of $\Sigma$ with
distance from the centre of the nearest C4 group.

\begin{figure*}
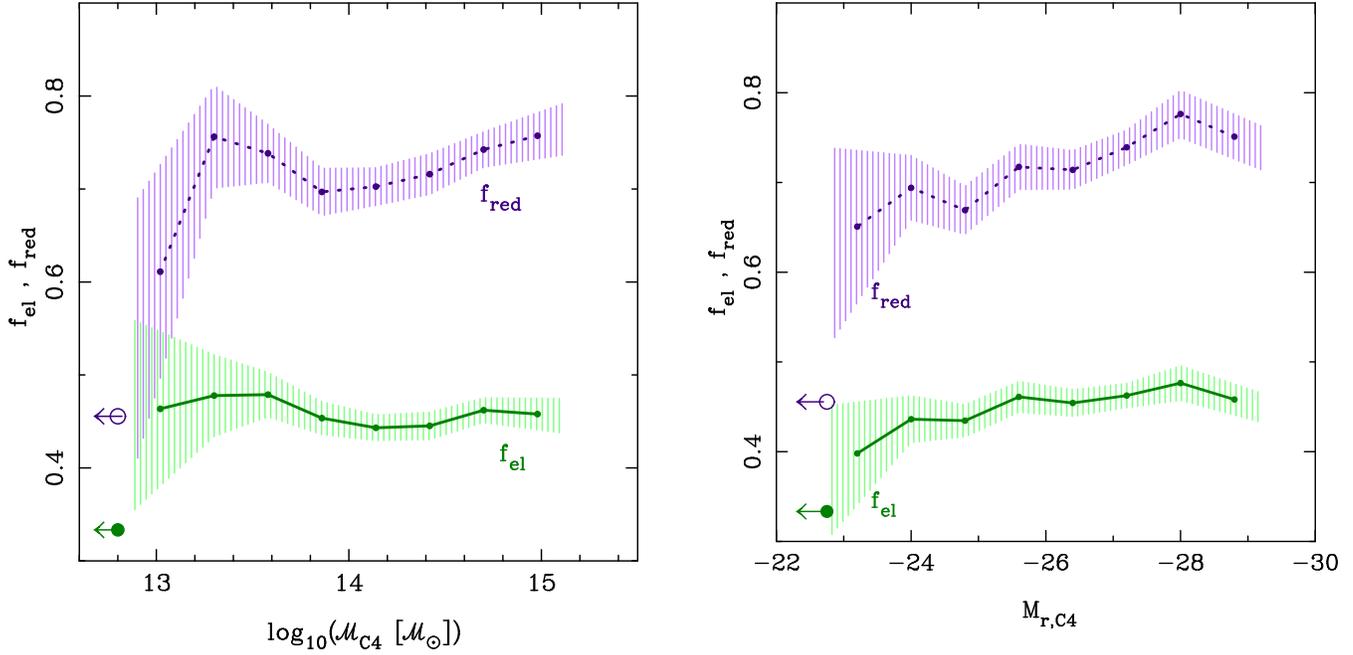

\centering
\includegraphics[height=0.475\textwidth,angle=270]{fig13_left.ps}
\hfill
\includegraphics[height=0.475\textwidth,angle=270]{fig13_right.ps}
\caption{\label{fig:morph-colour-c4mass-c4lum} Early-type (green,
  solid line) and red (purple, dotted line) fractions versus (left
  panel) group mass and (right panel) integrated $r$-band group
  luminosity for members of C4 groups ($D_{\rmn{C4}} < 1
  R_{\rmn{vir}}$) in our \sample{luminosity-limited sample}. Only bins containing
  more than 50 galaxies from at least 5 groups are shown. The shaded
  region indicates the $2\sigma$ statistical error on each
  equally-spaced bin.  The large circles with arrows indicate the mean
  early-type (filled point) and red (open point) fractions for
  galaxies that are not members of a C4 group.  The early-type
  fraction is discussed alone in
  Sec.~\ref{sec:morph-env}, and compared with the red fraction in
  Sec.~\ref{sec:morph-colour-env}.}
\end{figure*}

The variation of morphological type fraction versus group mass for C4
group members is plotted in the left panel of
Fig.~\ref{fig:morph-colour-c4mass-c4lum}.  There is no evidence for a
change in the early-type fraction over a group mass range of
$10^{13}$~--~$10^{15}$~$\mathcal{M}_{\odot}$.  Another estimate of
group scale is the integrated red luminosity of the group members
\cite[e.g.,][]{2006ApJ...650L..99L}.  As a check we thus additionally
show the type fractions as a function of the summed $r$-band
luminosity of all group members with spectroscopic redshifts.  There
is a slight trend of increasing early-type fraction with increasing
group luminosity.  However, as versus group mass, this trend is small
compared with the trends versus local density.  Galaxies at all
stellar masses considered in this paper display similarly flat trends
versus group mass, simply with offsets due to the dependence of
morphology on stellar mass.

\begin{figure*}
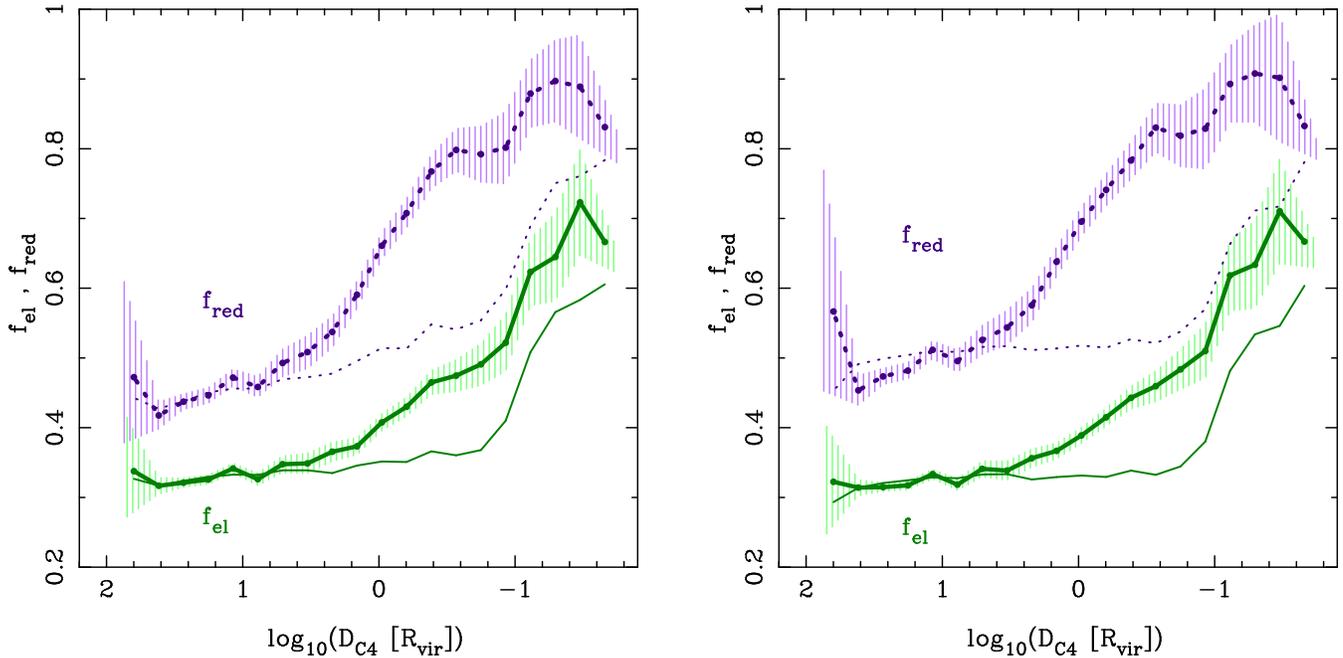

\centering
\includegraphics[height=0.475\textwidth,angle=270]{fig14_left.ps}
\hfill
\includegraphics[height=0.475\textwidth,angle=270]{fig14_right.ps}
\caption{\label{fig:morph-colour-c4normdist} Early-type (green, thick,
  solid line) and red (purple, thick, dotted line) fractions versus
  distance to the nearest C4 group normalised by its virial radius,
  for (left) galaxies in our \sample{luminosity-limited sample} and
  (right) our \sample{$V_{\rmn{max}}$-weighted mass-limited sample},
  complete for stellar masses
  $\log(\mathcal{M}_{\ast}/\mathcal{M}_{\odot}) > 9.8$, and
  constructed by applying $1/V_{\rmn{max}}$ weightings to our
  \sample{magnitude-limited sample} (see Sec.~\ref{sec:stellarmasses}).
  The shaded regions indicate the $2\sigma$ statistical error on each
  equally-spaced $\log{D_{\rmn{C4}}}$ bin. Note that in this and
  subsequent plots the $\log{D_{\rmn{C4}}}$ axis is plotted with
  positive values on the left and negative values on the right for
  easier comparison with the plots versus $\Sigma$. The right-most
  $\log{D_{\rmn{C4}}}$ bin additionally includes all galaxies with
  lower $\log{D_{\rmn{C4}}}$. Thin lines show the contribution to each
  relation attributable to variation in the stellar mass function with
  environment (see Sec.~\ref{sec:c4dist-massdep}). The
  early-type fraction is discussed alone
  in Sec.~\ref{sec:morph-env}, and compared with the red fraction in
  Sec.~\ref{sec:morph-colour-env}.}
\end{figure*}

In Fig.~\ref{fig:morph-colour-c4normdist} we show the variation of
morphological type fraction versus the distance to the nearest C4
group for our usual \sample{luminosity-limited sample} and \sample{$V_{\rmn{max}}$-weighted
  mass-limited sample}. A strong dependence is seen, with the fraction
of early-type galaxies increasing for galaxies closer to a group
centre. This dependence is particularly strong within the group viral
radius, $D_{\rmn{C4}} \la R_{\rmn{vir}}$, although a slight trend
appears to continue out to larger distances, particularly for the
\sample{luminosity-limited sample}. As with density, the relation
appears consistent with a smooth function of groupocentric distance,
rather than a broken line as advocated by \citet{2003MNRAS.346..601G}.

\subsubsection{Stellar mass dependence}
\label{sec:c4dist-massdep}

As was done in Sec.~\ref{sec:sigma-massdep} for local density, we can
determine the contribution to the dependence of morphology on
groupocentric distance from the environmental variation in the stellar
mass function. The stellar mass distribution versus groupocentric
distance is illustrated in the right panel of
Fig.~\ref{fig:mstar-density-c4normdist}. There is a gradual increase
in stellar mass with decreasing groupocentric distance, until
$D_{\rmn{C4}} \sim 0.2 R_{\rmn{vir}}$, within which the stellar mass
distribution rapidly shifts to masses $\sim 3$ times larger. The exact
radius at which this occurs may depend somewhat upon our `Fingers of
God' correction (see Appendix~\ref{sec:FoG}). However, it is clear
that galaxies in the cores of groups are typically more massive than
those in the field.

We determine the relation between early-type fraction and stellar mass
at large groupocentric distances ($\log{D_{\rmn{C4}}} > 1$) in each
sample, and combine this with the dependence of stellar mass on
groupocentric distance. The result is a relation between early-type
fraction and groupocentric distance that is due only to the empirical
dependence of stellar-mass distribution on environment and morphology
on stellar mass, without any direct dependence of morphology on
environment. This is plotted as the thin green lines in
Fig.~\ref{fig:morph-colour-c4normdist}. Any small trend at large
groupocentric distances is attributable to the varying mass function.
Below $D_{\rmn{C4}} \sim 3 R_{\rmn{vir}}$, however, the early-type
fraction increases faster than the mass function can explain. In the
cores of groups, $D_{\rmn{C4}} \le 0.2 R_{\rmn{vir}}$, the quickly
varying stellar mass distribution causes a similar rise in the
early-type fraction, sufficient to account for most of the difference
relative to the field. However at intermediate groupocentric
distances, the varying mass function is unable to account for any of
the increase in early-type fraction. This clearly points to a separate
process occurring throughout galaxy groups which increases the
early-type fraction at a given stellar mass.

\begin{figure*}
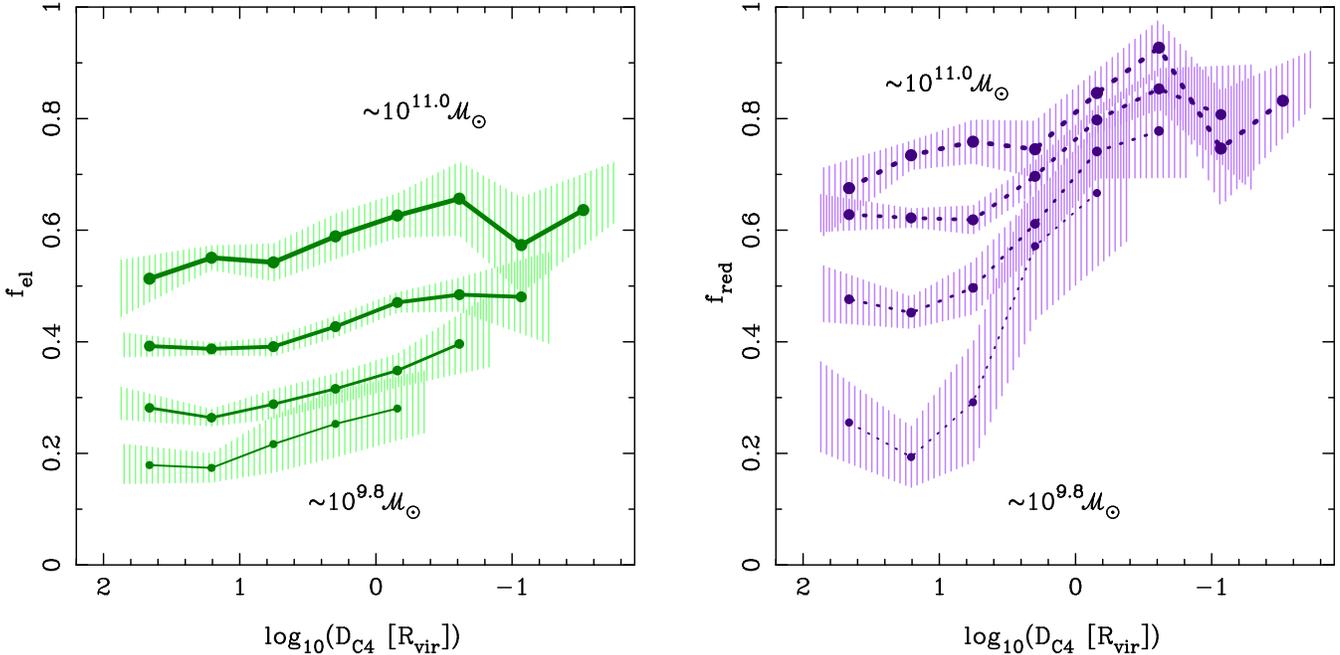

\centering
\includegraphics[height=0.475\textwidth,angle=270]{fig15_left.ps}
\hfill
\includegraphics[height=0.475\textwidth,angle=270]{fig15_right.ps}
\caption{\label{fig:morph-colour-normdist_mstarbins} (Left) early-type
  and (right) red fraction versus distance to the nearest C4 group
  normalised by its virial radius, for galaxies in narrow bins of
  stellar mass: $\log(\mathcal{M}_{\ast}/\mathcal{M}_{\odot}) =$
  9.7--9.9, 10.1--10.3, 10.5--10.7, 10.9--11.1, in order from the
  lowest to highest average $f_\rmn{el}$.  The shaded regions indicate
  the $2\sigma$ statistical error for each $\log{D_{\rmn{C4}}}$ bin.
  For each stellar mass range, the right-most $\log{D_{\rmn{C4}}}$ bin
  additionally includes all galaxies with lower $\log{D_{\rmn{C4}}}$.}
\end{figure*}

We divide the morphology--groupocentric distance relation into bins of
stellar mass in Fig.~\ref{fig:morph-colour-normdist_mstarbins}.  Due to the
smaller number of objects for which we have reliable groupocentric
distances, the stellar mass range we can probe with good statistics is
less than that for local density. Nevertheless, we can still consider
over an order of magnitude in stellar mass.  Each stellar mass bin
displays a similar trend, with offsets due to the dependence of
morphology on stellar mass.  At fixed stellar mass there is a small,
steady increase in early-type fraction at $D_{\rmn{C4}} \le 3
R_{\rmn{vir}}$, but no evidence for a trend at larger radii.

The trends in morphology with groupocentric distance are very similar
to those with respect to local density.  As explained in
Sec.~\ref{sec:data_env}, these two ways of characterising environment
are measured by different methods and are subject to different
interpretational issues.  The agreement in the relations of morphology
versus both these measures of environment demonstrates that these
trends are general and robust.

\section{Comparing morphology and colour}
\label{sec:morph-colour-env}

It is informative to compare the trends of morphology versus
environment with those of colour versus environment.  As discussed in
Sec.~\ref{sec:intro}, morphology is generally considered to be
determined by a combination of a galaxy's dynamical state and
star-formation history.  Colour is an indicator of a galaxy's recent
($\la 1$ Gyr) star-formation history, with no direct dependence on the
spatial distribution of its stars.  Differences in the behaviour of
colour and morphology with environment will thus give indications of
the environmental mechanisms at work.

Several previous studies have addressed this issue using automatic
morphology proxies. Those which consider structural measures of
morphology, such as concentration or Sersic index, find that these
quantities are much less dependent upon environment than they are on
galaxy mass \citep{2004MNRAS.353..713K,2004ApJ...601L..29H}. In
contrast, colour and star-forming fractions are strongly dependent
upon both environment and mass \citep{2006MNRAS.373..469B}. However,
\citet{2008arXiv0801.1995V} has shown that using a more sophisticated
morphology proxy, which accounts for the presence of `bumpiness' in
the surface-brightness distribution, leads one to conclude that
morphology is more dependent on environment than mass. Different
morphological and structural measurements thus lead to apparently
opposing conclusions. With the Galaxy Zoo dataset we can address these
issues using traditional visual morphologies, without recourse to
automatic proxies. As discussed in Sec.~\ref{sec:intro}, visual
morphology is primarily concerned with azimuthal structure, i.e.
spiral arms, which are not well characterised by proxies, and which
contains different information to that provided by radial structural
measurements.

In order to compare how colour and morphology depend on environment we
must adopt a measure of colour and a criterion for dividing objects
into red and blue samples.  We follow \citet{2006MNRAS.373..469B},
using the optimal, stellar-mass-dependent, divider they determine to
separate galaxies based on the bimodality of the $u-r$ colour
distribution.  This utilises the SDSS model magnitudes
\citep{2002AJ....123..485S}, and therefore the colour is effectively
centrally weighted.  We simply class objects above and below the
\citeauthor{2006MNRAS.373..469B} divider as `red' and `blue',
respectively.  We do not attempt to work with likelihoods of whether
individual galaxies lie on the red or blue sequence, though this is
possible by assuming the Gaussian fits of
\citeauthor{2006MNRAS.373..469B}.

Figure~\ref{fig:morph-colour-density} plots both the early-type and
red galaxy fractions versus local galaxy density, for our
\sample{luminosity-limited sample} (complete for $M_r < -20.17$), on the left, and \sample{$V_{\rmn{max}}$-weighted
  mass-limited sample} (complete for
$\log(\mathcal{M}_{\ast}/\mathcal{M}_{\odot}) > 9.8$), on the right.
We see that the qualitative trends for morphology and colour are
similar, but the morphology--density and colour--density relations are
offset from one another. Averaging over our \sample{luminosity-limited
  sample} the fraction of red galaxies which have early-type
morphology is 67 per cent. However, the differences in the shape of
the early-type and red fraction trends in
Fig.~\ref{fig:morph-colour-density} demonstrate that this fraction is
a function of environment. It is clear that morphology and colour are
not equivalent ways of classifying galaxies, and that they are not
equally affected by environment. We shall explore this further below.

Moving onto group properties, the left panel of
Fig.~\ref{fig:morph-colour-c4mass-c4lum} shows that, as was seen for
early-type fraction alone in the previous section, there is no clear
trend of red-fraction with group mass.  The right panel indicates a
small trend versus the summed $r$-band luminosity of group members,
though this is small compared with the variation with environment and,
as we shall see below, stellar mass. \citet{2007ApJ...664..791B} find
a strong variation of red fraction with group luminosity.  However,
this is driven by lower group masses than we consider here.  At high
group masses their red fractions match ours.  This also agrees with
\citet{2006ApJ...642..188P} who find that for SDSS groups with $\ga
10^{13.4} \mathcal{M}_{\odot}$ (assuming the velocity dispersion to
mass conversion given by eqn. 2 of \citealt{2006A&A...456...23B})
there is no mass-dependence of the star-forming fraction, while below
this there is a clear dependence.  Interestingly, they also show that
the plateau of star-forming fraction in high-mass groups has only set
in since $z \sim 0.6$.

Earlier we showed that the early-type fraction is closely related to
the groupocentric distance.  In Fig.~\ref{fig:morph-colour-c4normdist}
we show that this is also true for the red fraction.  The fraction of
red galaxies with early-type morphology varies strongly with
groupocentric distance.  As for density, however, there is a
considerable difference in the the red and early-type relations versus
groupocentric distance, in terms of both an offset and the shape of
the dependence.

As was done earlier for morphology (see Sec.~\ref{sec:sigma-massdep}),
we determine the contribution to the relations between colour and
environment due to trends in the distribution of galaxy stellar masses
versus environment. The relations between red fraction and stellar
mass for low density environments in our data is shown in
Fig.~\ref{fig:morph-colour-Mstar}, for both our
\sample{luminosity-limited} and \sample{$V_{\rmn{max}}$-weighted
  mass-limited} samples. The trends of colour versus environment
expected from these relations, combined with the environmental
dependence of stellar mass (Fig.~\ref{fig:mstar-density-c4normdist}),
are indicated by the thin, dotted lines in
Fig.\ref{fig:morph-colour-density} \&
\ref{fig:morph-colour-c4normdist}, for density and groupocentric
distance, respectively. We find that, as for morphology, the varying
stellar mass function is inadequate to explain the rapid increase in
the red fraction at intermediate to high local densities, and at
intermediate groupocentric distances. Again, this is plain evidence
for an environmental process at work in groups.

In low density environments we have seen that any
morphology--environment relation can be explained by the dependence of
the stellar mass function on environment.  In contrast, for colour
there is a strong indication of a residual trend versus both density
and groupocentric distance.  The slope of the colour--environment
relations expected due to the varying stellar mass function is not as
steep as that observed.  This strongly suggests that, while morphology
is unaltered, colour is significantly affected by environmental
variation at low densities.  Studies of galaxies in void environments
have previously found colour, and star formation, to vary with local
galaxy density
\citep{2004ApJ...617...50R,2005ApJ...624..571R,2008arXiv0805.0790C}.
Figures~\ref{fig:morph-colour-density} \&
\ref{fig:morph-colour-c4normdist} clearly demonstrate that this effect
is in addition to that expected from the varying stellar mass
function, and that it is not related to changes in the ratio of
early-types to spirals.

\subsection{Stellar mass dependence}

The colour--density and colour--groupocentric distance relations for
narrow bins of stellar mass are shown in the right panels of
Fig.~\ref{fig:morph-colour-density_mstarbins} \&
\ref{fig:morph-colour-normdist_mstarbins}, respectively. As
\citet{2006MNRAS.373..469B} have shown, the red fraction depends
strongly on environment at fixed stellar mass, particularly for low
masses ($< 10^{10} \mathcal{M}_{\odot}$), for which the red fraction
varies from $< 0.2$ to $>0.8$ between low and high-density
environments. This is in excellent agreement with
\citet{2006ApJ...647L..21H} who find that the relation between
star-formation history and environment is very different for giant and
dwarf galaxies (roughly above and below $\sim 10^{9.5}
\mathcal{M}_{\odot}$, respectively). However, rather than a sharp division
between giant and dwarf galaxies, we see a continuous, though
rapid, decrease in the sensitivity of red fraction to environment with
increasing stellar mass.

The dependence of red fraction on environment and stellar mass
contrasts with the trends for early-type fraction shown in the left
panels of these figures. The early-type and red fractions vary
similarly with environment at high masses, but for low masses colour
is much more sensitive to environment. The shapes of the environmental
dependences also differ, with the morphology relations displaying
their largest gradients at the highest densities and smallest
groupocentric distances, whereas the colour relations rise most
rapidly at intermediate densities and groupocentric distances, and
flatten off in the densest environments.

As mentioned earlier for morphology alone, the agreement of the
relations versus groupocentric distance and local galaxy density
demonstrate the robustness of these trends in galaxy properties with
respect to a general definition of environment.  We will consider the
detailed differences between the morphological and colour trends with
respect to these two environmental measures in a subsequent paper.

In this paper, we focus our investigation on the transformation of
galaxies due to environmental mechanisms. However, it is apparent that
many massive galaxies are red and have early-type morphology
independent of their environment. For some reason their star formation
has ceased, and in many cases their stellar dynamics have become
dominated by random motions, without an obvious environmental cause.
Some of these galaxies may be fossil groups
\citep{2003MNRAS.343..627J}, and hence only appear to be in
low-density environments. However, in any case, such massive, isolated
galaxies are likely to be centred in a gaseous halo, which should be
supplying them with fuel for star formation. Preventing gas-cooling,
and hence star formation, in massive galaxies remains a puzzle,
although active galactic nuclei (AGN) may play a significant role
\citep{2006MNRAS.365...11C,2006MNRAS.370..645B,2008MNRAS.386.2285C}.
There is even a possibility that AGN activity could directly affect
the morphology of galaxies \citep{2008arXiv0809.4574F}. However, AGN
are only powerful enough to have a significant impact in massive
galaxies. Furthermore, AGN activity is unlikely to depend directly
upon environment, although in principle AGN activity may vary as a
result of the same environmental processes that affect star formation.
\citet{2007MNRAS.381....7H} investigate the relative importance of AGN
versus environmental mechanisms for preventing star formation in
galaxies of different masses.

It is often implicitly assumed that the red population corresponds to
objects with early-type morphology, and that blue corresponds to
spiral.  While this is correct for the majority of objects, making the
assumption that these correspondences hold true in general can be
highly misleading.  The differences between the red and early-type
fractions imply the existence of substantial populations of
`unconventional' galaxies: red spirals and blue early-types.
Furthermore, the contrast between their dependences versus
both environment and stellar mass require that these `unconventional'
populations are not a constant contaminant, but vary strongly.  One
cannot, therefore, make inferences about the environment and mass
dependences of morphology from a consideration of colour alone.

We now examine the dependence of the red spiral and blue early-type
populations on environment and stellar mass.

\subsection{Red spirals and blue early-types}

\begin{figure*}
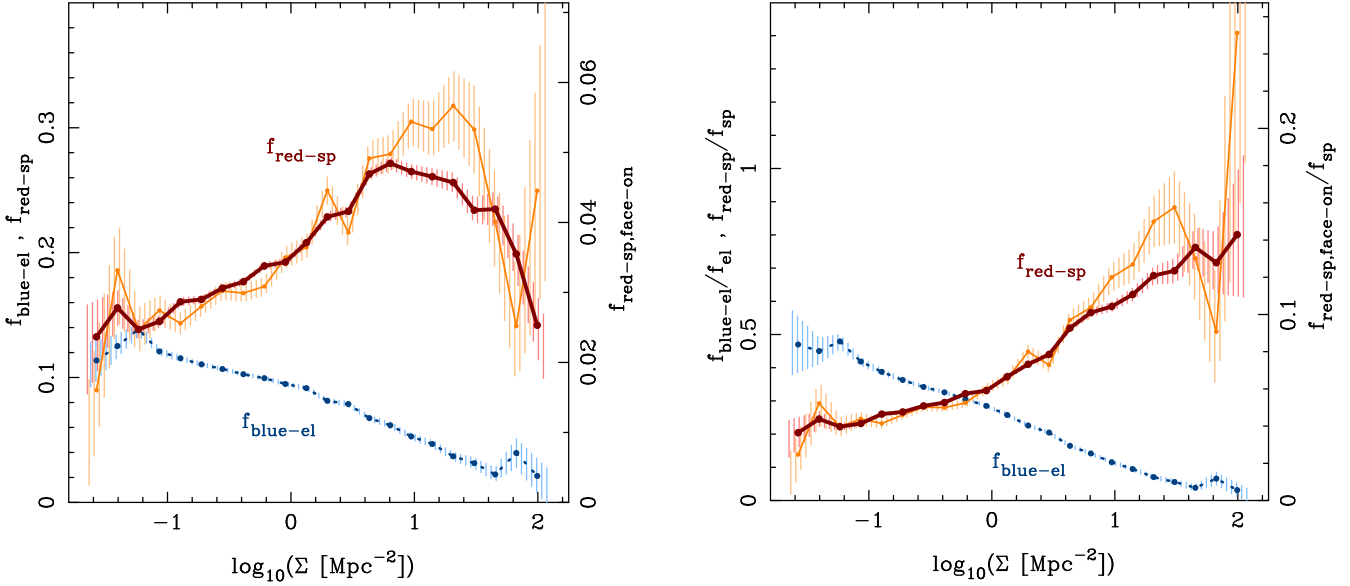

\centering 
\includegraphics[height=0.475\textwidth,angle=270]{fig16_left.ps}
\hfill
\includegraphics[height=0.475\textwidth,angle=270]{fig16_right.ps}
\caption{\label{fig:odd-density} The number of red spirals (red,
  thick, solid line) and blue early-types (blue, dotted line) in our \sample{luminosity-limited
  sample} versus local galaxy density (left) as fractions of our whole sample
  and (right) as fractions of all spirals and early-types,
  respectively.  Also shown for comparison is the fraction of
  face-on red spirals (orange, thin, solid line), scaled to have the
  same mean fraction as the full red spirals sample, as indicated by
  the scale on the right of the plot.  The shaded regions indicate the
  2-sigma statistical uncertainties on each $\log{\Sigma}$ bin. }
\end{figure*}


\begin{figure*}
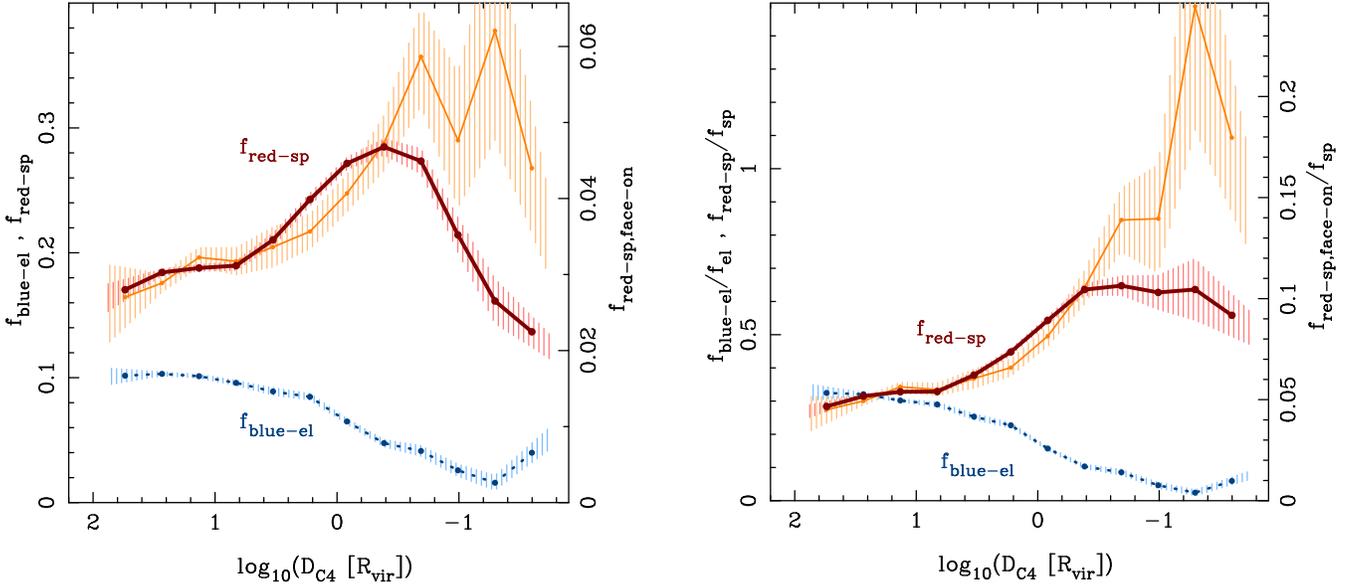

\centering
\includegraphics[height=0.475\textwidth,angle=270]{fig17_left.ps}
\hfill
\includegraphics[height=0.475\textwidth,angle=270]{fig17_right.ps}
\caption{\label{fig:odd-c4normdist} The number of red spirals (red,
  thick, solid line) and blue early-types (blue, dotted line) in our \sample{luminosity-limited
  sample} versus distance to the nearest C4 group normalised by its virial radius
  (left) as fractions of our whole sample and
  (right) as fractions of all spirals and early-types, respectively.
  Also shown for comparison is the fraction of face-on red spirals
  (orange, thin, solid line), scaled to have the same mean fraction as
  the full red spirals sample, as indicated by the scale on the right
  of the plot.  The shaded regions indicate the 2-sigma statistical
  uncertainties on each $\log{D_{\rmn{C4}}}$ bin.  The right-most
  $\log{D_{\rmn{C4}}}$ bin
  additionally includes all galaxies with lower $\log{D_{\rmn{C4}}}$.}
\end{figure*}

In the left panels of Fig.~\ref{fig:odd-density} \&
\ref{fig:odd-c4normdist} we plot the fraction of galaxies that meet
our red spiral and blue early-type criteria as functions of local
galaxy density and groupocentric distance, respectively.  We continue
to derive morphological type fractions directly from the de-biased
Galaxy Zoo type likelihoods.  Here we simply modify these for `red
spirals' by reducing the likelihood to zero for those objects bluer
than the \citet{2006MNRAS.373..469B} divider.  Likewise, for `blue
early-types' we remove objects redder than the colour divider.

The fraction of red spirals displays a clear peak versus both local
density and groupocentric distance.
In low density regions and far from C4 groups, red spirals constitute
$\sim 16$ per cent of the local galaxy population brighter than $M_r <
-20.17$.  This fraction almost doubles with increasing local density
or decreasing groupocentric distance, rising to a peak of $\sim 28$ per cent
at $\sim 6$ galaxies Mpc$^{-2}$ or $\sim 0.4 R_{\rmn{vir}}$.  At
higher densities or in the cores of groups the red spiral fraction
declines sharply.

The fraction of blue early-types simply diminishes steadily from $\sim
12$ per cent to $\sim 2$ per cent from the least dense to most dense
environments.  The extreme members of this population are studied in
detail by \citet{Kevin_GZ}, and are also found to preferentially
inhabit low-density environments.  \citet{2007ApJS..173..512S} have
also previously shown that the fraction of blue early-types declines
with local galaxy density.  The blue colours of these galaxies
indicates that they have recently formed significant numbers of new
stars, in contrast to the usual picture of early-types as `red and
dead'.  By using UV$-$optical colours, \citet{2007ApJS..173..512S}
have shown that over 30 per cent of bright ($M_r < -21.5$) early-types
have recently formed stars.  Our fractions are lower, due to the lower
sensitivity of the $(u-r)$ colour used in this paper to low levels of
recent star formation, but still substantial, and trace the same
environmental dependence.

For both red spirals and blue early-types there is little evidence for
any dependence on groupocentric distance beyond $\sim 5
R_{\rmn{vir}}$.  As previously, we find very little dependence of
group members on the mass of the group in which they reside, over
the range $10^{13.4}$~--~$10^{15.2} M_{\odot}$.  The fraction C4 group
members which are red spirals or blue early-types is flat at $\sim 26$
and $\sim 6$ per cent, respectively.

What we have discussed so far, and plotted in the left panels of
Fig.~\ref{fig:odd-density} \& \ref{fig:odd-c4normdist}, is the
fraction of all galaxies that are either red and spiral or blue and
early-type.  This quantity indicates the environments in which these
objects contribute most significantly to the overall galaxy
population.  However, it is also informative to consider the fractions
of spirals that are red, and early-types that are blue, as the spiral
and early-type fractions are themselves strong functions of
environment.  These `type normalised' fractions are plotted in the
right panels of Figs.~\ref{fig:odd-density} \&
\ref{fig:odd-c4normdist} as functions of local density
and groupocentric distance, respectively.  

The environmental dependence of the fraction of spirals that are red
is simpler than the fraction of the whole galaxy population that are
red and spiral.  We now see that the decline in the fraction of red
spirals at high densities, and in the the cores of groups, is a result
of the declining spiral fraction.  The fraction of spirals that are
red increases steadily to the highest local densities.  The dependence
versus groupocentric distance is most interesting, as it appears
localised to a range of radii.  Between $0.4$ and $6 R_{\rmn{vir}}$
the fraction of spirals that are red doubles, while outside of this
range it is constant.  A similar behaviour is seen for the fraction of
early-types that are blue, with a two-thirds decline over the same
groupocentric distance range.

\begin{figure*}
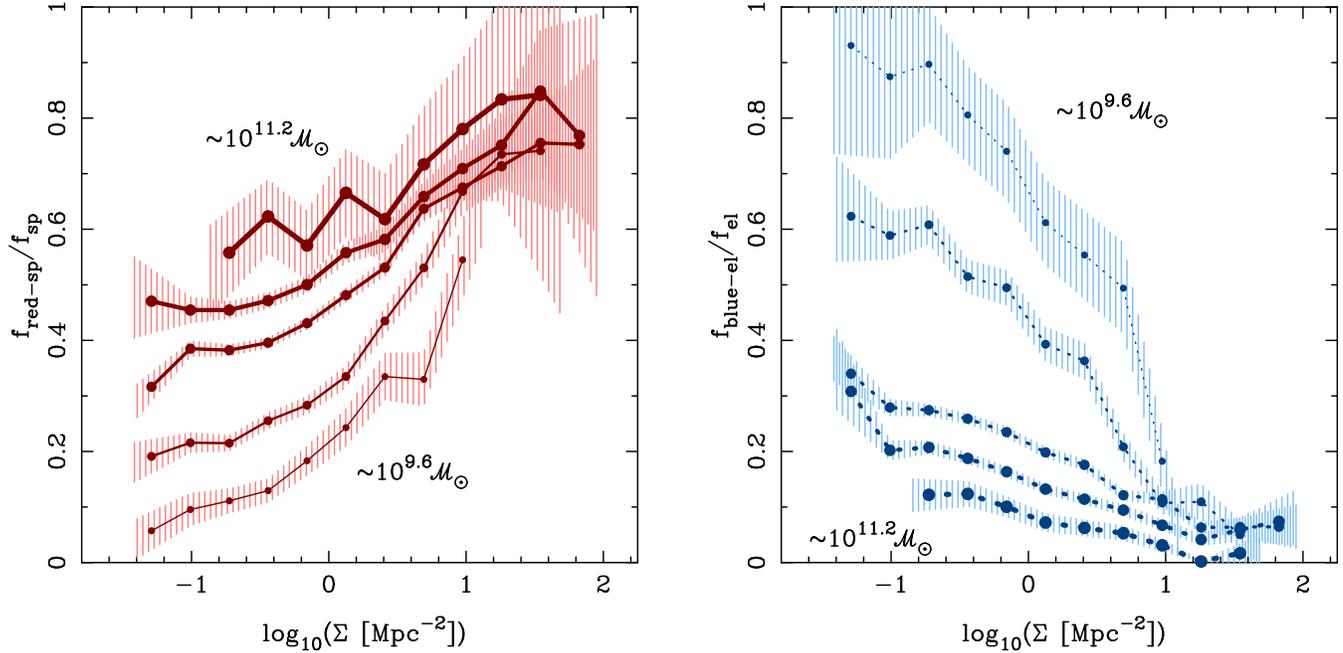

\centering
\includegraphics[height=0.475\textwidth,angle=270]{fig18_left.ps}
\hfill
\includegraphics[height=0.475\textwidth,angle=270]{fig18_right.ps}
\caption{\label{fig:odd-density_mstarbins_norm} The fraction of (left)
  spirals that are classified as red galaxies, and (right) early-types
  classified as blue galaxies, plotted
  as a function of local galaxy density, for our \sample{binned
    mass-limited sample}: $\log(\mathcal{M}_{\ast}/\mathcal{M}_{\odot}) =$ 9.5--9.7,
  9.9--10.1, 10.3--10.5, 10.7--10.9, 11.1--11.3.  The shaded regions
  indicate the 2-sigma statistical uncertainties on each
  $\log{\Sigma}$ bin. }
\end{figure*}

\begin{figure*}
\centering
\includegraphics[height=0.475\textwidth,angle=270]{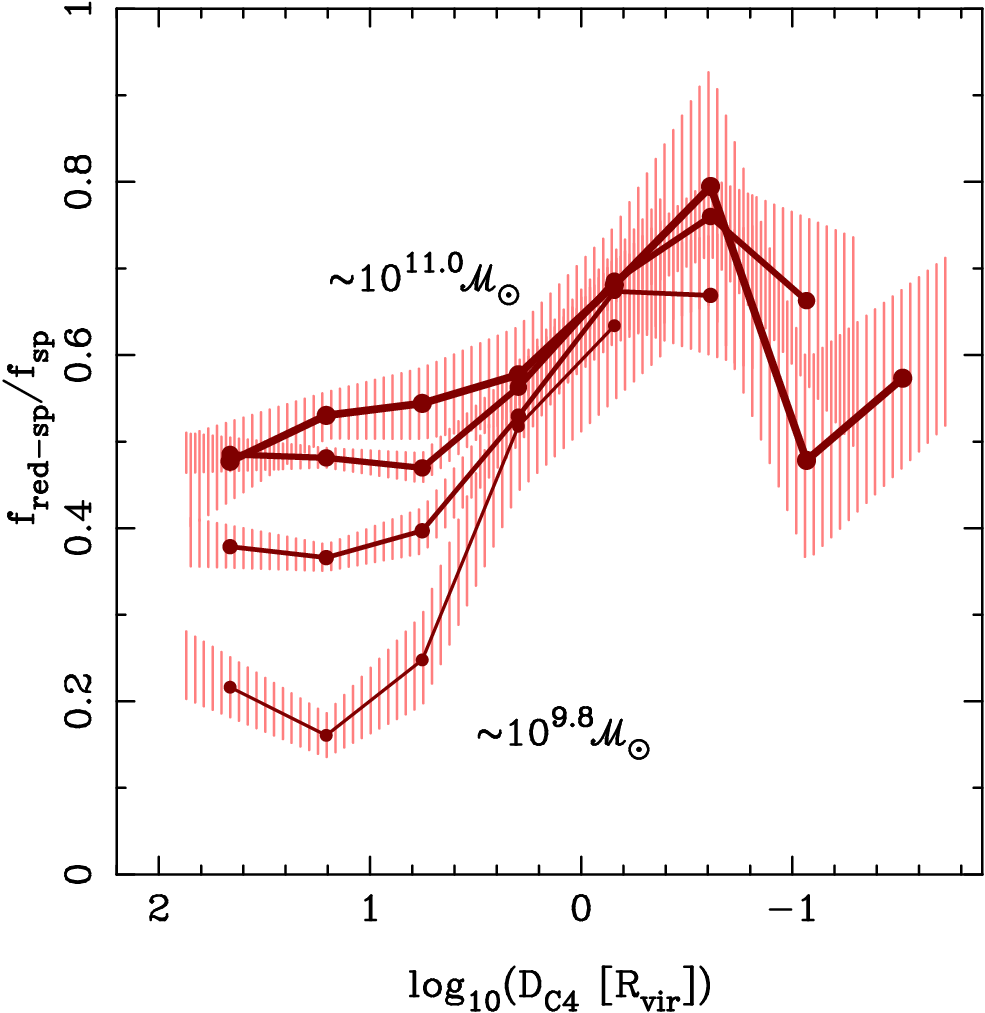}
\hfill
\includegraphics[height=0.475\textwidth,angle=270]{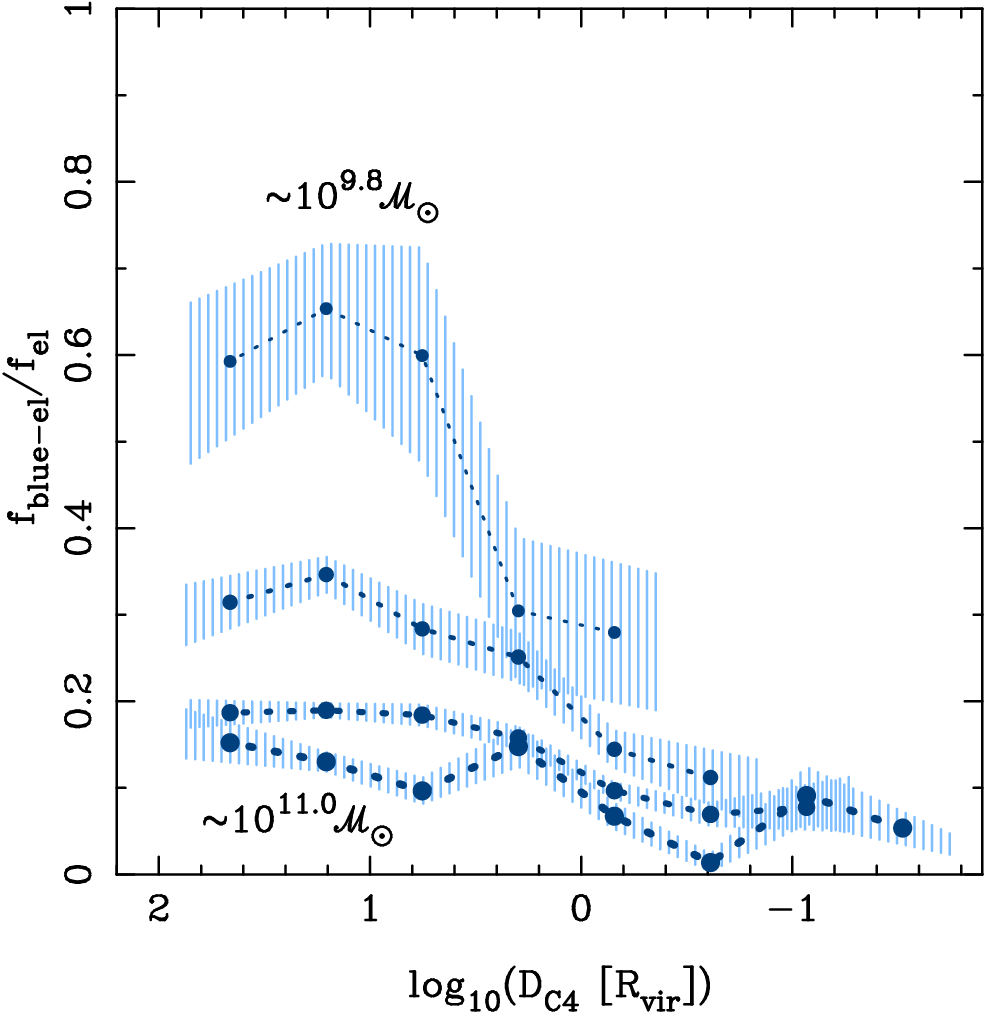}
\caption{\label{fig:odd-c4normdist_mstarbins_norm} The fraction of
  (left) spirals that are classified as red galaxies, and (right)
  early-types classified as blue galaxies, plotted as a function of
  groupocentric distance, in our \sample{binned mass-limited
  sample}: $\log(\mathcal{M}_{\ast}/\mathcal{M}_{\odot}) =$
  9.7--9.9, 10.1--10.3, 10.5--10.7, 10.9--11.1.  The shaded
  regions indicate the 2-sigma statistical uncertainties on each
  $\log{D_{\rmn{C4}}}$ bin. }
\end{figure*}

To gain further insight into the behaviour of the red spiral and blue
early-type populations with environment, we again utilise our \sample{binned
  mass-limited sample} and split our sample into complete, narrow bins
of stellar mass. Figures~\ref{fig:odd-density_mstarbins_norm} \&
\ref{fig:odd-c4normdist_mstarbins_norm} plot the fractions of spirals
that are red and early-types that are blue for these stellar mass
bins, versus local galaxy density and groupocentric distance,
respectively.

At all stellar masses there is a trend in the fractions of spirals
that are red and early-types that are blue versus environment. In low
density environments these fractions are strong functions of stellar
mass. In these environments, at high stellar masses,
$\mathcal{M}_{\ast} \ga 10^{11} \mathcal{M}_{\odot}$, the majority of
spirals, and almost all early-types, are red; at low stellar masses,
$\mathcal{M}_{\ast} \la 10^{9.6} \mathcal{M}_{\odot}$, almost all
galaxies are blue, irrespective of their morphology. This corresponds
with the finding of \citet{2006MNRAS.373.1389C} that the correlation
between galaxy colour and stellar mass does not strongly depend on
morphology. It does, however, depend strongly on environment. The
environmental dependence for low mass galaxies is much stronger than
for high mass galaxies, such that in high density environments
($\Sigma > 10$ Mpc$^{-2}$, $D_{\rmn{C4}} < 0.3 R_{\rmn{vir}}$) 80 per
cent of spirals and 90 per cent of early-types are red, irrespective
of their stellar mass. In dense environments the fractions of spirals
that are red and early-types that are blue are therefore only weak
functions of stellar mass. 

\subsubsection{S0 galaxies and red spirals}

We see a significant fraction of red spirals in the field,
particularly at high stellar masses.  This results from a lack of
correspondence between the morphology and colour bimodalities
\citep{2007AJ....134.1508V}, which is present across all environments.
Many massive galaxies appear on the red side of the colour bimodality,
irrespective of their morphology.  As shown in appendix A of
\citet{2008arXiv0802.4421J}, with typical selections one finds that
half to two-thirds of Sa galaxies are red. \citet{2008arXiv0801.3286M}
have also shown that many red (and concentrated) objects are truly
disk galaxies. These would often be mistakenly considered to be
early-types in studies using colour and concentration as proxies for
morphology.  About one-third of these objects are red simply due to
being orientated edge-on and thus suffering from significant dust
reddening.

A substantial fraction of the red spiral population we have identified
shows a strong dependence on environment, preferentially occurring in
intermediate local densities and within the virial radius, but not the
cores, of galaxy groups.  It is unclear whether our red spiral sample
comprises only similar objects, which display only limited
environmental variation, or whether it is made up from several
distinct populations with different environmental dependences.
However, we do not see any change in the fraction of red-spirals that
are classified as edge-on/unclear as a function of environment.  Thus,
the environmental trends cannot simply be attributed to a change in
the proportion of red-spirals that are selected without having clearly
visible spiral arms.

S0 galaxies possess discs and are generally found to have little star
formation and to be red in colour.  They are also known to
preferentially inhabit clusters, especially at low redshifts
\citep[e.g.,][]{1980ApJ...236..351D}.  This morphological type is not
explicitly identified in the current Galaxy Zoo classification scheme,
and S0s are thus distributed between the early-type and spiral classes
in an uncertain manner.  There is therefore a concern that the
environmental behaviour we have found for red spirals may simply be
due to the S0 population.  However, recall that
Fig.~\ref{fig:p_distribs} demonstrates that almost all the objects
classified as E/S0 or S0 by F07 have low spiral likelihoods in Galaxy
Zoo.  These objects therefore contribute very little to the spiral
fraction.  Objects classified by F07 as S0/Sa also generally have low
spiral likelihoods, and many of these, especially those with higher
spiral likelihoods, may be better classified as spiral rather than S0,
as they must have a suggestion of spiral arms.  It is thus apparent
that the red spirals we discuss above are an additional population to
the S0s considered in many other studies.

There remains the possibility that Galaxy Zoo, F07, and thus
presumably most other studies, mistakenly classify a significant
fraction of edge-on S0s as spirals.  In Galaxy Zoo we can ask the
question of whether objects were classified as spiral due to the
presence of visible spiral arms or simply due to a disky, edge-on
appearance.  We define as `edge-on/unclear' those objects for which
the majority of Galaxy Zoo classifiers could not individually discern
a spiral arm direction ($p_{CW} + p_{ACW} < p_{EU}$ where CW, ACW and
EU refer to clockwise, anti-clockwise and edge-on/unclear,
respectively).  Note that this is conservative, as if even a minority
of classifiers can discern spiral arms in a galaxy then it is very
likely to be a spiral.  The classification bias correction (see
Appendix~\ref{sec:bias}) increases the spiral likelihood for such
objects when they have apparent luminosities or sizes which may make
distinguishing them from early-types difficult.  The spin likelihoods
($p_{CW}$, $p_{ACW}$, $p_{EU}$) have not been corrected for such
biases, and simply reflect the proportion of user
classifications. Spiral galaxies may therefore include a surprisingly
high fraction of edge-on/unclear objects, given the above definition.

For all objects that are more likely to be spirals than anything else
($p_{\rmn{sp,adj}} > 0.5$), the fraction of edge-on/unclear objects is
$55$ per cent, while for red spirals it is $81$ per cent.  There are
three possible reasons for this discrepancy: (a) edge-on spirals
suffer from greater dust-reddening, so will be more prevalent in a red
sample; (b) due to the connection between star formation and spiral
arms, objects with low specific star-formation rates will tend to be
red and have lower spiral arm contrast, and are thus more likely to
be classified as edge-on/unclear; and (c) a fraction of objects
classified as edge-on/unclear spirals are actually S0 galaxies, with
typically redder colours.  For cases (b) and (c), there must be a
population of these objects viewed closer to face-on and hence
classified as early-type.  The environmental trends we see in a
fraction of the red spirals must correspond to some combination of
environmental variations in the (a) dust-content, (b) star-formation
rate, and (c) S0 contamination of the spiral galaxy population.

Edge-on/unclear spirals are, in the median, $0.29$~mag redder than
galaxies with clearly visible spiral arms (in $(u-r)_{\rmn{model}}$,
relative to the stellar-mass-dependent divider of
\citealt{2006MNRAS.373..469B}).  However, determining whether this is
due to extinction affecting otherwise normal spiral galaxies, or due
to older luminosity-weighted stellar populations, is beyond the scope
of this paper.

For 19 per cent of our red spirals significant spiral arms are seen by
a majority of classifiers, and the true fraction with spiral arms,
including those undetected in Galaxy Zoo, must be higher.  At least a
fifth of our red-spiral population must therefore be true spirals, not
S0 galaxies.  We can check that these face-on spirals, which we are
certain possess spiral arms, behave in the same manner as our whole
red spiral population.  This is shown by the orange lines in
Figs.~\ref{fig:odd-density} \& \ref{fig:odd-c4normdist}, which plot
the fractions for face-on red spirals only, scaled by
the ratio of all red spirals to face-on red spirals.  As a function of
local density, the fraction of face-on red spirals behaves very
similarly to the full red spiral sample.  Versus groupocentric
distance they also behave similarly, except within $D_{\rmn{C4}} \la
0.4 R_{\rmn{vir}}$, where there is an excess of face-on red spirals
compared with the full red spiral population.  This suggests that in
groups edge-on red spirals are less likely to be classified as spirals
than they are in the field.  This is interesting in itself, but here
we simply stress that this implies that our red spiral population is
not dominated by S0s in groups, rather there may be proportionally
fewer S0s in our red spiral sample in dense regions.  A substantial
fraction of our red spiral population thus appears to comprise
galaxies with significant spiral arm structure, even if many are seen
edge-on.

As explained in Sec.~\ref{sec:basicdata}, the colours used in this
study are based on SDSS model  magnitudes.  They are thus
sensitive to the dominance of the bulge component of a disc galaxy.
In contrast, the Galaxy Zoo morphologies are sensitive to the presence
of spiral arms or an edge-on disk, even when the galaxy light is
dominated by a bulge.  This is because the images were displayed for
classification using a non-linear (arcsinh) stretch, as described in
\citet{2004PASP..116..133L}.  Other photometric apertures, such as
Petrosian, may give colours that are more representative of the
overall galaxy, and so which may be more suitable for identifying
galaxies with red discs.  Measuring colour in an annular aperture may
be most appropriate for this purpose \citep{2007ApJ...658..898P}.  The
choice of other photometric bands (e.g., $g-r$) and dividing lines in
the colour--magnitude diagram may somewhat modify the trends shown.
However, we do not expect the overall behaviour of the red-spiral
population versus environment to depend dramatically upon our choice
of colour.

\subsubsection{Breaking down the dependence of red fraction on
  environment}

Having studied the relationship between morphology and environment,
and the environmental variations in the fractions of early-types and
spirals that are red, we are now in a position where we can identify
the various contributions to the dependence of red fraction on environment.

The red fraction can be written as a sum of contributions from
red spirals and red early-types:
\begin{equation}
f_{\rmn{red}} = \frac{f_{\rmn{red-sp}}}{f_{\rmn{sp}}}f_{\rmn{sp}} + \left(1 - \frac{f_{\rmn{blue-el}}}{f_{\rmn{el}}}\right)f_{\rmn{el}}\,. \label{eqn:fred}
\end{equation}
We have already determined $f_{\rmn{sp}}$, $f_{\rmn{el}}$,
$f_{\rmn{red-sp}}/f_{\rmn{sp}}$ and $f_{\rmn{blue-el}}/f_{\rmn{el}}$,
as functions of density.  To assess the red fraction resulting from
the morphology--density relation we simply set
$f_{\rmn{red-sp}}/f_{\rmn{sp}}$ and $f_{\rmn{blue-el}}/f_{\rmn{el}}$
to their average values in Eqn.~\ref{eqn:fred}, while allowing
$f_{\rmn{sp}}$ and $f_{\rmn{el}}$ to vary with density.  The
contribution from the changing fraction of spirals that are red is
found by holding all terms except
$f_{\rmn{red-sp}}/f_{\rmn{sp}}$ constant at their mean values.
Finally, by holding all terms except
$f_{\rmn{blue-el}}/f_{\rmn{el}}$ at their mean values we
determine the contribution from the environmental variation in the
fraction of early-types which are blue.

\begin{figure*}
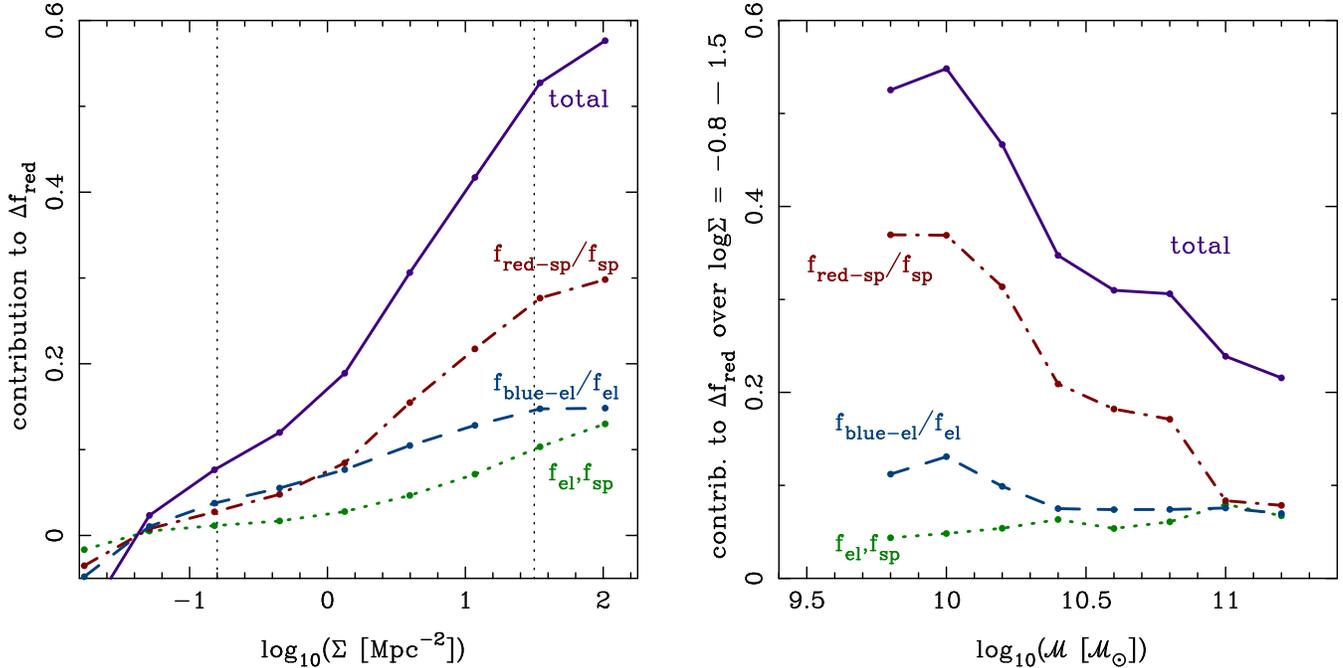

\centering
\includegraphics[height=0.475\textwidth,angle=270]{fig20_left.ps}
\hfill
\includegraphics[height=0.475\textwidth,angle=270]{fig20_right.ps}
\caption{\label{fig:density_f_red_contributions} (Left) the change in
  the red fraction of galaxies in our \sample{luminosity-limited sample}, divided
  into contributions due to the variation with local density of the
  morphological type fractions (green, dotted line), the fraction of
  spirals that are red (red, dot-dashed line) and the fraction of
  early-types that are blue (blue, dashed line).  The total
  colour--density trend is shown by the solid, purple line.  (Right)
  the contributions to the change in red fraction over the local
  density range $\log\Sigma = 0.8$--$1.5$ as a function of stellar
  mass (indicated by the vertical dotted lines in the left panel).}
\end{figure*}

These contributions to the relation between red fraction and local
density, for our \sample{luminosity-limited sample}, are plotted in the left panel
of Fig.~\ref{fig:density_f_red_contributions}.  In order to examine
only trends, rather than absolute values, we normalise each curve to
the mean value of the three lowest density bins.  The most striking
result of this exercise, though already partly clear from
Fig.~\ref{fig:morph-colour-density}, is that the morphology--density
relation is responsible for very little of the colour--density
relation.  The modest increase in the early-type fraction with local density
is not enough to explain the much larger increase in red fraction.
Furthermore, as the early-type and spiral populations both contain
significant fractions of red galaxies, which increase to comparable proportions in dense environments, the rise in the early-type fraction with density has
surprisingly little effect on the red fraction.  The colour--density
relation is driven by changes in the red fraction at fixed morphology,
i.e. the fraction of early-types and spirals that have red colours.
In low density environments the rise of red spirals and fall of blue
early-types contribute similarly.  At higher densities the rapidly
increasing red spiral fraction is the dominant contribution to the
colour--density relation.

Note that the variation of the stellar mass function with environment is
still present in Fig.~\ref{fig:density_f_red_contributions}.  Its influence
will be shared amongst the various contributions, for as we have seen,
early-type fraction and the fraction of early-types and spirals that are red
depend strongly on stellar mass
(Fig.~\ref{fig:morph-colour-density_mstarbins} and
Fig.~\ref{fig:odd-density_mstarbins_norm}).

The relative importance of the red spiral population varies with
stellar mass.  To show this we determine the contribution to the
colour--density relation over the range $\log\Sigma = 0.8$--$1.5$ in
narrow bins of stellar mass.  These are shown in the right panel of
Fig.~\ref{fig:density_f_red_contributions}.  Using narrow stellar mass
bins also has the advantage of removing the effect of environmental
variations in the stellar mass function.  Red spirals dominate the
colour--density relation at low stellar masses, but at high masses red
spirals, blue early-types, and the morphology--density relation
contribute equally.

\subsubsection{Why are galaxies red in dense environments?}

Our red spirals, and particularly the portion which are prevalent in
the outskirts of galaxy groups, bear a strong resemblance to
populations discussed by a number of previous studies.  These include
the anaemic spirals of \citet{1976ApJ...206..883V}, the passive
spirals identified by \citet{2003PASJ...55..757G} and confirmed by
\citet{2004MNRAS.352..815Y}, H$\alpha$ truncated
\citep[e.g.,][]{2004ApJ...613..866K} and H\textsc{I}-deficient spirals
\citep[e.g.,][]{2004AJ....127.3300V}.  More recently
\citet{2007MNRAS.378..716L} have studied a population of dusty, red
spirals associated with the intermediate density regions in a
supercluster, which \citet{2007AAS...211.6701W} and
\citet{STAGES_sf_red_spirals} find to still be hosting low levels of
highly-obscured star formation.  All of these populations are
obviously closely related, and their memberships must overlap
considerably.  However, the various selections may highlight objects
at different stages of an evolutionary sequence, or subject to the
action of different mechanisms.  In general, most of these red spiral
populations appear to result from the removal of gas from the haloes
and/or discs of normal spiral galaxies in cluster and group
environments.

Considering Fig.~\ref{fig:odd-density}, one might conclude that a more
general, rather than a group-specific, mechanism is responsible for
the environmental variation in the red spiral fraction.  This is
implied by the finding that the slope of the red-spiral fraction
versus local density relation declines smoothly and does not entirely
flatten-off at low densities.  For example, such a conclusion is
reached by \citet{2004MNRAS.348.1355B} due to a similar behaviour of
the star-forming fraction versus local density.  However, when
considering groupocentric distance as the environmental indicator, in
Fig.~\ref{fig:odd-c4normdist}, we see that almost all the variation in
the red-spiral fraction occurs within a few $R_{\rmn{vir}}$.  The mass
of the group involved does not significantly affect the red-spiral
fraction, at least above some threshold which lies below the range of
group masses considered in this paper ($\ga 10^{13}
\mathcal{M}_{\odot}$).

Recall from Fig.~\ref{fig:sigma_c4mass} that there is a spread of
local density for group members: and thus a sharp transition in
groupocentric distance will result in a smoother variation with
respect to local density.  This argument suggests that the majority of
the environmental dependence of red-spirals is due to group
membership, rather than local density, and hence that a significant
proportion of red spirals are a result of group-specific mechanisms.

In the cores of groups, many galaxies have early-type morphology and
red colours.  Much of this can be explained by a shift of the stellar
mass function to higher masses in these regions, together with the
higher early-type and red fractions for more massive galaxies
(Figs~\ref{fig:mstar-density-c4normdist} \&
\ref{fig:morph-colour-Mstar}).  However, this does not help to explain
the high early-type and red fractions in the outskirts of groups.  At
a given stellar mass the early-type fraction is higher in dense
environments.  This indicates that some process must have created
early-types preferentially in dense regions, either at the epoch of
galaxy formation or subsequently.  A recent direct study of evolution
in the morphology--density relation by \citet{2007ApJ...670..206V}
demonstrates that the the relation has not significantly changed out
to $z \sim 1$, when one considers a mass-selected sample.  The
evolution found by previous studies
\citep{2005ApJ...620...78S,2005ApJ...623..721P} is attributed to
effects of their luminosity-based selection.  The star-formation
properties of galaxies in dense regions appears to have changed, while
their morphologies have not.  This is at odds with the picture of
spirals transforming to S0s in clusters
\citep{1997ApJ...490..577D,2006MNRAS.373.1125B,2006A&A...458..101A,2007A&A...470..173B,2007ApJ...660.1151D}.
However, this picture remains contentious, with other studies
concluding that luminous S0s cannot have formed from faded spirals
because of their larger bulge sizes and luminosities, relative to
those of spirals
\citep{1979ApJ...234..435B,1980ApJ...236..351D,1980AJ.....85..623G,2004ApJ...616..192C,2006PASP..118..517B}. The
study of \citet{2007ApJ...670..206V} is limited to massive galaxies
($\mathcal{M}_{\ast} \la 10^{10.6} \mathcal{M}_{\odot}$), so there is
still room for the morphology--density relation to evolve at lower
masses.

The significant population of red spirals we identify, which
preferentially reside in the outskirts of groups, are predominantly of
intermediate stellar mass. Furthermore, the lowest mass objects appear
to be those which are most strongly affected by their environment, as
in the field they are almost all blue. The star formation in these
objects must have declined recently, within the past few Gyr, in order
for their spiral arms to remain with sufficient contrast to be
visible. Indeed, \citet{2007MNRAS.376L...1W} find that when one
considers stellar population age, rather than simply the colour
bimodality, the morphology--density relation may be explained by an
age--density relation and age-morphology relation (using more finely
divided morphologies than our study). Their results imply that, in
clusters, red spirals are younger than red early-types, and that both
the morphology of a galaxy and its distance to the centre of the
cluster are strongly related to the length of time since the galaxy's
star formation ceased, presumably when it entered the group
environment. The timescale implied by red colour but visible spiral
arms suggests that our red spirals are a likely candidate for the
population of relatively-faint, red galaxies that has built up in
clusters in recent times \citep{2007ApJ...660.1151D}. Unless these red
spirals are somehow `revived', the only clear option for them is to
evolve into galaxies resembling S0s. It is not a large step to presume
that the objects that began their star-formation decline earliest may
already appear to have S0 morphology. The existence of red spirals in
dense environments thus lends support to the hypothesis of spirals
evolving to S0s through environmental processes.

It seems likely that low mass S0s do form from spirals, while high
mass S0s were formed more directly at an earlier epoch, in a manner
similar to ellipticals
\citep{2001ApJ...563..118P,2003A&A...407..423M,2008arXiv0803.0305B}.
However, Fig.~\ref{fig:morph-colour-density_mstarbins} shows that a
morphology--density relation exists at all stellar masses.  Galaxies
of all stellar masses are thus subject to environmental processes,
though the mechanisms and the epoch at which they were effective may
vary with mass.  Candidate mechanisms, in order of declining
effectiveness, include major mergers, minor mergers
\citep{1994ApJ...425L..13M}, harassment \citep{1996Natur.379..613M}
and cluster tidal effects \citep{1983ApJ...264...24M}.  All of these
may heat or destroy a spiral's stellar disk, leading to an early-type
morphology.  Our results clearly show that galaxy colour is
transformed more readily than morphology, particularly for low mass
galaxies, so we should expect to see significant numbers of red
spirals, even if they will eventually evolve into S0s.

Figures~\ref{fig:odd-density_mstarbins_norm} \&
\ref{fig:odd-c4normdist_mstarbins_norm} provide conclusive evidence
for a process closely related to environment which causes galaxies
with identical stellar mass and morphology to be more likely to be red
in denser regions.  This affects both spiral and early-type galaxies,
though red spirals are more common than blue early-types.  It affects
low mass galaxies significantly more than those with high stellar
mass.  This process, or another, also results in an increase in the
early-type fraction in dense environments, but a transition from blue
to red is not generally accompanied by a transition from spiral to
early-type.  This rules out major mergers as the responsible
mechanism, as they generally result in an early-type morphology
(however, see \citet{2008arXiv0806.1739H}).  The most likely culprit
is starvation \citep{1980ApJ...237..692L}, where the gaseous halo of a
galaxy is removed by some environmental process, and therefore ceases
to supply cold gas in the disk.  As the current store is consumed by
star formation, there is no more gas available and star formation
declines and eventually terminates.  The main candidate mechanisms for
removing the halo gas are thermal evaporation
\citep{1977Natur.266..501C}, ram-pressure stripping
\citep{1972ApJ...176....1G}, viscous stripping
\citep{1982MNRAS.198.1007N}, and interaction with the cluster tidal
field \citep{1983ApJ...264...24M}.  Direct stripping of gas from the
disks of galaxies may also be important within groups
\citep{2006PASP..118..517B}.  Note that at higher redshift,
\citet{2008ApJ...684..888P} present signs that the environmental
dependences of star formation and morphology may be more strongly
linked than they are today.  They also find an indication that
star formation is enhanced in intermediate-density environments at $z
\sim 0.6$ (see also \citealt{2005MNRAS.361..109B}), in contrast to the
independence \citep[e.g.,][]{2004MNRAS.348.1355B} or suppression
\citep[e.g.,][]{2004ApJ...613..866K} of star formation with respect to
increasing local density seen nowadays. This suggests that the
dominant mechanism transforming galaxies in dense environments may
change with cosmic time.

The cessation of star formation in galaxies entering the halo of a
larger galaxy is a key ingredient of models of galaxy evolution, both
semi-analytic models \citep[e.g.][]{1993MNRAS.264..201K} and analytic
halo models \citep{2002PhR...372....1C,2008arXiv0805.0310S}. All of
these models make a critical distinction between the central and
satellite galaxies in a halo. The termination of star formation in
satellites, and the manner in which it has been achieved, has
generally been handled very simplistically in these models. Even so,
with the addition of various assumptions, these models have met with
considerable success in reproducing broad features of the galaxy
population. Recent refinements have led to significant improvements in
the quantitative agreement of model satellite colours with
observations \citep{2008MNRAS.389.1619F,2008arXiv0805.1233S}. However,
there are many aspects which remain to be reconciled with
observations. In particular, comparison between the morphology
dependences predicted by these models and observations has barely
begun. The dependence of morphology and colour on environment and
stellar mass outlined in this paper provide clear observations with
which to refine the next generation of models.

\section{Conclusions}
\label{sec:conclusions}

The Galaxy Zoo project has produced a catalogue of visual
morphological galaxy classifications more than an order of magnitude
larger than any previous catalogue.  We have illustrated the biases
present in this dataset, which are likely to have affected all
previous work using morphologies to some extent, but which are only
directly quantifiable in a dataset of this size.  More importantly, we
have demonstrated a procedure for measuring and correcting for these
biases, such that they may be reliably used for statistical studies of
morphological type fractions.  Without this correction procedure the
number of galaxies which may be used for such analyses is severely
limited.

With the de-biased Galaxy Zoo dataset we have examined the
relationship between morphological type fractions and environment,
characterised by both local galaxy density and the distance to the
nearest galaxy group.  We reproduce the trends seen by earlier
studies, but in much greater detail.  The early-type fraction rises
smoothly with increasing local density or decreasing groupocentric
distance.  Part of this trend is due to an increased proportion of
massive galaxies in dense environments, which are preferentially
early-types in any environment.  However, at fixed stellar mass, a
morphology--density relation is still present, with a similar change
in early-type fraction with environment at all stellar masses.

Remarkably, the morphological fraction of group members varies very
little with the mass of the group, either determined from the velocity
dispersion or using integrated light as a mass proxy.

In studies based on modern surveys, galaxies on either side of the
colour bimodality are often discussed in morphological language.
However, colour and morphological fractions depend differently on
environment. The fraction of red galaxies varies more strongly than
the early-type fraction does.  The result is that there is a
considerably higher fraction of galaxies with red colour but spiral
morphology in denser environments.  There is also a substantial
population of galaxies with blue colour but early-type morphology in
low-density environments.  The prevalence of red spirals peaks at
intermediate local densities and within a few times the virial radius
of galaxy groups.  These objects mostly have true spiral morphology,
and are additional to the S0 population. Their fraction decreases in
the group cores and at the highest densities, although this is largely
due to the rapidly declining fraction of all spirals in these regions.
The fraction of spirals that are red increases steadily with density.
Likewise, the fraction of early-types with blue colours declines
steadily with increasing environmental density.  High stellar mass
galaxies, both spiral and early-type, are significantly less affected
by their environment than galaxies with low stellar mass.  Most low
mass galaxies, of any morphology, are blue in the field and red in
dense environments.

Only a small fraction of the colour--density relation is a consequence
of the morphology--density relation.  It is primarily driven by the
environmental dependence of the fraction of red galaxies at fixed
morphology: red spirals and blue early-types.  Note that we consider
morphology in terms of spiral versus early-type in this paper.  It is
remains possible that there is as stronger link between the
colour-density and morphology-density relations in terms of an
environmental trend of morphology within the spiral class (i.e., Sa,
Sb, Sc, \dots).  However, such a trend must be due to more gentle
processes than those which effect a full transformation from spiral to
early-type.  We therefore rule out mechanisms sufficiently violent to
destroy galaxy disks, such as major mergers, as the primary cause of
the colour-density relation.

These results catalogue a complex relationship between the morphology,
colour, stellar mass and environment of galaxies, which successful models of
galaxy formation and evolution must explain.

\section*{Acknowledgements}

We are grateful to Ian Smail and David Wake for constructive comments
on an earlier draft of this paper, and thank Ramin Skibba for
insightful discussions.  We also thank the referee for helpful
comments which have improved the clarity of this paper.

This work has depended on the participation of many members of the
public in visually classifying SDSS galaxies on the Galaxy Zoo
website. We thank them for there extraordinary efforts in making this
project a success. We are also indebted to various members of the
media, both traditional and online, for helping to bring this project
to the public's attention.

SPB acknowledges support from STFC. KL was funded by a Glasstone
research fellowship, and further supported by Christ Church, Oxford.
CJL acknowledges support from the STFC Science in Society Program. KS
was supported by the Henry Skynner Junior Research Fellowship at
Balliol College, Oxford.

Funding for the Sloan Digital Sky Survey (SDSS) and SDSS-II has been
provided by the Alfred P. Sloan Foundation, the Participating
Institutions, the National Science Foundation, the U.S. Department of
Energy, the National Aeronautics and Space Administration, the
Japanese Monbukagakusho, and the Max Planck Society, and the Higher
Education Funding Council for England. The SDSS Web site is
http://www.sdss.org/.

The SDSS is managed by the Astrophysical Research Consortium (ARC) for
the Participating Institutions. The Participating Institutions are the
American Museum of Natural History, Astrophysical Institute Potsdam,
University of Basel, University of Cambridge, Case Western Reserve
University, The University of Chicago, Drexel University, Fermilab,
the Institute for Advanced Study, the Japan Participation Group, The
Johns Hopkins University, the Joint Institute for Nuclear
Astrophysics, the Kavli Institute for Particle Astrophysics and
Cosmology, the Korean Scientist Group, the Chinese Academy of Sciences
(LAMOST), Los Alamos National Laboratory, the Max-Planck-Institute for
Astronomy (MPIA), the Max-Planck-Institute for Astrophysics (MPA), New
Mexico State University, Ohio State University, University of
Pittsburgh, University of Portsmouth, Princeton University, the United
States Naval Observatory, and the University of Washington.

\bsp

\bibliographystyle{mn2e}
\bibliography{bamford_gz_mc}

\begin{thebibliography}{}

\bibitem[\protect\citeauthoryear{{Adelman-McCarthy} et~al.,}{{Adelman-McCarthy}
   et~al.}{2008}]{2007arXiv0707.3413A}
{Adelman-McCarthy} J.~K.,  et~al., 2008, \apjs, 175, 297

\bibitem[\protect\citeauthoryear{{Allen}, {Driver}, {Graham}, {Cameron},
  {Liske} \& {de Propris}}{{Allen} et~al.}{2006}]{2006MNRAS.371....2A}
{Allen} P.~D.,  {Driver} S.~P.,  {Graham} A.~W.,  {Cameron} E.,  {Liske} J.,
  {de Propris} R.,  2006, \mnras, 371, 2

\bibitem[\protect\citeauthoryear{{Ann}, {Park} \& {Choi}}{{Ann}
  et~al.}{2008}]{2008arXiv0805.0637A}
{Ann} H.~B.,  {Park} C.,    {Choi} Y.-Y.,  2008, \mnras, 389, 86

\bibitem[\protect\citeauthoryear{{Arag{\'o}n-Salamanca}, {Bedregal} \&
  {Merrifield}}{{Arag{\'o}n-Salamanca} et~al.}{2006}]{2006A&A...458..101A}
{Arag{\'o}n-Salamanca} A.,  {Bedregal} A.~G.,    {Merrifield} M.~R.,  2006,
  \aap, 458, 101

\bibitem[\protect\citeauthoryear{{Baldry}, {Balogh}, {Bower}, {Glazebrook} \&
  {Nichol}}{{Baldry} et~al.}{2004}]{2004AIPC..743..106B}
{Baldry} I.~K.,  {Balogh} M.~L.,  {Bower} R.,  {Glazebrook} K.,    {Nichol}
  R.~C.,  2004, in {Allen} R.~E.,  {Nanopoulos} D.~V.,   {Pope} C.~N.,  eds,
  The New Cosmology: Conference on Strings and Cosmology Vol.~743 of Am. Inst.
  of Phys. Conf. Series, {Color bimodality: Implications for galaxy evolution}.
pp 106--119

\bibitem[\protect\citeauthoryear{{Baldry}, {Balogh}, {Bower}, {Glazebrook},
  {Nichol}, {Bamford} \& {Budavari}}{{Baldry}
  et~al.}{2006}]{2006MNRAS.373..469B}
{Baldry} I.~K.,  {Balogh} M.~L.,  {Bower} R.~G.,  {Glazebrook} K.,  {Nichol}
  R.~C.,  {Bamford} S.~P.,    {Budavari} T.,  2006, \mnras, 373, 469

\bibitem[\protect\citeauthoryear{{Ball}, {Loveday} \& {Brunner}}{{Ball}
  et~al.}{2008}]{2008MNRAS.383..907B}
{Ball} N.~M.,  {Loveday} J.,    {Brunner} R.~J.,  2008, \mnras, 383, 907

\bibitem[\protect\citeauthoryear{{Ball}, {Loveday}, {Fukugita}, {Nakamura},
  {Okamura}, {Brinkmann} \& {Brunner}}{{Ball}
  et~al.}{2004}]{2004MNRAS.348.1038B}
{Ball} N.~M.,  {Loveday} J.,  {Fukugita} M.,  {Nakamura} O.,  {Okamura} S.,
  {Brinkmann} J.,    {Brunner} R.~J.,  2004, \mnras, 348, 1038

\bibitem[\protect\citeauthoryear{{Balogh} et~al.,}{{Balogh}
  et~al.}{2004}]{2004MNRAS.348.1355B}
{Balogh} M.,  et~al., 2004, \mnras, 348, 1355

\bibitem[\protect\citeauthoryear{{Balogh}, {Christlein}, {Zabludoff} \&
  {Zaritsky}}{{Balogh} et~al.}{2001}]{2001ApJ...557..117B}
{Balogh} M.~L.,  {Christlein} D.,  {Zabludoff} A.~I.,    {Zaritsky} D.,  2001,
  \apj, 557, 117

\bibitem[\protect\citeauthoryear{{Bamford}, {Milvang-Jensen},
  {Arag{\'o}n-Salamanca} \& {Simard}}{{Bamford}
  et~al.}{2005}]{2005MNRAS.361..109B}
{Bamford} S.~P.,  {Milvang-Jensen} B.,  {Arag{\'o}n-Salamanca} A.,    {Simard}
  L.,  2005, \mnras, 361, 109

\bibitem[\protect\citeauthoryear{{Barr}, {Bedregal}, {Arag{\'o}n-Salamanca},
  {Merrifield} \& {Bamford}}{{Barr} et~al.}{2007}]{2007A&A...470..173B}
{Barr} J.~M.,  {Bedregal} A.~G.,  {Arag{\'o}n-Salamanca} A.,  {Merrifield}
  M.~R.,    {Bamford} S.~P.,  2007, \aap, 470, 173

\bibitem[\protect\citeauthoryear{{Bedregal}, {Arag{\'o}n-Salamanca} \&
  {Merrifield}}{{Bedregal} et~al.}{2006}]{2006MNRAS.373.1125B}
{Bedregal} A.~G.,  {Arag{\'o}n-Salamanca} A.,    {Merrifield} M.~R.,  2006,
  \mnras, 373, 1125

\bibitem[\protect\citeauthoryear{{Bedregal}, {Arag{\'o}n-Salamanca},
  {Merrifield} \& {Cardiel}}{{Bedregal} et~al.}{2008}]{2008arXiv0803.0305B}
{Bedregal} A.~G.,  {Arag{\'o}n-Salamanca} A.,  {Merrifield} M.~R.,    {Cardiel}
  N.,  2008, \mnras, 387, 660

\bibitem[\protect\citeauthoryear{{Biviano}, {Murante}, {Borgani}, {Diaferio},
  {Dolag} \& {Girardi}}{{Biviano} et~al.}{2006}]{2006A&A...456...23B}
{Biviano} A.,  {Murante} G.,  {Borgani} S.,  {Diaferio} A.,  {Dolag} K.,
  {Girardi} M.,  2006, \aap, 456, 23

\bibitem[\protect\citeauthoryear{{Blakeslee} et~al.,}{{Blakeslee}
  et~al.}{2006}]{2006ApJ...644...30B}
{Blakeslee} J.~P.,  et~al., 2006, \apj, 644, 30

\bibitem[\protect\citeauthoryear{{Blanton} \& {Berlind}}{{Blanton} \&
  {Berlind}}{2007}]{2007ApJ...664..791B}
{Blanton} M.~R.,  {Berlind} A.~A.,  2007, \apj, 664, 791

\bibitem[\protect\citeauthoryear{{Blanton}, {Eisenstein}, {Hogg}, {Schlegel} \&
  {Brinkmann}}{{Blanton} et~al.}{2005}]{2005ApJ...629..143B}
{Blanton} M.~R.,  {Eisenstein} D.,  {Hogg} D.~W.,  {Schlegel} D.~J.,
  {Brinkmann} J.,  2005, \apj, 629, 143

\bibitem[\protect\citeauthoryear{{Blanton} et~al.,}{{Blanton}
  et~al.}{2003a}]{2003ApJ...594..186B}
{Blanton} M.~R.,  et~al., 2003a, \apj, 594, 186

\bibitem[\protect\citeauthoryear{{Blanton} et~al.,}{{Blanton}
  et~al.}{2003b}]{2003ApJ...592..819B}
{Blanton} M.~R.,  et~al., 2003b, \apj, 592, 819

\bibitem[\protect\citeauthoryear{{Blanton} \& {Roweis}}{{Blanton} \&
  {Roweis}}{2007}]{2007AJ....133..734B}
{Blanton} M.~R.,  {Roweis} S.,  2007, \aj, 133, 734

\bibitem[\protect\citeauthoryear{{Boselli} \& {Gavazzi}}{{Boselli} \&
  {Gavazzi}}{2006}]{2006PASP..118..517B}
{Boselli} A.,  {Gavazzi} G.,  2006, \pasp, 118, 517

\bibitem[\protect\citeauthoryear{{Bower}, {Benson}, {Malbon}, {Helly}, {Frenk},
  {Baugh}, {Cole} \& {Lacey}}{{Bower} et~al.}{2006}]{2006MNRAS.370..645B}
{Bower} R.~G.,  {Benson} A.~J.,  {Malbon} R.,  {Helly} J.~C.,  {Frenk} C.~S.,
  {Baugh} C.~M.,  {Cole} S.,    {Lacey} C.~G.,  2006, \mnras, 370, 645

\bibitem[\protect\citeauthoryear{{Burstein}}{{Burstein}}{1979}]{1979ApJ...234.%
.435B}
{Burstein} D.,  1979, \apj, 234, 435

\bibitem[\protect\citeauthoryear{{Butcher} \& {Oemler} Jr.}{{Butcher} \&
  {Oemler}}{1984}]{1984ApJ...285..426B}
{Butcher} H.,  {Oemler} Jr. A.,  1984, \apj, 285, 426

\bibitem[\protect\citeauthoryear{{Cameron} \& {Driver}}{{Cameron} \&
  {Driver}}{2007}]{2007MNRAS.377..523C}
{Cameron} E.,  {Driver} S.~P.,  2007, \mnras, 377, 523

\bibitem[\protect\citeauthoryear{{Ceccarelli}, {Padilla} \&
  {Lambas}}{{Ceccarelli} et~al.}{2008}]{2008arXiv0805.0790C}
{Ceccarelli} L.,  {Padilla} N.,    {Lambas} D.~G.,  2008, \mnras, 390, L9

\bibitem[\protect\citeauthoryear{{Christlein} \& {Zabludoff}}{{Christlein} \&
  {Zabludoff}}{2004}]{2004ApJ...616..192C}
{Christlein} D.,  {Zabludoff} A.~I.,  2004, \apj, 616, 192

\bibitem[\protect\citeauthoryear{{Conselice}}{{Conselice}}{2003}]{2003ApJS..14%
7....1C}
{Conselice} C.~J.,  2003, \apjs, 147, 1

\bibitem[\protect\citeauthoryear{{Conselice}}{{Conselice}}{2006}]{2006MNRAS.37%
3.1389C}
{Conselice} C.~J.,  2006, \mnras, 373, 1389

\bibitem[\protect\citeauthoryear{{Cooray} \& {Sheth}}{{Cooray} \&
  {Sheth}}{2002}]{2002PhR...372....1C}
{Cooray} A.,  {Sheth} R.,  2002, \physrep, 372, 1

\bibitem[\protect\citeauthoryear{{Couch}, {Barger}, {Smail}, {Ellis} \&
  {Sharples}}{{Couch} et~al.}{1998}]{1998ApJ...497..188C}
{Couch} W.~J.,  {Barger} A.~J.,  {Smail} I.,  {Ellis} R.~S.,    {Sharples}
  R.~M.,  1998, \apj, 497, 188

\bibitem[\protect\citeauthoryear{{Cowie} \& {Songaila}}{{Cowie} \&
  {Songaila}}{1977}]{1977Natur.266..501C}
{Cowie} L.~L.,  {Songaila} A.,  1977, \nat, 266, 501

\bibitem[\protect\citeauthoryear{{Croton} et~al.,}{{Croton}
  et~al.}{2005}]{2005MNRAS.356.1155C}
{Croton} D.~J.,  et~al., 2005, \mnras, 356, 1155

\bibitem[\protect\citeauthoryear{{Croton} et~al.,}{{Croton}
  et~al.}{2006}]{2006MNRAS.365...11C}
{Croton} D.~J.,  et~al., 2006, \mnras, 365, 11

\bibitem[\protect\citeauthoryear{{Croton} \& {Farrar}}{{Croton} \&
  {Farrar}}{2008}]{2008MNRAS.386.2285C}
{Croton} D.~J.,  {Farrar} G.~R.,  2008, \mnras, 386, 2285

\bibitem[\protect\citeauthoryear{{de Jong}}{{de
  Jong}}{1995}]{1995PhDT.......129D}
{de Jong} R.,  1995, PhD thesis, PhD thesis.~Univ.~Groningen

\bibitem[\protect\citeauthoryear{{Desai} et~al.,}{{Desai}
  et~al.}{2007}]{2007ApJ...660.1151D}
{Desai} V.,  et~al., 2007, \apj, 660, 1151

\bibitem[\protect\citeauthoryear{{Dressler}}{{Dressler}}{1980a}]{1980ApJS...42%
..565D}
{Dressler} A.,  1980a, \apjs, 42, 565

\bibitem[\protect\citeauthoryear{{Dressler}}{{Dressler}}{1980b}]{1980ApJ...236%
..351D}
{Dressler} A.,  1980b, \apj, 236, 351

\bibitem[\protect\citeauthoryear{{Dressler} et~al.,}{{Dressler}
  et~al.}{1997}]{1997ApJ...490..577D}
{Dressler} A.,  et~al., 1997, \apj, 490, 577

\bibitem[\protect\citeauthoryear{{Dressler} \& {Gunn}}{{Dressler} \&
  {Gunn}}{1992}]{1992ApJS...78....1D}
{Dressler} A.,  {Gunn} J.~E.,  1992, \apjs, 78, 1

\bibitem[\protect\citeauthoryear{{Faber} et~al.,}{{Faber}
  et~al.}{2007}]{2007ApJ...665..265F}
{Faber} S.~M.,  et~al., 2007, \apj, 665, 265

\bibitem[\protect\citeauthoryear{{Fan}, {Lapi}, {De Zotti} \& {Danese}}{{Fan}
  et~al.}{2008}]{2008arXiv0809.4574F}
{Fan} L.,  {Lapi} A.,  {De Zotti} G.,    {Danese} L.,  2008, ArXiv e-prints

\bibitem[\protect\citeauthoryear{{Font} et~al.,}{{Font}
  et~al.}{2008}]{2008MNRAS.389.1619F}
{Font} A.~S.,  et~al., 2008, \mnras, 389, 1619

\bibitem[\protect\citeauthoryear{{Fukugita} et~al.,}{{Fukugita}
  et~al.}{2007}]{2007AJ....134..579F}
{Fukugita} M.,  et~al., 2007, \aj, 134, 579

\bibitem[\protect\citeauthoryear{{Gallazzi}, {Charlot}, {Brinchmann}, {White}
  \& {Tremonti}}{{Gallazzi} et~al.}{2005}]{2005MNRAS.362...41G}
{Gallazzi} A.,  {Charlot} S.,  {Brinchmann} J.,  {White} S.~D.~M.,
  {Tremonti} C.~A.,  2005, \mnras, 362, 41

\bibitem[\protect\citeauthoryear{{Gavazzi}, {Franzetti}, {Scodeggio}, {Boselli}
  \& {Pierini}}{{Gavazzi} et~al.}{2000}]{2000A&A...361..863G}
{Gavazzi} G.,  {Franzetti} P.,  {Scodeggio} M.,  {Boselli} A.,    {Pierini} D.,
   2000, \aap, 361, 863

\bibitem[\protect\citeauthoryear{{Gisler}}{{Gisler}}{1980}]{1980AJ.....85..623%
G}
{Gisler} G.~R.,  1980, \aj, 85, 623

\bibitem[\protect\citeauthoryear{{Glazebrook} et~al.,}{{Glazebrook}
  et~al.}{2004}]{2004Natur.430..181G}
{Glazebrook} K.,  et~al., 2004, \nat, 430, 181

\bibitem[\protect\citeauthoryear{{G{\'o}mez} et~al.,}{{G{\'o}mez}
  et~al.}{2003}]{2003ApJ...584..210G}
{G{\'o}mez} P.~L.,  et~al., 2003, \apj, 584, 210

\bibitem[\protect\citeauthoryear{{Goto} et~al.,}{{Goto}
  et~al.}{2003}]{2003PASJ...55..757G}
{Goto} T.,  et~al., 2003, \pasj, 55, 757

\bibitem[\protect\citeauthoryear{{Goto}, {Yamauchi}, {Fujita}, {Okamura},
  {Sekiguchi}, {Smail}, {Bernardi} \& {Gomez}}{{Goto}
  et~al.}{2003}]{2003MNRAS.346..601G}
{Goto} T.,  {Yamauchi} C.,  {Fujita} Y.,  {Okamura} S.,  {Sekiguchi} M.,
  {Smail} I.,  {Bernardi} M.,    {Gomez} P.~L.,  2003, \mnras, 346, 601

\bibitem[\protect\citeauthoryear{{Gunn} \& {Gott}}{{Gunn} \&
  {Gott}}{1972}]{1972ApJ...176....1G}
{Gunn} J.~E.,  {Gott} J.~R.~I.,  1972, \apj, 176, 1

\bibitem[\protect\citeauthoryear{{Haines}, {Gargiulo}, {La Barbera},
  {Mercurio}, {Merluzzi} \& {Busarello}}{{Haines}
  et~al.}{2007}]{2007MNRAS.381....7H}
{Haines} C.~P.,  {Gargiulo} A.,  {La Barbera} F.,  {Mercurio} A.,  {Merluzzi}
  P.,    {Busarello} G.,  2007, \mnras, 381, 7

\bibitem[\protect\citeauthoryear{{Haines}, {La Barbera}, {Mercurio}, {Merluzzi}
  \& {Busarello}}{{Haines} et~al.}{2006}]{2006ApJ...647L..21H}
{Haines} C.~P.,  {La Barbera} F.,  {Mercurio} A.,  {Merluzzi} P.,
  {Busarello} G.,  2006, \apjl, 647, L21

\bibitem[\protect\citeauthoryear{{Hogg} et~al.,}{{Hogg}
  et~al.}{2004}]{2004ApJ...601L..29H}
{Hogg} D.~W.,  et~al., 2004, \apjl, 601, L29

\bibitem[\protect\citeauthoryear{{Hopkins}, {Cox}, {Younger} \&
  {Hernquist}}{{Hopkins} et~al.}{2008}]{2008arXiv0806.1739H}
{Hopkins} P.~F.,  {Cox} T.~J.,  {Younger} J.~D.,    {Hernquist} L.,  2008,
  ArXiv e-prints

\bibitem[\protect\citeauthoryear{{Hubble} \& {Humason}}{{Hubble} \&
  {Humason}}{1931}]{1931ApJ....74...43H}
{Hubble} E.,  {Humason} M.~L.,  1931, \apj, 74, 43

\bibitem[\protect\citeauthoryear{{Hubble}}{{Hubble}}{1922}]{1922ApJ....56..162%
H}
{Hubble} E.~P.,  1922, \apj, 56, 162

\bibitem[\protect\citeauthoryear{{Huchra}, {Davis}, {Latham} \&
  {Tonry}}{{Huchra} et~al.}{1983}]{1983ApJS...52...89H}
{Huchra} J.,  {Davis} M.,  {Latham} D.,    {Tonry} J.,  1983, \apjs, 52, 89

\bibitem[\protect\citeauthoryear{{Huchra} \& {Geller}}{{Huchra} \&
  {Geller}}{1982}]{1982ApJ...257..423H}
{Huchra} J.~P.,  {Geller} M.~J.,  1982, \apj, 257, 423

\bibitem[\protect\citeauthoryear{{James}, {Knapen}, {Shane}, {Baldry} \& {de
  Jong}}{{James} et~al.}{2008}]{2008arXiv0802.4421J}
{James} P.~A.,  {Knapen} J.~H.,  {Shane} N.~S.,  {Baldry} I.~K.,    {de Jong}
  R.~S.,  2008, \aap, 482, 507

\bibitem[\protect\citeauthoryear{{Jones}, {Ponman}, {Horton}, {Babul},
  {Ebeling} \& {Burke}}{{Jones} et~al.}{2003}]{2003MNRAS.343..627J}
{Jones} L.~R.,  {Ponman} T.~J.,  {Horton} A.,  {Babul} A.,  {Ebeling} H.,
  {Burke} D.~J.,  2003, \mnras, 343, 627

\bibitem[\protect\citeauthoryear{{Kaiser}}{{Kaiser}}{1987}]{1987MNRAS.227....1%
K}
{Kaiser} N.,  1987, \mnras, 227, 1

\bibitem[\protect\citeauthoryear{{Kauffmann} et~al.,}{{Kauffmann}
  et~al.}{2003}]{2003MNRAS.341...33K}
{Kauffmann} G.,  et~al., 2003, \mnras, 341, 33

\bibitem[\protect\citeauthoryear{{Kauffmann}, {White} \&
  {Guiderdoni}}{{Kauffmann} et~al.}{1993}]{1993MNRAS.264..201K}
{Kauffmann} G.,  {White} S.~D.~M.,    {Guiderdoni} B.,  1993, \mnras, 264, 201

\bibitem[\protect\citeauthoryear{{Kauffmann}, {White}, {Heckman}, {M{\'e}nard},
  {Brinchmann}, {Charlot}, {Tremonti} \& {Brinkmann}}{{Kauffmann}
  et~al.}{2004}]{2004MNRAS.353..713K}
{Kauffmann} G.,  {White} S.~D.~M.,  {Heckman} T.~M.,  {M{\'e}nard} B.,
  {Brinchmann} J.,  {Charlot} S.,  {Tremonti} C.,    {Brinkmann} J.,  2004,
  \mnras, 353, 713

\bibitem[\protect\citeauthoryear{{Koopmann} \& {Kenney}}{{Koopmann} \&
  {Kenney}}{2004}]{2004ApJ...613..866K}
{Koopmann} R.~A.,  {Kenney} J.~D.~P.,  2004, \apj, 613, 866

\bibitem[\protect\citeauthoryear{{Lahav} et~al.,}{{Lahav}
  et~al.}{1995}]{1995Sci...267..859L}
{Lahav} O.,  et~al., 1995, Science, 267, 859

\bibitem[\protect\citeauthoryear{{Land} et~al.,}{{Land}
  et~al.}{2008}]{Kate_GZ}
{Land} K.,  et~al., 2008, \mnras, 388, 1686

\bibitem[\protect\citeauthoryear{{Lane}, {Gray}, {Arag{\'o}n-Salamanca}, {Wolf}
  \& {Meisenheimer}}{{Lane} et~al.}{2007}]{2007MNRAS.378..716L}
{Lane} K.~P.,  {Gray} M.~E.,  {Arag{\'o}n-Salamanca} A.,  {Wolf} C.,
  {Meisenheimer} K.,  2007, \mnras, 378, 716

\bibitem[\protect\citeauthoryear{{Larson}, {Tinsley} \& {Caldwell}}{{Larson}
  et~al.}{1980}]{1980ApJ...237..692L}
{Larson} R.~B.,  {Tinsley} B.~M.,    {Caldwell} C.~N.,  1980, \apj, 237, 692

\bibitem[\protect\citeauthoryear{{Lewis} et~al.,}{{Lewis}
  et~al.}{2002}]{2002MNRAS.334..673L}
{Lewis} I.,  et~al., 2002, \mnras, 334, 673

\bibitem[\protect\citeauthoryear{{Lin}, {Mohr}, {Gonzalez} \& {Stanford}}{{Lin}
  et~al.}{2006}]{2006ApJ...650L..99L}
{Lin} Y.-T.,  {Mohr} J.~J.,  {Gonzalez} A.~H.,    {Stanford} S.~A.,  2006,
  \apjl, 650, L99

\bibitem[\protect\citeauthoryear{{Lintott} et~al.,}{{Lintott}
  et~al.}{2008}]{Chris_GZ}
{Lintott} C.~J.,  et~al., 2008, \mnras, 389, 1179

\bibitem[\protect\citeauthoryear{{Lupton}, {Blanton}, {Fekete}, {Hogg},
  {O'Mullane}, {Szalay} \& {Wherry}}{{Lupton}
  et~al.}{2004}]{2004PASP..116..133L}
{Lupton} R.,  {Blanton} M.~R.,  {Fekete} G.,  {Hogg} D.~W.,  {O'Mullane} W.,
  {Szalay} A.,    {Wherry} N.,  2004, \pasp, 116, 133

\bibitem[\protect\citeauthoryear{{Maller}}{{Maller}}{2008}]{2008arXiv0801.4568%
M}
{Maller} A.~H.,  2008, ArXiv: 0801.4568, 801

\bibitem[\protect\citeauthoryear{{Maller}, {Berlind}, {Blanton} \&
  {Hogg}}{{Maller} et~al.}{2008}]{2008arXiv0801.3286M}
{Maller} A.~H.,  {Berlind} A.~A.,  {Blanton} M.~R.,    {Hogg} D.~W.,  2008,
  ArXiv: 0801.3286, 801

\bibitem[\protect\citeauthoryear{{Mehlert}, {Thomas}, {Saglia}, {Bender} \&
  {Wegner}}{{Mehlert} et~al.}{2003}]{2003A&A...407..423M}
{Mehlert} D.,  {Thomas} D.,  {Saglia} R.~P.,  {Bender} R.,    {Wegner} G.,
  2003, \aap, 407, 423

\bibitem[\protect\citeauthoryear{{Merritt}}{{Merritt}}{1983}]{1983ApJ...264...%
24M}
{Merritt} D.,  1983, \apj, 264, 24

\bibitem[\protect\citeauthoryear{{Mihos} \& {Hernquist}}{{Mihos} \&
  {Hernquist}}{1994}]{1994ApJ...425L..13M}
{Mihos} J.~C.,  {Hernquist} L.,  1994, \apjl, 425, L13

\bibitem[\protect\citeauthoryear{{Miller} et~al.,}{{Miller}
  et~al.}{2005}]{2005AJ....130..968M}
{Miller} C.~J.,  et~al., 2005, \aj, 130, 968

\bibitem[\protect\citeauthoryear{{Mo} \& {White}}{{Mo} \&
  {White}}{2002}]{2002MNRAS.336..112M}
{Mo} H.~J.,  {White} S.~D.~M.,  2002, \mnras, 336, 112

\bibitem[\protect\citeauthoryear{{Mo}, {Yang}, {van den Bosch} \& {Jing}}{{Mo}
  et~al.}{2004}]{2004MNRAS.349..205M}
{Mo} H.~J.,  {Yang} X.,  {van den Bosch} F.~C.,    {Jing} Y.~P.,  2004, \mnras,
  349, 205

\bibitem[\protect\citeauthoryear{{Moore}, {Katz}, {Lake}, {Dressler} \&
  {Oemler}}{{Moore} et~al.}{1996}]{1996Natur.379..613M}
{Moore} B.,  {Katz} N.,  {Lake} G.,  {Dressler} A.,    {Oemler} A.,  1996,
  \nat, 379, 613

\bibitem[\protect\citeauthoryear{{Nakamura}, {Fukugita}, {Brinkmann} \&
  {Schneider}}{{Nakamura} et~al.}{2004}]{2004AJ....127.2511N}
{Nakamura} O.,  {Fukugita} M.,  {Brinkmann} J.,    {Schneider} D.~P.,  2004,
  \aj, 127, 2511

\bibitem[\protect\citeauthoryear{{Nulsen}}{{Nulsen}}{1982}]{1982MNRAS.198.1007%
N}
{Nulsen} P.~E.~J.,  1982, \mnras, 198, 1007

\bibitem[\protect\citeauthoryear{{Padmanabhan} et~al.,}{{Padmanabhan}
  et~al.}{2008}]{2007astro.ph..3454P}
{Padmanabhan} N.,  et~al., 2008, \apj, 674, 1217

\bibitem[\protect\citeauthoryear{{Park}, {Choi}, {Vogeley}, {Gott} \&
  {Blanton}}{{Park} et~al.}{2007}]{2007ApJ...658..898P}
{Park} C.,  {Choi} Y.-Y.,  {Vogeley} M.~S.,  {Gott} J.~R.~I.,    {Blanton}
  M.~R.,  2007, \apj, 658, 898

\bibitem[\protect\citeauthoryear{{Pimbblet}, {Smail}, {Kodama}, {Couch},
  {Edge}, {Zabludoff} \& {O'Hely}}{{Pimbblet}
  et~al.}{2002}]{2002MNRAS.331..333P}
{Pimbblet} K.~A.,  {Smail} I.,  {Kodama} T.,  {Couch} W.~J.,  {Edge} A.~C.,
  {Zabludoff} A.~I.,    {O'Hely} E.,  2002, \mnras, 331, 333

\bibitem[\protect\citeauthoryear{{Poggianti} et~al.,}{{Poggianti}
  et~al.}{2001}]{2001ApJ...563..118P}
{Poggianti} B.~M.,  et~al., 2001, \apj, 563, 118

\bibitem[\protect\citeauthoryear{{Poggianti} et~al.,}{{Poggianti}
  et~al.}{2006}]{2006ApJ...642..188P}
{Poggianti} B.~M.,  et~al., 2006, \apj, 642, 188

\bibitem[\protect\citeauthoryear{{Poggianti} et~al.,}{{Poggianti}
  et~al.}{2008}]{2008ApJ...684..888P}
{Poggianti} B.~M.,  et~al., 2008, \apj, 684, 888

\bibitem[\protect\citeauthoryear{{Poggianti}, {Smail}, {Dressler}, {Couch},
  {Barger}, {Butcher}, {Ellis} \& {Oemler}}{{Poggianti}
  et~al.}{1999}]{1999ApJ...518..576P}
{Poggianti} B.~M.,  {Smail} I.,  {Dressler} A.,  {Couch} W.~J.,  {Barger}
  A.~J.,  {Butcher} H.,  {Ellis} R.~S.,    {Oemler} A.~J.,  1999, \apj, 518,
  576

\bibitem[\protect\citeauthoryear{{Postman} et~al.,}{{Postman}
  et~al.}{2005}]{2005ApJ...623..721P}
{Postman} M.,  et~al., 2005, \apj, 623, 721

\bibitem[\protect\citeauthoryear{{Postman} \& {Geller}}{{Postman} \&
  {Geller}}{1984}]{1984ApJ...281...95P}
{Postman} M.,  {Geller} M.~J.,  1984, \apj, 281, 95

\bibitem[\protect\citeauthoryear{{Rojas}, {Vogeley}, {Hoyle} \&
  {Brinkmann}}{{Rojas} et~al.}{2004}]{2004ApJ...617...50R}
{Rojas} R.~R.,  {Vogeley} M.~S.,  {Hoyle} F.,    {Brinkmann} J.,  2004, \apj,
  617, 50

\bibitem[\protect\citeauthoryear{{Rojas}, {Vogeley}, {Hoyle} \&
  {Brinkmann}}{{Rojas} et~al.}{2005}]{2005ApJ...624..571R}
{Rojas} R.~R.,  {Vogeley} M.~S.,  {Hoyle} F.,    {Brinkmann} J.,  2005, \apj,
  624, 571

\bibitem[\protect\citeauthoryear{{Schawinski} et~al.,}{{Schawinski}
  et~al.}{2007}]{2007ApJS..173..512S}
{Schawinski} K.,  et~al., 2007, \apjs, 173, 512

\bibitem[\protect\citeauthoryear{{Schawinski} et~al.,}{{Schawinski}
  et~al.}{2008}]{Kevin_GZ}
{Schawinski} K.,  et~al., 2008, submitted to \mnras

\bibitem[\protect\citeauthoryear{{Schawinski}, {Thomas}, {Sarzi}, {Maraston},
  {Kaviraj}, {Joo}, {Yi} \& {Silk}}{{Schawinski}
  et~al.}{2007}]{2007MNRAS.382.1415S}
{Schawinski} K.,  {Thomas} D.,  {Sarzi} M.,  {Maraston} C.,  {Kaviraj} S.,
  {Joo} S.-J.,  {Yi} S.~K.,    {Silk} J.,  2007, \mnras, 382, 1415

\bibitem[\protect\citeauthoryear{{Sheth} \& {Tormen}}{{Sheth} \&
  {Tormen}}{1999}]{1999MNRAS.308..119S}
{Sheth} R.~K.,  {Tormen} G.,  1999, \mnras, 308, 119

\bibitem[\protect\citeauthoryear{{Simard} et~al.,}{{Simard}
  et~al.}{2002}]{2002ApJS..142....1S}
{Simard} L.,  et~al., 2002, \apjs, 142, 1

\bibitem[\protect\citeauthoryear{{Skibba}, {Sheth}, {Connolly} \&
  {Scranton}}{{Skibba} et~al.}{2006}]{2006MNRAS.369...68S}
{Skibba} R.,  {Sheth} R.~K.,  {Connolly} A.~J.,    {Scranton} R.,  2006,
  \mnras, 369, 68

\bibitem[\protect\citeauthoryear{{Skibba}}{{Skibba}}{2008}]{2008arXiv0805.1233%
S}
{Skibba} R.~A.,  2008, ArXiv: 0805.1233, 805

\bibitem[\protect\citeauthoryear{{Skibba} \& {Sheth}}{{Skibba} \&
  {Sheth}}{2008}]{2008arXiv0805.0310S}
{Skibba} R.~A.,  {Sheth} R.~K.,  2008, ArXiv e-prints

\bibitem[\protect\citeauthoryear{{Slosar} et~al.,}{{Slosar}
  et~al.}{2008}]{Anze_GZ}
{Slosar} A.,  et~al., 2008, submitted to \mnras

\bibitem[\protect\citeauthoryear{{Smail}, {Dressler}, {Couch}, {Ellis},
  {Oemler}, {Butcher} \& {Sharples}}{{Smail}
  et~al.}{1997}]{1997ApJS..110..213S}
{Smail} I.,  {Dressler} A.,  {Couch} W.~J.,  {Ellis} R.~S.,  {Oemler} A.~J.,
  {Butcher} H.,    {Sharples} R.~M.,  1997, \apjs, 110, 213

\bibitem[\protect\citeauthoryear{{Smith}, {Treu}, {Ellis}, {Moran} \&
  {Dressler}}{{Smith} et~al.}{2005}]{2005ApJ...620...78S}
{Smith} G.~P.,  {Treu} T.,  {Ellis} R.~S.,  {Moran} S.~M.,    {Dressler} A.,
  2005, \apj, 620, 78

\bibitem[\protect\citeauthoryear{{Stoughton} et~al.,}{{Stoughton}
  et~al.}{2002}]{2002AJ....123..485S}
{Stoughton} C.,  et~al., 2002, \aj, 123, 485

\bibitem[\protect\citeauthoryear{{Strauss} et~al.,}{{Strauss}
  et~al.}{2002}]{2002AJ....124.1810S}
{Strauss} M.~A.,  et~al., 2002, \aj, 124, 1810

\bibitem[\protect\citeauthoryear{{Torki} et~al.,}{{Torki}
  et~al.}{2008}]{Torki_C4}
{Torki} M.,  et~al., 2008, submitted to \mnras

\bibitem[\protect\citeauthoryear{{Treu}, {Ellis}, {Kneib}, {Dressler}, {Smail},
  {Czoske}, {Oemler} \& {Natarajan}}{{Treu} et~al.}{2003}]{2003ApJ...591...53T}
{Treu} T.,  {Ellis} R.~S.,  {Kneib} J.-P.,  {Dressler} A.,  {Smail} I.,
  {Czoske} O.,  {Oemler} A.,    {Natarajan} P.,  2003, \apj, 591, 53

\bibitem[\protect\citeauthoryear{{van den Bergh}}{{van den
  Bergh}}{1976}]{1976ApJ...206..883V}
{van den Bergh} S.,  1976, \apj, 206, 883

\bibitem[\protect\citeauthoryear{{van den Bergh}}{{van den
  Bergh}}{2007}]{2007AJ....134.1508V}
{van den Bergh} S.,  2007, \aj, 134, 1508

\bibitem[\protect\citeauthoryear{{van den Bosch}, {Pasquali}, {Yang}, {Mo},
  {Weinmann}, {McIntosh} \& {Aquino}}{{van den Bosch}
  et~al.}{2008}]{2008arXiv0805.0002V}
{van den Bosch} F.~C.,  {Pasquali} A.,  {Yang} X.,  {Mo} H.~J.,  {Weinmann} S.,
   {McIntosh} D.~H.,    {Aquino} D.,  2008, ArXiv: 0805.0002, 805

\bibitem[\protect\citeauthoryear{{van der Wel}}{{van der
  Wel}}{2008}]{2008arXiv0801.1995V}
{van der Wel} A.,  2008, \apjl, 675, L13

\bibitem[\protect\citeauthoryear{{van der Wel} et~al.,}{{van der Wel}
  et~al.}{2007}]{2007ApJ...670..206V}
{van der Wel} A.,  et~al., 2007, \apj, 670, 206

\bibitem[\protect\citeauthoryear{{Vogt}, {Haynes}, {Giovanelli} \&
  {Herter}}{{Vogt} et~al.}{2004}]{2004AJ....127.3300V}
{Vogt} N.~P.,  {Haynes} M.~P.,  {Giovanelli} R.,    {Herter} T.,  2004, \aj,
  127, 3300

\bibitem[\protect\citeauthoryear{{von der Linden}, {Best}, {Kauffmann} \&
  {White}}{{von der Linden} et~al.}{2007}]{2007MNRAS.379..867V}
{von der Linden} A.,  {Best} P.~N.,  {Kauffmann} G.,    {White} S.~D.~M.,
  2007, \mnras, 379, 867

\bibitem[\protect\citeauthoryear{{Wake} et~al.,}{{Wake}
  et~al.}{2006}]{2006MNRAS.372..537W}
{Wake} D.~A.,  et~al., 2006, \mnras, 372, 537

\bibitem[\protect\citeauthoryear{{Weinmann}, {van den Bosch}, {Yang} \&
  {Mo}}{{Weinmann} et~al.}{2006}]{2006MNRAS.366....2W}
{Weinmann} S.~M.,  {van den Bosch} F.~C.,  {Yang} X.,    {Mo} H.~J.,  2006,
  \mnras, 366, 2

\bibitem[\protect\citeauthoryear{{Whitmore} \& {Gilmore}}{{Whitmore} \&
  {Gilmore}}{1991}]{1991ApJ...367...64W}
{Whitmore} B.~C.,  {Gilmore} D.~M.,  1991, \apj, 367, 64

\bibitem[\protect\citeauthoryear{{Whitmore}, {Gilmore} \& {Jones}}{{Whitmore}
  et~al.}{1993}]{1993ApJ...407..489W}
{Whitmore} B.~C.,  {Gilmore} D.~M.,    {Jones} C.,  1993, \apj, 407, 489

\bibitem[\protect\citeauthoryear{{Wolf} et~al.,}{{Wolf}
  et~al.}{2007}]{2007AAS...211.6701W}
{Wolf} C.,  et~al., 2007 Vol.~211 of AAS Meeting Abstracts, {Optically Passive
  Infall Spirals In Stages: Star Formation Only Semi-quenched}.
p. 67.01

\bibitem[\protect\citeauthoryear{{Wolf} et~al.,}{{Wolf}
  et~al.}{2008}]{STAGES_sf_red_spirals}
{Wolf} C.,  et~al., 2008, submitted to \mnras

\bibitem[\protect\citeauthoryear{{Wolf}, {Gray}, {Arag{\'o}n-Salamanca}, {Lane}
  \& {Meisenheimer}}{{Wolf} et~al.}{2007}]{2007MNRAS.376L...1W}
{Wolf} C.,  {Gray} M.~E.,  {Arag{\'o}n-Salamanca} A.,  {Lane} K.~P.,
  {Meisenheimer} K.,  2007, \mnras, 376, L1

\bibitem[\protect\citeauthoryear{{Yamauchi} \& {Goto}}{{Yamauchi} \&
  {Goto}}{2004}]{2004MNRAS.352..815Y}
{Yamauchi} C.,  {Goto} T.,  2004, \mnras, 352, 815

\bibitem[\protect\citeauthoryear{{Yoon}, {Schawinski}, {Sheen}, {Ree} \&
  {Yi}}{{Yoon} et~al.}{2008}]{2008ApJS..176..414Y}
{Yoon} J.~H.,  {Schawinski} K.,  {Sheen} Y.-K.,  {Ree} C.~H.,    {Yi} S.~K.,
  2008, \apjs, 176, 414

\bibitem[\protect\citeauthoryear{{Zehavi} et~al.,}{{Zehavi}
  et~al.}{2002}]{2002ApJ...571..172Z}
{Zehavi} I.,  et~al., 2002, \apj, 571, 172

\end{thebibliography}

\appendix

\clearpage

\section{Quantifying classification bias}
\label{sec:bias}

\begin{figure}
\centering
\includegraphics[height=0.475\textwidth,angle=270]{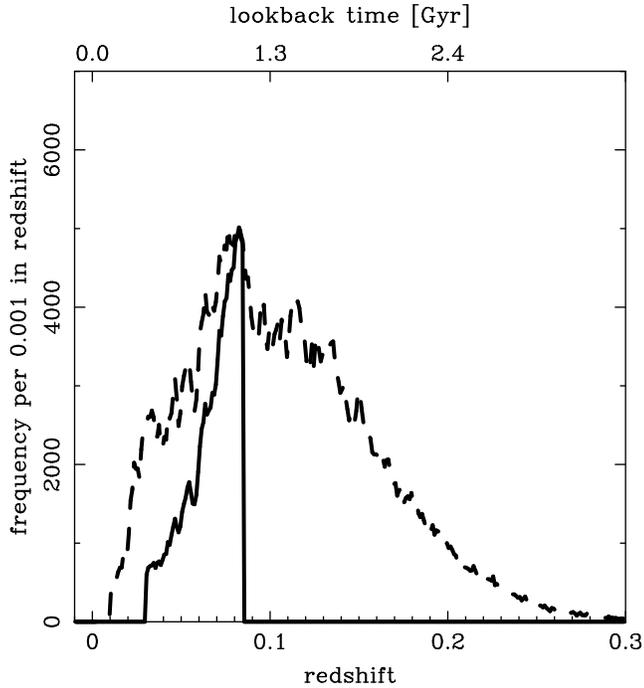}
\caption{\label{fig:z_dist} The redshift distribution of the
  \sample{luminosity-limited
 sample} analysed in this paper (solid line) and the SDSS Main
Galaxy Sample (dashed line) from which it is drawn. }
\end{figure}

\subsection{Demonstrating the bias}

\begin{figure*}
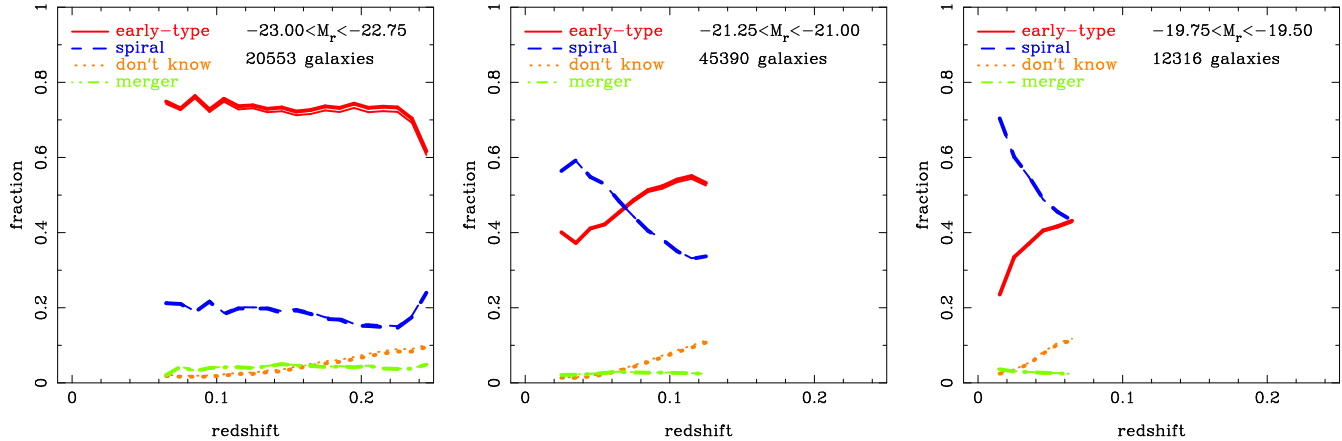

\centering
\includegraphics[height=0.32\textwidth,angle=270]{figA2_left.ps}
\hfill
\includegraphics[height=0.32\textwidth,angle=270]{figA2_middle.ps}
\hfill
\includegraphics[height=0.32\textwidth,angle=270]{figA2_right.ps}
\caption{\label{fig:zbins_magbins} Morphological type fraction versus
  redshift for galaxies in three example bins of absolute magnitude
  from our \sample{full sample}.
  The thick and thin lines corresponds to the weighted and unweighted
  samples, respectively.}
\end{figure*}

\begin{figure*}
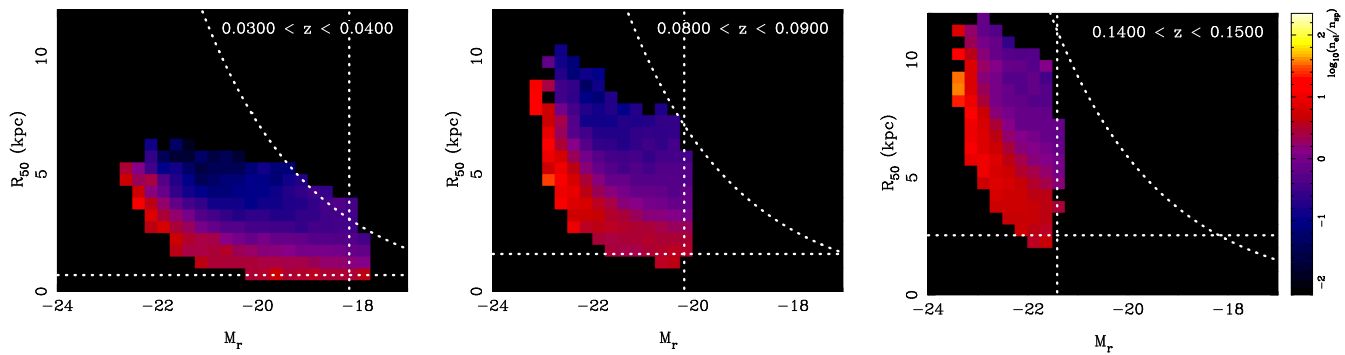

\centering
\includegraphics[scale=0.275,angle=270]{figA3_left.ps}
\hfill
\includegraphics[scale=0.275,angle=270]{figA3_middle.ps}
\hfill
\includegraphics[scale=0.275,angle=270]{figA3_right.ps}
\caption{\label{fig:ratio_grid_zbins} Ratio of the early-type to
  spiral measured type likelihoods as a function of absolute magnitude
  and physical size, for galaxies in three example redshift bins.  The
  logarithmic colour scale is shown by the bar on the right.
  The dotted lines indicate the $r=17.77$ apparent magnitude limit,
  $\mu_{50,r}=23$ mag~arcsec$^{-2}$ apparent surface brightness limit,
  and the physical scale corresponding to an angle of $1$ arcsec, at
  the central redshift of each bin. }
\end{figure*}

In the absence of evolution in the galaxy population, the true
distribution of galaxy morphologies does not change with redshift.
However, the apparent brightness and size of a given galaxy depends on
the distance at which it is viewed, and so its measured morphology may
vary with redshift.  We term this \emph{classification bias}.

For example, a spiral galaxy viewed nearby may be easy to classify
correctly, but if viewed at higher redshift the lower signal-to-noise
and smaller size, relative to the atmospheric seeing, would make the
same galaxy appear as a `fuzzy blob'.  Classifiers with a good
understanding of these issues will tend to recognise these limitations
and assign such an object to the unclassified category, though not
always.  Non-experts, such as the vast majority of Galaxy Zoo
participants, will more frequently classify such an object as an
early-type.  This is exacerbated by the scaling of images to always
present objects at a similar size on-screen, as was done for Galaxy
Zoo.  However, it is likely that all morphological studies are
affected by classification bias to some extent, even including those
studies performed by expert classifiers with the ability to vary the
characteristics of the image display.  This is the first study with a
sample of morphologically classified objects that is sufficiently large 
for this bias to be directly quantified.

The redshift range probed by the majority of our Galaxy Zoo \sample{full
  sample}, $0.03 \la z \la 0.15$, (see Fig.~\ref{fig:z_dist})
corresponds to an interval of $1.5$~Gyr. This is $10$ per cent of the
age of the universe, during its most quiescent period. Over this
period the galaxy population has not changed greatly. The latest
studies of evolution in the luminosity functions of red and blue
galaxies find a $\sim 2$ increase in the ratio of red versus blue
galaxies since $z \sim 1$ \citep{2007ApJ...665..265F}, which roughly
corresponds to $\sim 30$ per cent over the redshift range of the
Galaxy Zoo sample. However, this figure is fairly uncertain as it is
not clear when during the $z = 0$ -- $1$ period most of this evolution
occurred. In demonstrating and quantifying classification bias in the
Galaxy Zoo data we will often assume negligible evolution. If this
assumption is not valid, then the classification bias we measure will
also include a contribution due to galaxy evolution. Ultimately, our
aim for this paper is to correct the raw Galaxy Zoo classifications so
that objects over the range of redshifts sampled may be combined.
Removing any evolution, in addition to classification bias, will help
us to achieve this aim.

Observational limitations can introduce a variation in the population
of galaxies that is sampled at different redshifts. Quantities
measured from the sample may therefore vary with redshift, even if the
galaxy population does not. This is termed \emph{selection bias}. For
example, intrinsically faint objects can only be detected nearby,
while luminous objects are rare and so only appear in the sample at
larger distances where the sampled volume is larger. The apparent
magnitude limit and volume effect dominate the variation in sample
selection with redshift. Figure \ref{fig:zbin_counts_weights} (in the
main part of the paper) indicates that redshift-dependent biases are
present in the GZ data, but does not discriminate between
classification and selection biases. However, by considering objects
in narrow bins of intrinsic luminosity, i.e. absolute magnitude, we
can remove the effects of this component of selection bias.

Figure \ref{fig:zbins_magbins} illustrates the variation in
morphological type fraction as a function of redshift for galaxies in
three absolute magnitude intervals.  If evolution is negligible, there
are no selection biases with respect to quantities other than apparent
magnitude, and in the absence of classification bias, we would expect
these type fractions to be constant with redshift.  Note that the
fractions change between different absolute magnitude bins due to
selection effects and the inherent correlation between morphology and
luminosity.

Intrinsically bright objects are included in the Galaxy Zoo selection
limits over a fairly wide redshift range.  Their type fractions remain
fairly constant with redshift, as expected if there is no
classification bias.  However, the majority of objects with this
luminosity are truly early-types, and hence free from the
classification bias effect.  The primary trend is a gradual increase
in the fraction of galaxies classified as `don't know', mostly at
the expense of spirals.

At more typical galaxy luminosities and fainter, a gradient in the
type fractions versus redshift is obvious. Towards higher redshifts a
greater fraction of objects are classified as early-type rather than
spiral. A strong classification bias therefore appears to be present
in the data, with apparently faint galaxies being preferentially
classified as early-type, presumably because fewer details may be
discerned given the noise. The presence of classification bias is also
indicated by the thin lines in Fig.~\ref{fig:zbin_weights_corrected}
(in the main part of the paper), which shows type fraction trends
based on the raw type likelihoods with selection effects removed below
a given redshift by imposing a faint magnitude limit.

In Fig.~\ref{fig:zbins_magbins}, the turn-up in the spiral fraction at
the highest redshifts probed in each luminosity bin is a reversal of
the more general bias, and its origin is unclear.  Plausible
possibilities affecting the faintest galaxies are that the users
realise they are seeing a blurred image and attempt to compensate with
their classifications, that noise artifacts are more frequently
interpreted as structure, or that early-types are being lost as a
result of the effective size selection limit.  However, it does not
appear to be a significant effect for the redshift ranges considered
in the main part of this paper.

In addition to apparent magnitude, the other main factor influencing a
Galaxy Zoo classifier's ability to accurately determine a galaxy's
morphological type is its apparent size.  However, if we consider the
measured type fractions versus redshift in bins of physical size (in
kpc), then the apparent magnitude selection effects are not accounted
for and dominate, together with the intrinsic correlation between
luminosity and size \citep[e.g.][]{1995PhDT.......129D,2007MNRAS.377..523C}.
 
\begin{figure*}
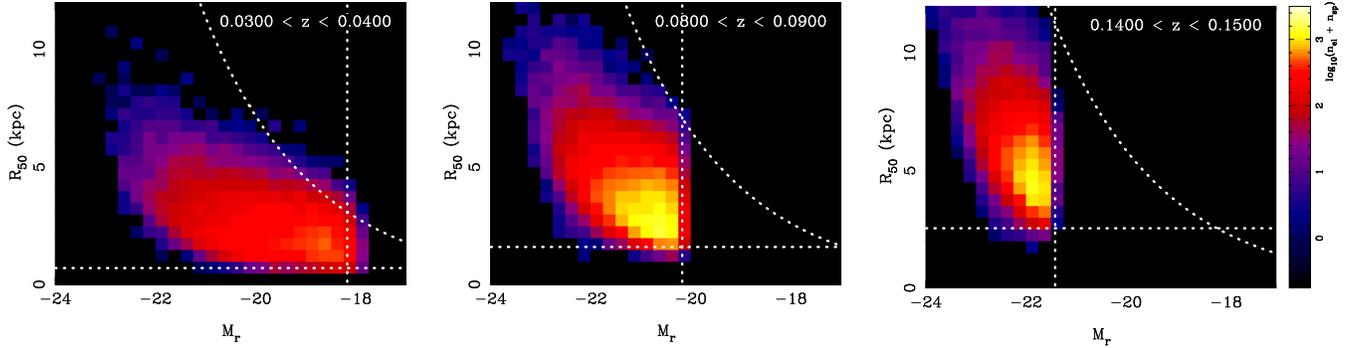

\centering
\includegraphics[scale=0.275,angle=270]{figA4_left.ps}
\hfill
\includegraphics[scale=0.275,angle=270]{figA4_middle.ps}
\hfill
\includegraphics[scale=0.275,angle=270]{figA4_right.ps}
\caption{\label{fig:count_grid_zbins} Number of early-type and spiral
  galaxies in each bin of absolute magnitude and physical size, for
  three example redshift bins.  The logarithmic colour scale is shown
  on the right.  Dotted lines are as in
  Fig.~\ref{fig:ratio_grid_zbins}.}
\end{figure*}

We therefore need to consider the measured type fractions as a
bivariate function of both luminosity and size\footnote{We could have
  chosen to use surface brightness in place of either luminosity or
  size, but we find that surface brightness alone is not sufficient to
  describe the behaviour of the bias.}.  In this paper we are
concerned with the early-type and spiral fractions.  Studying the
merger candidates will be an interesting topic for future studies, but
their classification is complicated by additional biases.  In any case
the numbers of objects classified as mergers are low.  The `don't
know' option provides an indication of where the classifiers
themselves feel they are unable to determine accurate classifications.
This supplies supplementary information for judging the unbiased
region of parameter space, but ideally we wish to determine this
directly from the biases in the measured type likelihoods.  We
therefore concentrate on the ratio of early-type to spiral likelihoods
for the remainder of this paper.  This ratio is plotted for three
example redshift bins in Fig.~\ref{fig:ratio_grid_zbins}.  The number
of galaxies in each luminosity--size--redshift bin are shown in
Fig.~\ref{fig:count_grid_zbins}.

Assuming evolution is negligible, and that any selection biases are
functions of only apparent magnitude and size (including surface
brightness, which depends on both), then in the absence of
classification bias, we would expect the ratio of early-types to
spirals to be constant with redshift.  

In simple terms, in Fig.~\ref{fig:ratio_grid_zbins} changes in the
colour (type ratio) of a particular square (luminosity--size bin)
between redshift slices indicate a redshift dependent classification
bias.  This bias is due to the decreasing signal-to-noise and
resolution for similar galaxies viewed at increasing redshifts.  A
movie is available in the online supplementary material, which steps
through this plot in redshift slices and demonstrates the effect more
clearly.

\subsection{Constructing a baseline correction}
To quantify how classification bias changes with redshift, as a
function of luminosity and size, we first construct an estimate of the
$n_{el}/n_{sp}$ versus luminosity and size in the absence of any
redshift-dependent classification bias.  To do this, for each
luminosity--size bin we find the lowest redshift bin that contains at
least a certain number of galaxies, and assume that the ratio in this
bin is accurate.  We take the minimum number of galaxies required in a
bin to be 30.

From inspecting Fig.~\ref{fig:ratio_grid_zbins} this approach appears
to be reasonable.  At low redshift the \sample{full sample} includes faint, small
galaxies.  Those well inside the apparent magnitude, size and surface
brightness limits should be unbiased.  At higher redshifts, these
limits become more restrictive, and biases grow at the faint, small
end of the galaxy distribution.  The scarcity of bright, large
galaxies means that they are only seen in significant numbers at
higher redshifts, where the survey encompasses a larger volume.
However, at the redshifts at which $n_{el}/n_{sp}$ can be estimated
for these galaxies, they are still well within the luminosity and size
limits, helped by the intrinsic size--luminosity relation.  On the
other hand, the surface brightness limit may cause residual biases for
the largest galaxies at a given luminosity.  In order to avoid these
biases, we further restrict the bins considered to those which are
$1$~mag~arcsec$^{-1}$ brighter than the surface brightness limit,
$1$~mag brighter than the magnitude limit, and with size greater than
twice the angular resolution at each redshift.  The resulting plot
should be unbiased, or as close as can be reasonably achieved given
the available data.

\begin{figure}
\centering
\includegraphics[height=0.475\textwidth,angle=270]{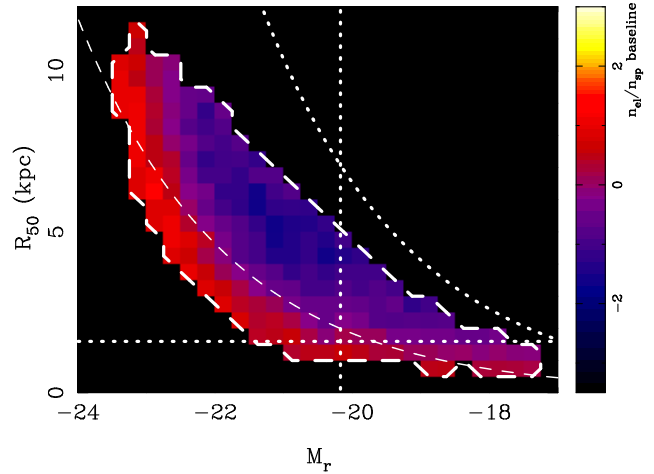}
\caption{\label{fig:ratio_baseline} Baseline estimate of the local
  unbiased early-type to spiral ratio versus luminosity and size. The
  dotted lines indicate the $r=17.77$ apparent magnitude limit,
  $\mu_{50,r}=23$ mag~arcsec$^{-2}$ apparent surface brightness limit,
  and the physical scale corresponding to an angle of $1$ arcsec, at
  the $z = 0.085$ upper redshift limit of the analysis samples (see
  Sec.~\ref{sec:basicdata}). The
  thin dashed curve indicates an apparent surface brightness of
  $\mu_{50,r}=20.25$ mag~arcsec$^{-2}$.  This approximately separates
  early-types and spirals, but clearly a simple surface brightness
  dependence is not sufficient to describe all the behaviour in this
  plot.}
\end{figure}

Taking this approach we obtain a baseline estimate of the unbiased
early-type to spiral ratio versus luminosity and size at $z \sim 0$,
shown in Fig.~\ref{fig:ratio_baseline}.  This figure clearly shows a
region of luminosity--size space dominated by early-types (lower-left)
and another dominated by spirals (upper-right), with a fairly sharp,
curved transition between the two.  There are a small number of
galaxies in the Galaxy Zoo \sample{full sample} which are located in an area of
luminosity--size space for which we do not have a direct estimate of
the local early-type to spiral ratio, due to a combination of the low
numbers of objects in these areas and their excision to avoid
including potentially biased regions in the baseline estimate.  In
order to extrapolate to these regions, as well as generally removing
noise and reducing the impact of residual bias, it is advantageous to
fit a smooth function to the binned baseline estimate.  

Motivated by the observed behaviour of the early-type to spiral ratio
versus luminosity and size, after trying a variety of functions we
choose the following to fit the local baseline estimate:
\begin{equation}
  \frac{n_{\rmn{el}}}{n_{\rmn{sp}}} = \frac{p_1}{1 + \exp\left(
      \frac{s_1(R_{50}) - M_r}{s_2(R_{50})}\right)} + p_2,
\end{equation}
This gives a smooth step function in $M_r$ with position and width varying
with $R_{50}$ as
\begin{eqnarray}
s_1(R_{50}) &=& q_1^{-(q_2 + q_3 {R_{50}}^{q_4})} + q_5,\nonumber \\
\rmn{and} &&\\
s_2(R_{50}) &=& r_1 + r_2 (s_1(R_{50}) - q_5), \nonumber
\end{eqnarray}
respectively.

\begin{figure}
\centering
\includegraphics[height=0.475\textwidth,angle=270]{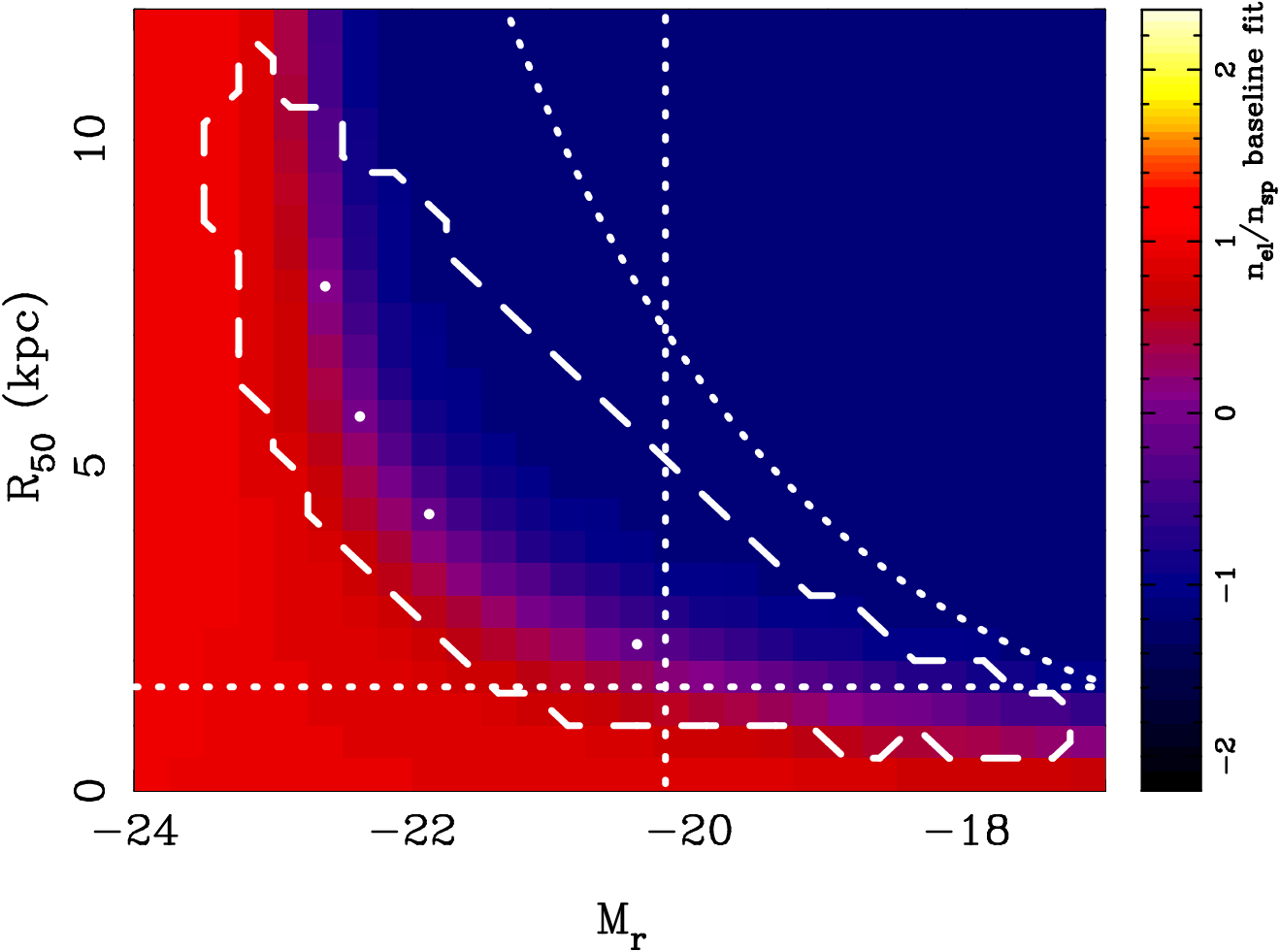}
\caption{\label{fig:ratio_baseline_fit} Fitted baseline estimate of
  the local unbiased early-type to spiral ratio versus luminosity and
  size.  Dotted lines are as in Fig.~\ref{fig:ratio_baseline}.  The
  four white dots indicate the bins plotted in
  Fig.~\ref{fig:baseline_correction_zfunc}.}
\end{figure}

This fit, shown in Fig.~\ref{fig:ratio_baseline_fit}, provides a smooth
local baseline estimate against we can compare the raw
early-type to spiral ratio versus luminosity and size as a function of
redshift.  In this manner we can derive a correction that will remove
the majority of the classification bias, including all of its redshift
dependence over our sample.  This method will also remove any
evolution effects in this quantity, if present over the limited
redshift range of the sample.

To estimate the correction, $C(M_r, R_{50}, z)$, we simply determine
the difference between $n_{\rmn{el}}/n_{\rmn{sp}}$ in each redshift
bin and the baseline just determined.  This correction is plotted for
three example redshift bins in
Fig.~\ref{fig:baseline_correction_zbins}.  The dependence of the
correction on redshift is shown for a few example bins in
Fig.~\ref{fig:baseline_correction_zfunc}.

\begin{figure*}
\centering
\includegraphics[scale=0.275,angle=270]{figA7_left.ps}
\hfill
\includegraphics[scale=0.275,angle=270]{figA7_middle.ps}
\hfill
\includegraphics[scale=0.275,angle=270]{figA7_right.ps}
\caption{\label{fig:baseline_correction_zbins} Difference between the
  raw $\log_{10}(n_{el}/n_{sp})$ in three example redshift bins and
  the baseline.  This is the adjustment that must be applied to the value
  in each luminosity--size--redshift bin to remove the classification
  bias.  The colour scale is shown by the bar on the right.  The
  region enclosed by the dashed line is determined from the data,
  while outside this region the adjustment is estimated from the
  nearest well-determined bins.  Only the region containing significant
  galaxy counts is coloured.}
\end{figure*}

\begin{figure}
\centering
\includegraphics[height=0.475\textwidth,angle=270]{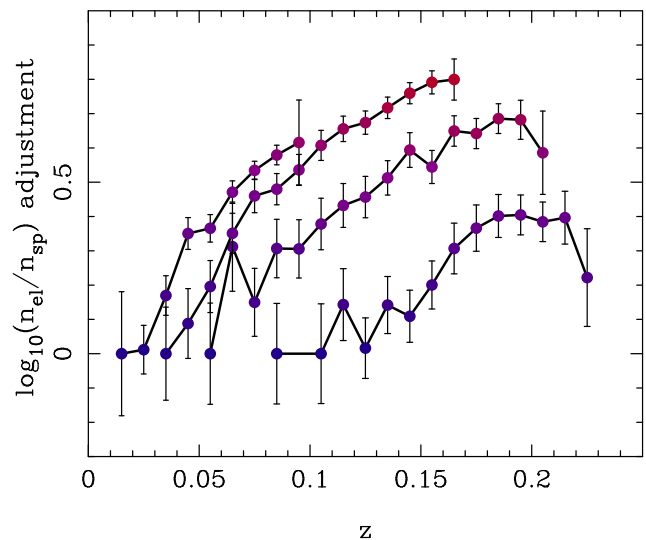}
\caption{\label{fig:baseline_correction_zfunc} Redshift dependence of
  the correction to $\log_{10}(n_{el}/n_{sp})$ for a few example
  luminosity--size bins, at $(M_r, R_{50}) = (-22.625, 7.75),
  (-22.375, 5.75), (-21.875, 4.25), (-20.375, 2.25)$, as indicated by
  the white dots in Fig.~\ref{fig:ratio_baseline_fit}. The colours of
  the points corresponds to the colour scale
  Fig.~\ref{fig:baseline_correction_zbins}.}
\end{figure}

At this point we have a baseline correction in bins of luminosity,
size and redshift, $C(M_r, R_{50}, z)$. A small fraction of galaxies
lie in regions of parameter space containing insufficient objects to
determine a direct correction.  For galaxies in these bins we use the
mean correction for the nearest neighbours within 1.2 times the
distance to the nearest bin centre.  This has the advantage of
averaging over bins when there are several at similar distances.  This
extrapolation is only valid close to the region which has a directly
determined correction, but it only needs to be, as there are very few
galaxies far from this region.

\subsection{De-biased Galaxy Zoo samples}
\label{sec:debiased_sample}

The baseline correction derived above can now be directly applied to
de-bias the raw type likelihoods.  To obtain the de-biased type
likelihoods we adjust the raw likelihoods of each galaxy as
\begin{eqnarray}
p_{\rmn{el,adj}} &=& \frac{1}{1\big/\big(\frac{p_{\rmn{el}}}{p_{\rmn{sp}}}\big)_{\rmn{adj}} + \frac{p_x}{p_{\rmn{el}}} + 1},\\
p_{\rmn{sp,adj}} &=& \frac{1}{\big(\frac{p_{\rmn{el}}}{p_{\rmn{sp}}}\big)_{\rmn{adj}} + \frac{p_x}{p_{\rmn{sp}}} + 1}, \nonumber
\end{eqnarray}
where
\begin{equation}
{\scriptstyle\left(\frac{p_{\rmn{el}}}{p_{\rmn{sp}}}\right)_{\rmn{\!adj}}} = {\scriptstyle\left(\frac{p_{\rmn{el}}}{p_{\rmn{sp}}}\right)_{\rmn{\!raw}}} \Big/ 10^{C(M_r, R_{50}, z)} \nonumber
\end{equation}
and
$p_x = 1 - p_{\rmn{el}} - p_{\rmn{sp}}$.

\begin{figure}
\centering
\includegraphics[height=0.475\textwidth,angle=270]{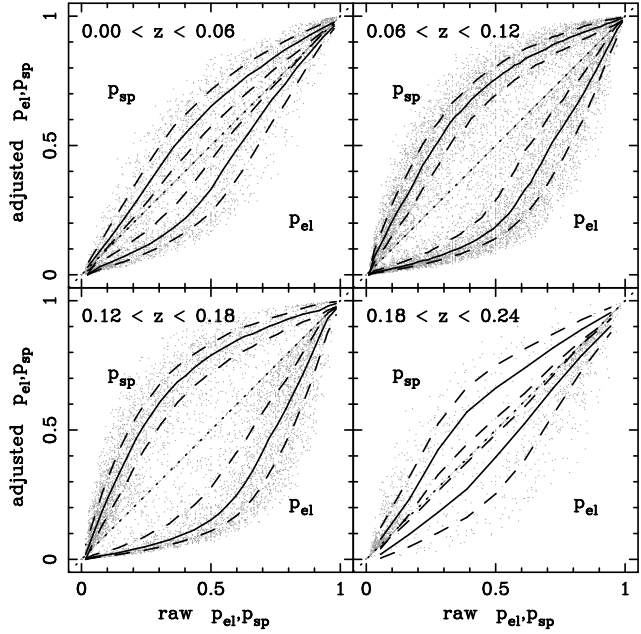}
\caption{\label{fig:p_correction} De-biased type likelihoods
  versus the raw values direct from the Galaxy Zoo classifications.
  The dotted diagonal line indicates the one-to-one relation.  Points
  above and below this line refer to the spiral and early-type
  likelihoods, respectively.  The solid and dashed lines trace the
  median and quartiles or the points.}
\end{figure}

This effect of this adjustment is shown in
Fig.~\ref{fig:p_correction}.  The overall effect is to lower the
early-type likelihoods, particularly for objects where there is some
indication that the object shows spiral features in the raw
likelihood.  The largest effect is for galaxies around the median
redshift of our \sample{full sample}.  For lower redshifts most galaxies are
well classified and need little adjustment.  For higher redshifts only
the most luminous objects are selected, the majority of which are
truly early-types and therefore do not suffer from the bias.

The success of our de-biasing procedure is demonstrated by
Figs.~\ref{fig:zbin_weights_corrected} \&
\ref{fig:morph-density_zbins} in the main body of this paper, which
show that the method works as expected.

\section{'Fingers of God' correction}
\label{sec:FoG}

The high velocity dispersions in galaxy groups lead one to grossly
overestimate the line-of-sight distances between the group centre and
its member galaxies, when a cosmological conversion between redshift
and distance is assumed.  We correct for this effect by reducing the
distance to the group centre for galaxies with redshifts consistent
with the group redshift and velocity dispersion, and close to the
group in projected distance.  Specifically, we determine the
line-of-sight comoving distance between the galaxy and group centre as
\begin{equation}
d_{\rmn{los}} =  
d_{\rmn{c}} \left(1 - e^{-A (|z_{\rmn{gal}} - z_{\rmn{grp}}|/{\zeta})^B} 
                      e^{-C (\max(R_{\rmn{vir}}, d_{\rmn{pro}})-R_{\rmn{vir}})^D}\right),
\end{equation}
where $A$, $B$, $C$ and $D$ are chosen to ensure the following
behaviour.  For galaxies with projected distances $\le R_{\rmn{vir}}$:
$d_{\rmn{los}} = 0$ when $z_{\rmn{gal}} = z_{\rmn{grp}}$;
$d_{\rmn{los}} = R_{\rmn{vir}}$ when $|z_{\rmn{gal}} - z_{\rmn{grp}}|
= \zeta$; and the adjustment is less than 1 per cent when
$|z_{\rmn{gal}} - z_{\rmn{grp}}| \ge 2 \zeta$.  Here, $\zeta = 3
\sigma_{\rmn{grp}} (1+z_{\rmn{grp}}) / c_0$ is the redshift difference
corresponding to three times the group velocity dispersion,
$d_{\rmn{c}}$ is the comoving distance between the galaxy and group
redshifts assuming only cosmological motions, and $d_{\rmn{pro}}$ is
the projected distance between the galaxy and the group's brightest
member assuming they both lie at the group cosmological redshift.  For
galaxies with $d_{\rmn{pro}}> R_{\rmn{vir}}$ the adjustment is
decreased to 90 per cent of its $\le R_{\rmn{vir}}$ value by
$d_{\rmn{pro}} = 2 R_{\rmn{vir}}$ and 1 per cent of its $\le
R_{\rmn{vir}}$ value by $d_{\rmn{pro}} = 5 R_{\rmn{vir}}$.

\begin{figure}
\centering
\includegraphics[height=0.475\textwidth,angle=270]{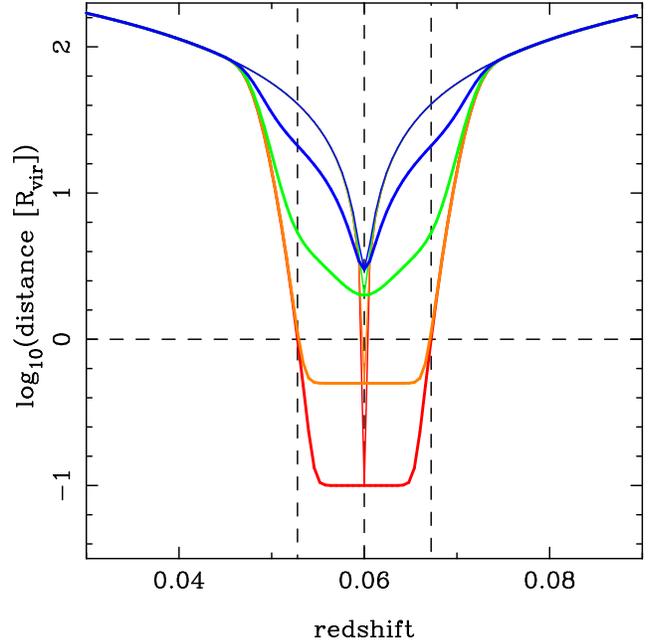}
\caption{\label{fig:normdist_correction} Illustration of the `Fingers
  of God' suppression method.  Coloured lines show the relation
  between redshift and the logarithmic distance, in virial radii, to a
  group with properties typical of those in the C4 catalogue: $z =
  0.06$, $\sigma = 680$ km~s$^{-1}$ and $R_{\rmn{vir}} = 0.74$ Mpc.
  Thin, solid lines plot the relation assuming no peculiar velocities.
  In this case, small departures from the cluster redshift imply large
  distances.  However, in reality, peculiar velocities impart
  significant redshift differences on galaxies that are actually
  members of the group, and so at small distances from one another.
  Thick solid lines show the relation with the `Fingers of God'
  suppression method, which attempts to account for this.  The red,
  orange, green and blue lines indicate the effect on galaxies at
  projected separations of 0.1, 0.5, 2, and 3 $R_{\rmn{vir}}$,
  respectively.  }
\end{figure}

\begin{figure}
\centering
\includegraphics[height=0.475\textwidth,angle=270]{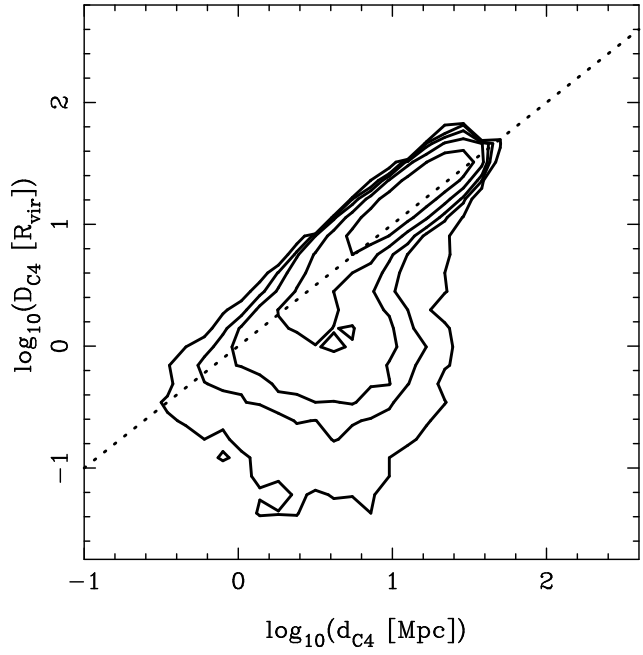}
\caption{\label{fig:c4normdist_c4dist} Redshift-space distance of each
  galaxy to its nearest C4 group, $d_{\rmn{C4}}$, versus a measure of
  this distance incorporating a correction for the `Fingers of God'
  effect and normalised by virial radius, $D_{\rmn{C4}}$.  Contours
  enclose 50, 75, 90, 95 and 99 per cent of galaxies in the
  \sample{luminosity-limited sample}. \label{lastpage}}
\end{figure}

We furthermore normalise the corrected distances by $R_{\rmn{vir}}$ of
the nearest group to account for scaling of any potential influence
with group size.  Figure~\ref{fig:normdist_correction} illustrates the
effect of the correction for model galaxies in a typical C4 group.  A
comparison of the redshift-space distances, $d_{\rmn{C4}} =
\sqrt{d_{\rmn{c}}^2 + d_{\rmn{pro}}^2}$ and the normalised, corrected
distances, $D_{\rmn{C4}} = \sqrt{d_{\rmn{los}}^2 +
  d_{\rmn{pro}}^2} \Big/ R_{\rmn{vir}}$, is shown in
Fig.~\ref{fig:c4normdist_c4dist}.

\end{document}